\documentclass[letterpaper, preprint, fleqn, floatfix, superscriptaddress]{achemso}

\usepackage{euscript, units, amsfonts, graphicx, dcolumn, fancyhdr,  mathrsfs, upgreek, amssymb, amsmath, pifont, epstopdf, color, subfigure}
\usepackage{hyperref}

\graphicspath{{./Figs2/}} %Setting the graphicspath

\usepackage{tikz,pgfplots}
\usepackage{tikz-3dplot}
\usepackage{rotating}
\usetikzlibrary{matrix,positioning,arrows,shapes,patterns,fpu}

\usepackage{placeins} %for floatbarrier

\usepackage{algorithm}
\usepackage{algpseudocode}

\newcommand{\revtt}[1]{{\color{black}#1}}

\def\J{{\mathcal{J}}}
\def\I{{\mathcal{I}}}

\floatname{algorithm}{Algorithm}

\newcommand\omicron{o}

\newcommand{\alphaap}{{\alpha_{a,+}}}
\newcommand{\alphaan}{{\alpha_{a,-}}}
\newcommand{\alphaaphat}{{\widehat{\alpha}_{a,+}}}
\newcommand{\alphaanhat}{{\widehat{\alpha}_{a,-}}}

\setkeys{acs}{maxauthors=10}
\setkeys{acs}{etalmode=truncate}

\usepackage{natbib}

\usepackage{xr}
\title{Learning Optimal Forms of Constitutive Relations Characterizing Ion Intercalation from Data in Mathematical Models of Lithium-ion Batteries}

\author{Lindsey Daniels}
\affiliation{Department of Mathematics and Statistics, McMaster University, Hamilton, Ontario, Canada L8S 4K1}

\author{Smita Sahu}
\affiliation{The Faraday Institution, Quad One, Becquerel Avenue, Harwell Campus, Didcot, OX11 0RA, United Kingdom}
\altaffiliation{School of Mathematics and Physics, University of Portsmouth, Lion Terrace, PO1 3HF, United Kingdom}

\author{Kevin J. Sanders}
\affiliation{Department of Chemistry, McMaster University, Hamilton, Ontario, L8S 4L8, Canada}

\author{Gillian R. Goward}
\affiliation{Department of Chemistry, McMaster University, Hamilton, Ontario, L8S 4L8, Canada}

\author{Jamie M. Foster}
\affiliation{The Faraday Institution, Quad One, Becquerel Avenue, Harwell Campus, Didcot, OX11 0RA, United Kingdom}
\altaffiliation{School of Mathematics and Physics, University of Portsmouth, Lion Terrace, PO1 3HF, United Kingdom}

\author{Bartosz Protas}
\affiliation{Department of Mathematics and Statistics, McMaster University, Hamilton, Ontario, Canada L8S 4K1}
\email{bprotas@mcmaster.ca  *}

\date{\today}

\SectionNumbersOn

\begin{document}

%%%%%%%%%%%%%%%%%%%%%%%%%%%%%%%%%%%%%%%%%%%
%Abstract
%%%%%%%%%%%%%%%%%%%%%%%%%%%%%%%%%%%%%%%%%%%
\begin{abstract}
  Most mathematical models of the transport of charged species in
  battery electrodes require a constitutive relation describing
  intercalation of Lithium, which is a reversible process taking place
  on the interface between the electrolyte and active particle. The
  most commonly used model is the Butler-Volmer relation, which gives
  the current density as a product of two expressions: {one, the
    exchange current,} depends on Lithium concentration only whereas
  the other expression depends on both Lithium concentration and on
  the overpotential.  We consider an inverse problem where an optimal
  form of the exchange current density is inferred, subject to minimum
  assumptions, from experimental voltage curves.  This inverse problem
  is recast as an optimization problem in which the least-squares
  error functional is minimized with a suitable Sobolev gradient
  approach. The proposed method is thoroughly validated and we also
  quantify the reconstruction uncertainty.  Finally, we identify the
  universal features in the constitutive relations inferred from data
  obtained during charging and discharging at different C-rates and
  discuss how these features differ from the behaviour predicted by
  the standard Butler-Volmer relation. {We also identify possible
    limitations of the proposed approach, mostly related to
    uncertainties inherent in the material properties assumed known in
    the inverse problem.}  Our approach can be used to systematically
  improve the accuracy of mathematical models employed to describe
  Li-ion batteries as well as other systems relying on the
  Butler-Volmer relation.
\end{abstract}

\maketitle \thispagestyle{plain}

%\tableofcontents
%\newpage

%%%%%%%%%%%%%%%%%%%%%%%%%%%%%%%%%%%%%%%%%%%
%%%%%%%%%%%%%%%%%%%%%%%%%%%%%%%%%%%%%%%%%%%
%Introduction

%%%%%%%%%%%%%%%%%%%%%%%%%%%%%%%%%%%%%%%%%%%
\section{Introduction}
\label{sec:intro}

The modelling of many physical processes relies on fundamental
understanding of the interface between two {interacting
  materials (or phases)}. It may also require the knowledge of
specific material properties, which can be challenging to obtain or
measure \cite{wang2022review}.  Interfacial models can be complex due
to the coupling of different physical processes at the interface which
makes probing their parametric dependence more difficult. In an effort
to aid in the parametrization challenge, we propose an inverse
modelling technique to optimally reconstruct constitutive
relationships describing an important class of interfacial processes
from data. Inverse modelling is an approach used to learn optimal forms of
material properties from data by combining a mathematical model of the
system with experimental measurements.  

The present work proposes a computational approach alternative
  to machine learning techniques which have recently become
  popular\cite{lv2022machine,houchins2020accurate,chemali2018state}.
  Instead, we use tools from calculus of variations to design a
  numerical procedure to recover an optimal form of the constitutive
  relation, where ``optimal'' means that the reconstructed
  constitutive relation achieves the best possible fit to the data.
  Such methods have already proved useful for solving a range of
  inverse problems in electrochemistry, which is an area abounding
  with systems whose material properties are notoriously difficult to
  measure
  directly\cite{kgnhllf12,sethurajan2015accurate,richardson2018effect,sethurajan2019dendrites,escalante2020discerning,escalante2021uncertainty,wang2022review}.
  Many of these applications are driven by recent advances in the
  mathematical modelling of transport processes in Lithium-ion
  (Li-ion) batteries, which are crucial to the development of electric
  vehicles and customer
  electronics\cite{chen2019review,blomgren2016development}.  While
  most of the work on inverse modelling in electrochemistry has
  focused on inferring material properties defined in the bulk {of the
    electrolyte or active material}, we extend this approach to
  constitutive relations characterizing interfacial phenomena. More
  specifically, we focus on constitutive relations describing the
  intercalation of ions from the electrolyte into the active material,
  which play a central role in modelling electrochemical
  systems\cite{nt04}, but also find applications in other area of
  chemistry and physics as well as in biology. For example, such
  interfacial models have been used to relate the current density to
  the overpotential in bio-anode polarization
  curve\cite{hamelers2011butler}, to calculate the reaction rate from
  voltammograms\cite{compton2011voltammetry}, to describe the
  electrode dynamics in supercapacitors \cite{oyarzun2021unraveling},
  and to describe the electrochemistry of a neuron
  \cite{hodgkin1952quantitative}.

An inverse problem is usually formulated to minimize the least-squares
error between the values predicted by the model and the experimentally
measured quantities.  When studying a distributed system (with
dependence on space and/or time) and when the reconstructed property
is a function, rather than a set of scalars, there are two main types
of inverse problems. In the more traditional problems, the
reconstructed material property is assumed to be a function of {\em
  independent} variables, usually space coordinate and/or
time\cite{banks2012estimation,Tarantola2005,i98}.  On the other hand,
in less standard modern inverse
problems\cite{bvp11,sethurajan2015accurate,doi:10.1002/jcc.25759,bukshtynov2013889,escalante2020discerning,escalante2021uncertainty},
the inferred properties are functions of the {\em dependent} variables
in the system.  The interfacial inverse problem we consider here
belongs to this second category.

Inverse problems tend to be ill-posed, as they usually do not admit a
unique exact solution, but rather a large number (often infinite) of
approximate solutions\cite{Tarantola2005}.  Variation in experimental
measurements together with weak dependence of the model on some
parameters can result in significant changes in the reconstructed
solution which can be amplified by experimental and numerical errors.
Yet, inverse modelling can help to elucidate optimal forms of
constitutive relations and assess the validity of different models, as
solutions violating thermodynamic principles can point to potential
inconsistencies in the models
used\cite{richardson2018effect,escalante2020discerning}.

The structure of the paper is as follows: in the next section we
discuss the Butler-Volmer model; in Section \ref{expsec}, the
experimental setup and the measurement data are introduced; Section
\ref{mathmodel} details the mathematical framework and optimization
approach, and outlines the sensitivity analysis; the reconstruction
results obtained with the inverse modelling approach are presented in
Section \ref{sec:results}; and finally, the conclusions of this work
are presented in Section \ref{conclusion}, whereas some additional
technical material is collected in three appendices.

\section{The Butler-Volmer Relation and its Generalization}
\label{sec:BV}

Below we state the Butler-Volmer relation in a form that is commonly
used, whereas the multiplicity of forms in which this relation is
encountered is discussed in a recent review\cite{dickinson2020butler}.
A complete derivation of the Butler-Volmer relation can be found in
the
literature\cite{latz2013thermodynamic,bard2001electrochemical,nt04}.
Hereafter, the subscripts/superscripts ``s'' and ``e'' will denote
quantities defined in the solid phase and in the electrolyte,
respectively.

The {net} current density $j$ is the difference between the backward and
forward current densities at the interface, and the dependence of this
quantity on electrochemical and thermodynamic parameters is given by
the Butler-Volmer relation
\begin{equation}
j =  Fk_0 c_{\rm{e}}^{\alpha_{\rm{a},m}} \left( c_{{\rm{s}},i,{\rm{max}}} - c_{{\rm{s}},i} \right)^{\alpha_{\rm{a},m}} c_{{\rm{s}},i}^{\alpha_{\rm{c},m}} \, \left[ e^{\frac{\alpha_{\rm{a},m} F}{RT}  \eta} - e^{\frac{-\alpha_{\rm{c},m} F}{RT}  \eta} \right], 
\quad \textrm{where} \quad \eta = \phi_{\rm{s}} - \phi_{\rm{e}} -  {U_{\rm{eq}}^i}(c_{{\rm{s}},i}), \label{eq:BV0}
\end{equation}
in which $\eta$ is the overpotential, $c_{{\rm{s}},i}$ and
$c_{\rm{e}}$ are the Lithium concentrations in the solid phase and in
the electrolyte with $i={a,c}$ indicating the anode or cathode
respectively, $\phi_{\rm{s}}$ and $\phi_{\rm{e}}$ are the potentials
in the solid phase and in the electrolyte (both measured with respect
to a Li reference electrode), whereas
${U_{\rm{eq}}^i}(c_{{\rm{s}},i})$ is the concentration-dependent
equilibrium potential of the electrode, $\alpha_{{a},m}$ and
$\alpha_{{c},m} = 1-\alpha_{{a},m}$ are the anodic and cathodic
charge transfer coefficients with $m = -,+$ indicating
  the anode or cathode respectively (we do not use {$a$} and {$c$}
  here to avoid repeated identical indices), $k_0$ is the reaction
rate {constant}, and $c_{{\rm{s}},i,{\rm{max}}}$ is the maximum
Lithium concentration in the solid.
% It is important to note that the Butler-Volmer relation \eqref{eq:BV0}
% does not account for more general interfacial phenomena, such as
% Lithium plating on the negative electrode surface or the diffusion and
% convection of intermediate reaction
% products\cite{richardson2020charge,bard2001electrochemical,nt04}.
As is evident from \eqref{eq:BV0}, the Butler-Volmer relation can be
naturally split into an algebraic factor depending on the Lithium
concentration only, typically referred to as the exchange current, and
the remaining exponential factor depending on both the potential
difference $\Delta\phi = \phi_{\rm{s}} - \phi_{\rm{e}}$ and the Lithium
concentration in the electrolyte, the latter dependence through the
equilibrium potential ${U_{\rm{eq}}^i}(c_{{\rm{s}},i})$.  This allows us to factor the
Butler-Volmer relation \eqref{eq:BV0} as
\begin{subequations}
\label{eq:BV1}
\begin{align}
j & = i_0(c_{{\rm{s}},i}) \, \psi(\Delta \phi, c_{{\rm{s}},i}),  \qquad \textrm{where} \label{eq:BV1f} \\
i_0(c_{{\rm{s}},i}) & = Fk_0 c_{\rm{e}}^{\alpha_{{a},m}} \left( c_{{\rm{s}},i,{\rm{max}}} - c_{{\rm{s}},i} \right)^{\alpha_{{a},m}} c_{{\rm{s}},i}^{\alpha_{{c},m}}, \label{eq:i0} \\
\psi(\Delta \phi,c_{{\rm{s}},i}) & = \exp\left[\frac{\alpha_{{a},m} F}{RT}\left(\Delta\phi - {U_{\rm{eq}}^i}(c_{{\rm{s}},i}) \right)\right] - \exp\left[-\frac{\alpha_{{c},m} F}{RT}\left(\Delta\phi - {U_{\rm{eq}}^i}(c_{{\rm{s}},i}) \right)\right]. \label{eq:psi}
\end{align}
\end{subequations}
While the exponential term \eqref{eq:psi} has had a long and
well-documented history in electrochemistry
\cite{fletcher2009tafel,latz2013thermodynamic,bard2001electrochemical},
additional assumptions are required to justify the form of the
concentration-dependent algebraic factor
\eqref{eq:i0}\cite{dickinson2020butler}. As will be discussed
  in more detail below, we focus on the dependence of the exchange
  current on $c_{{\rm{s}},i}$ and its dependence on $c_{\rm{e}}$ in
  \eqref{eq:i0} will be omitted so that our results
  are most applicable to situations in which the electrolyte
  concentration is approximately uniform (e.g., at relatively low
  currents or for thin electrodes).

In the present study, we thus consider a generalization of the
Butler-Volmer relation \eqref{eq:BV1} given
\begin{equation}
  j = \iota(c_{{\rm{s}},i}) \, \psi(\Delta \phi, c_{{\rm{s}},i}),
  \label{eq:BV}
\end{equation}
where $\psi(\Delta \phi, c_{{\rm{s}},i})$ is assumed known and given by
\eqref{eq:psi}, whereas the factor $i_0(c_{{\rm{s}},i})$, cf.~\eqref{eq:i0}, is
replaced by a general function $\iota(c_{{\rm{s}},i})$ whose optimal (in a
mathematically precise sense to be specified below) form will be
inferred {from} standard electrochemical measurements. This inverse
problem will be solved using a computationally robust and
thermodynamically consistent numerical optimization procedure where
optimal functional forms of the factor $\iota(c_{{\rm{s}},i})$ will be sought in a
very general form subject to minimum only assumptions. These
assumptions are\cite{richardson2020charge}:
\begin{itemize}
\item[A1.] the function $\iota = \iota(c_{{\rm{s}},i})$ must be sufficiently smooth
  (this will be made precise in Section \ref{sec:inverse}),

\item[A2.] when the Lithium concentration in the electrode goes to
  zero ($c_{{\rm{s}},i} \rightarrow 0$), the Lithium flux out of the active
  material must vanish to prevent nonphysical negative Lithium
  concentrations in the electrode, % while still allowing the electrode to relithiate, 
  and

\item[A3.] when the electrode is saturated ($c_{{\rm{s}},i} \rightarrow
  c_{{\rm{s}},i,{\rm{max}}}$), the Lithium flux into the active material must also
  vanish to prevent nonphysical overfilling of Lithium.%, while still allowing the electrode to delithiate.
\end{itemize}

These represent constraints on the behaviour of the factor
$\iota(c_{{\rm{s}},i})$ for small ($c_{{\rm{s}},i} \rightarrow 0$) and
large ($c_{{\rm{s}},i} \rightarrow c_{{\rm{s}},i,{\rm{max}}}$)
concentrations.  We note that this limiting behaviour is
  identical as in the standard Butler-Volmer model, cf.~\eqref{eq:i0}.
  However, since the equilibrium potential
  $U_{\rm{eq}}^i(c_{{\rm{s}},i})$ is unbounded in these limits,
  cf.~Supporting Information (Appendix \ref{sec:appBC}), the minimum
  and maximum concentrations of 0 and $c_{{\rm{s}},i,{\rm{max}}}$ are
  in practice never attained\cite{richardson2020charge}.  Application
of the proposed approach to {voltage measurements performed during
  charge and discharge experiments} produces optimal forms of the
exchange current $\iota(c_{{\rm{s}},i})$, which are consistent between
{charge and discharge} and reveal features making them quite different
from expression \eqref{eq:i0}. We also characterize the uncertainty of
the reconstructions and compare our results to a simpler approach
where the inverse problem is solved by calibrating a small number of
scalar parameters in expression \eqref{eq:i0}.

%\end{section}

%%%%%%%%%%%%%%%%%%%%%%%%%%%%%%%%%%%%%%%%%%%
%%%%%%%%%%%%%%%%%%%%%%%%%%%%%%%%%%%%%%%%%%%
%Experimental

%%%%%%%%%%%%%%%%%%%%%%%%%%%%%%%%%%%%%%%%%%%
\section{Experimental Datasets}
\label{expsec}

%In the formulation of our inverse problems, we will utilize several
%experimental datasets with measurements of the total voltage as a
%function of time obtained in electrochemical cells {similar} to those used in actual Li-ion cell {batteries}.  
In the formulation of our inverse problem, we will use several datasets
from experiments performed on Li-ion batteries under galvanostatic
conditions during both charge and discharge at different
rates\cite{sanders2021transient,chen2020development}.  Below, we first
describe the two experimental setups we considered and present the
data obtained.  Then, we demonstrate that, for all datasets, the
fraction of the signal (the time-dependent voltage) containing
information about the interfacial reactions is sufficiently large both
to be used in inverse modelling and for the main assumptions
underlying our mathematical model to be satisfied. Within the
electrolyte, there is a potential drop between the anode and the
cathode, which is accounted for when solving the complete
Doyle-Fuller-Newman (DFN) model\cite{nt04}.  {However,
  this potential drop is often either neglected in the standard
  Single-Particle model (SPM), cf.~Section \ref{sec:SPM}, based on the
  assumption that it is small when compared to the contributions from
  the electrode kinetics or is accounted for through a linearized
  model based on the assumption that the electrolyte can be considered
  ohmic if the current is
  small\cite{richardson2020generalised,escalante2021uncertainty}. }
The SPM is {typically} an accurate reduction of the DFN
model for C-rates up to approximately $1$C
\cite{escalante2021uncertainty} and given this well-known limitation
of the SPM, we focus here on small and moderate charge/discharge rates
only.

In order to assess whether the experimental datasets are consistent
with the SPM, we determine the corresponding time-dependent
concentration and potential profiles in the cells based on solutions
of the complete DFN model\cite{nt04} which are obtained using the
suite DandeLiion\cite{korotkin2021dandeliion}.  This allows us to
determine the relevant voltage drops for each component of the cell.
Since the potential drop across the electrolyte is a function of both
space (i.e., the distance $x$ from the anode) and the (dis)charge
time, we first compute the total drop as $\phi_{\rm{e}}(L,t) -\phi_{\rm{e}}(0,t)$,
where $L$ is the total length of the cell {thereby
  taking a maximal estimate of the potential drop across the
  electrolyte,} and then take the time average.

\subsection{Slow Charge Rates}
\label{sec:dataslow}

\begin{figure}[h!]
\includegraphics[scale = 0.3]{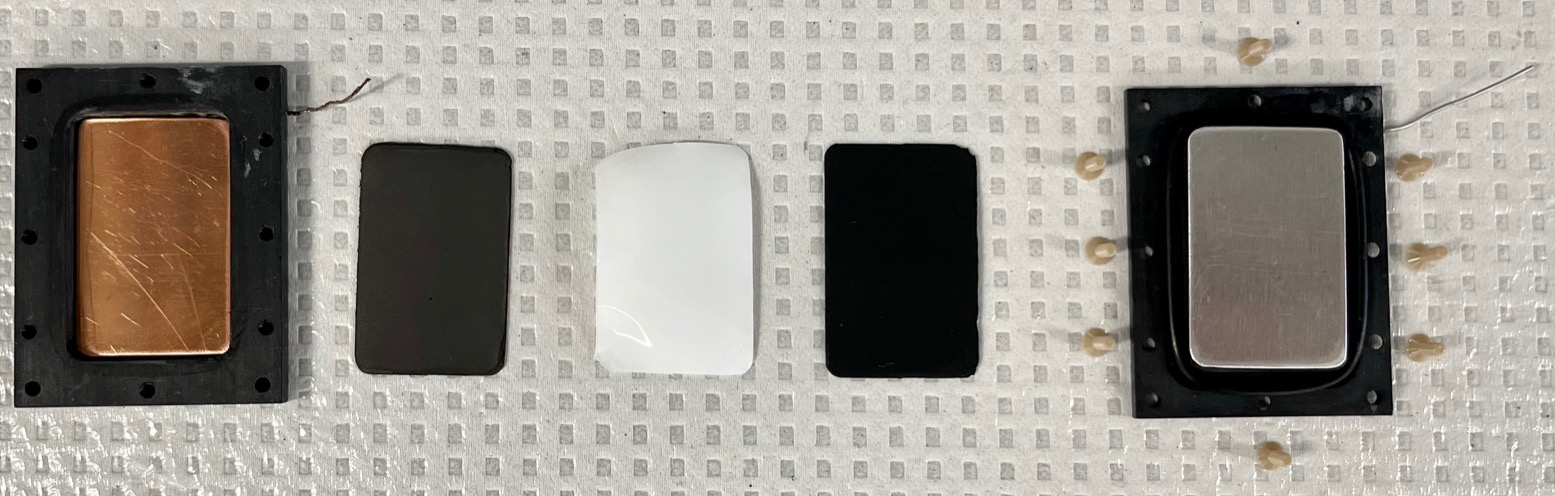}
\caption{Image of the cell assembly.  From left to right: bottom part
  of the cartridge with copper (Cu) metal current collector plate,
  graphite anode, Celgard $2325$ separator, NMC$622$ cathode, and top
  part of the cartridge with aluminum (Al) current collector
  plate. The EPDM o-ring used to provide a gas-tight seal and $8$ M$2$
  PEEK screws used to close the cartridge are also shown around the
  top part on the right-hand side.
  \label{Kevin_Experiment_pics} }
\end{figure}

For slow charge rates, we consider two datasets corresponding to
C-rates of C/10 and C/3.  These datasets were obtained using an
electrochemical cell constructed from an unused commercial Lithium-ion
battery containing a double-sided coated graphite anode and
LiNi$_{0.6}$Mn$_{0.2}${Co}$_{0.2}$O$_2$ (NMC$622$)
cathode \cite{sanders2021transient}. The commercial battery was
disassembled, the electrodes were washed, cut, and dried, and
reassembled in a replaceable cartridge-like battery container that
mimics the single-layer pouch cell configuration.  The assembled cell
consists of a copper (Cu) current collector plate, graphite anode,
Celgard $2325$ separator, NMC$622$ cathode, and an aluminum (Al)
current collector plate.  The assembled cell was rested for $24$ hours
before being subject to cycling protocols and experimental
measurements.  Figure \ref{Kevin_Experiment_pics} shows each
individual component of the cell before assembly. Full experimental
details and set up can be found by Sanders et al
\cite{sanders2021transient}.

\begin{figure}[h!]
  \mbox{
    \subfigure[]{\includegraphics[width=0.5\textwidth]{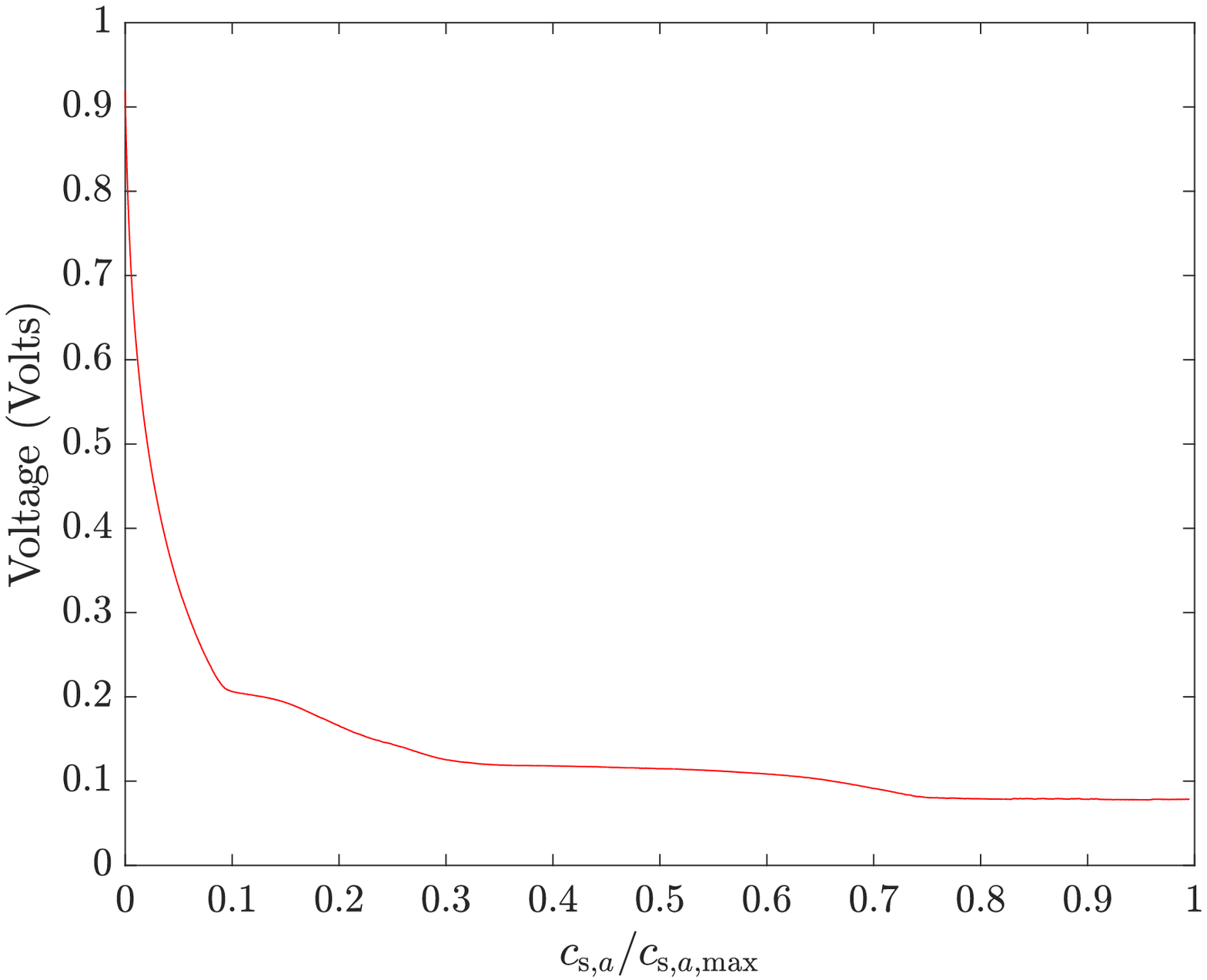}}\quad
\subfigure[]{\includegraphics[width=0.5\textwidth]{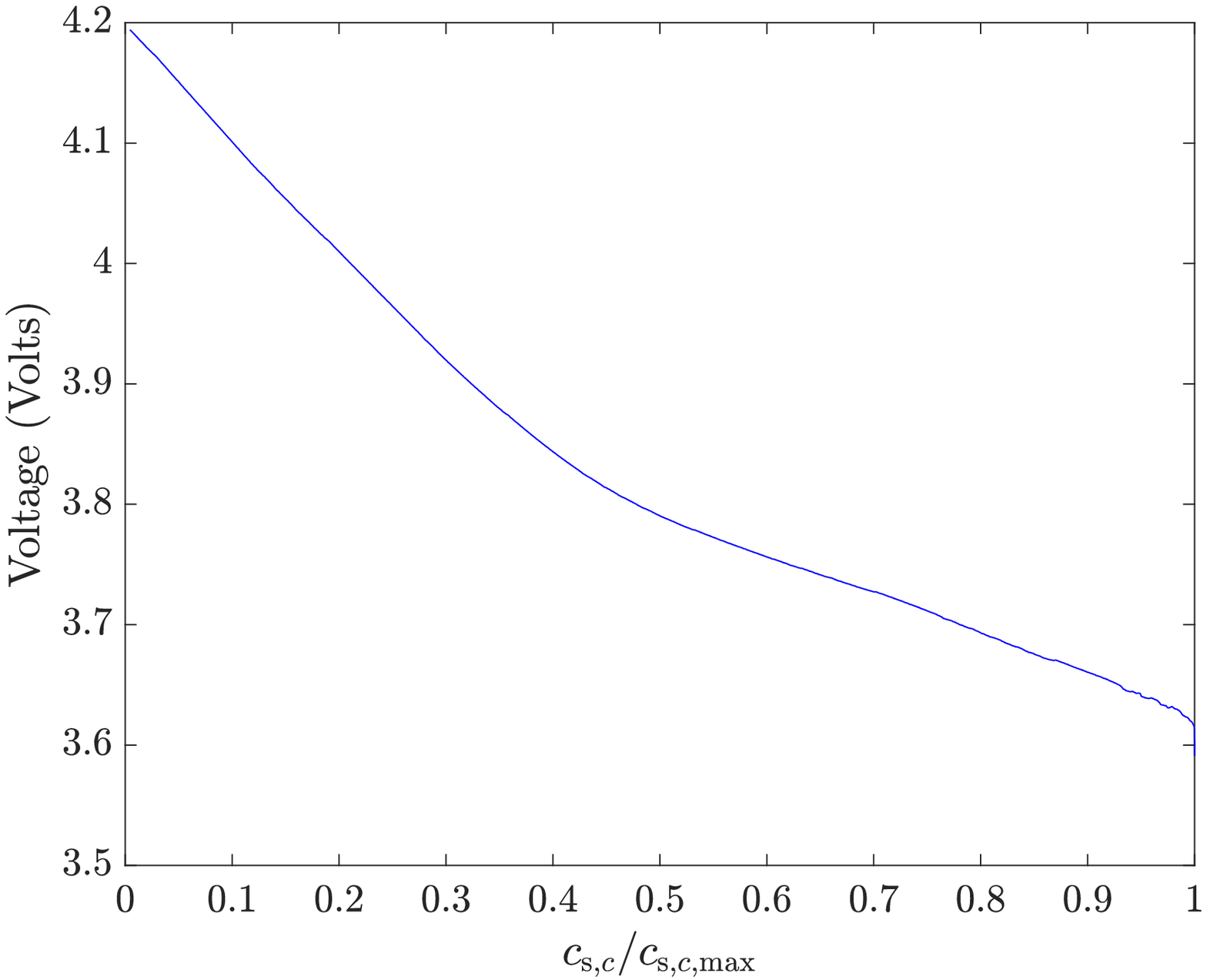}}} 
\caption{The equilibrium voltage curve for (a) the graphite anode and
  (b) for the NMC$622$ cathode in the cell used in the experiments
  described in Section \ref{sec:dataslow}\cite{sanders2021transient}.
  \label{KevinUeq} }
\end{figure}

In Figure \ref{KevinUeq}, the experimental voltage curves for the
anode and cathode equilibrium voltages are plotted against the
{normalized} concentration in the two electrodes.  The graphite voltage
curve in Figure \ref{KevinUeq}a was obtained by charging a half-cell
with a graphite electrode to the maximum voltage at a rate of C$/140$.
The voltage curve for the NMC$622$ electrode in Figure \ref{KevinUeq}b
was obtained by charging a half-cell with an NMC$622$ electrode to
$4.2$V at a rate of C$/140$. These measurements will be used as the
equilibrium potentials in our model.  {The voltage curves obtained with
the C-rates of C$/10$ and C$/3$ will be shown in Section
\ref{sec:results} where we discuss our results.}
%{In Figure \ref{KevinUdrop}, the experimental voltage curves for C-rates of C$/10$ and C$/3$, and the drop across the electrodes, c.f. Figure \ref{KevinUeq}. As expected, the experimental voltage curves have a shape similar to the equilibrium voltage drop.}
%\begin{figure}[h!]
%\includegraphics[scale = 0.5]{Kevin_voltages_vs_norm_time.eps}
%\caption{{The measured cell voltage at C-rates of C$/10$ and C$/3$, and the associated equilibrium drop across the electrodes used in the experiments
 % described in Section \ref{sec:dataslow}\cite{sanders2021transient}.}   \label{KevinUdrop} }
%\end{figure}

\FloatBarrier Table \ref{Kevintable} details the percentage breakdown
of the time-averaged voltage drops obtained by solving numerically the
complete DFN model using the material parameters listed in Tables
\ref{tab:kevinparams1} and \ref{tab:kevinparams2} in Supporting
Information (Appendix \ref{sec:expkevinpara}). In all cases, the
overpotentials $\eta_a$ and $\eta_c$ have approximately the same share
of the total potential as the time-averaged drop across the
electrolyte.  %These overpotentials are however much smaller, albeit nonnegligible, relative to the equilibrium potentials ($U_{\rm{eq}}$).
Table \ref{Kevintable} also shows the ratios between the different
potential drops corresponding to the two C-rates considered. We see
that for all potential drops (except for ${V_c - V_a}$, as expected)
these ratios are close to the ratio of the two C-rates, which is
$10/3$, confirming that the response is close to linear at small
C-rates.
 %This confirms the consistency of the analysis.

%\begin{figure}[h!]
%\includegraphics[scale=0.45]{Kevin_bar_june29.eps}
%\caption{Breakdown of the averaged anode and cathode overpotentials, $\eta_a$
%  and $\eta_c$, and the drop across the electrolyte in the experiments described in Section
%  \ref{sec:dataslow}\cite{sanders2021transient}. Since it is quite large, the share of the total voltage
%  corresponding to the total equilibrium potential ($V_c - V_a$) is not
%  shown to help elucidate the other potential drops. \label{Kevinpie}}
%\end{figure}

\begin{table}[t!]
%\begin{table}
\begin{tabular}{c c c c c c c c}
Charge rate & $V_c - V_a$ & $\eta_a$ & $\eta_c$ & Electrolyte  \\ \hline
 C/$10$ & $      99.1755$ &   $ 0.2368$ &   $ 0.3344$ & $ 0.2533$ \\
 C/$3$ &   $    97.1927$ &   $1.0208$ &   $ 1.0195$ &  $   0.7670$ \\ \hline
 Ratio & & & & \\ \hline
  C/$3$:C/$10$ & $   0.9800$ & $4.3103$ &   $ 3.0492$ &  $  3.0281$
\end{tabular}
\caption{Percentage breakdowns of the time-averaged potential drops
  corresponding to the main components of the cell used in the
  experiments described in Section
  \ref{sec:dataslow}\cite{sanders2021transient}.  The results were
  obtained by solving the complete DFN model using the material
  parameters listed in Tables \ref{tab:kevinparams1} and
  \ref{tab:kevinparams2} in Supporting Information (Appendix \ref{sec:expkevinpara}).
  \label{Kevintable} }
\end{table}

\subsection{Moderate Discharge Rates}
\label{sec:datamoderate}

Experimental data for discharge rates of C$/2$ and $1$C were obtained
from Chen et al.\cite{chen2020development}. Their experiments utilized
a cylindrical LG M$50$$21700$ (LGM$50$) cell containing an
LiNi$_{0.8}$Mn$_{0.1}$O$_1$ (NMC$811$) cathode and a graphite anode.
A commercial cell was taken apart and assembled into two half cells to
characterize the respective electrode and cell parameters.  The
obtained electrodes were then assembled into a three-electrode full
cell configuration with a Celgard $2325$ separator and subjected to
cycling protocols under a variety of environments.  Details can be found in the work by
Chen et al.\cite{chen2020development}

Table \ref{Ferrantable} details the percentage breakdown of the
time-averaged voltage drops associated with the main components of the
cell which were obtained by solving numerically the complete DFN model
using the material parameters listed in Tables \ref{tab:Ferranparams1}
and \ref{tab:Ferranparams2} in Supporting Information (Appendix
\ref{sec:expferranpara}).  Table \ref{Ferrantable} also shows the
ratios between the different potential drops corresponding to the two
C-rates considered.  We see that when the C-rate increases from C$/2$
to $1$C, these ratios are approximately $2$ for all potential drops
(except for the equilibrium potential ${V_c - V_a}$, as expected),
which is consistent with the linear behaviour expected at low C-rates.
Chen et al.\cite{chen2020development} also obtained results at a
higher discharge rate of 2C, but since this regime is outside the
range of validity of the SPM, we will not consider this data in our
study.

% However, when the C-rate increases from $1$C to $2$C,
% these ratios are approximately $1.3$, indicating a deviation from the
% linear
% behaviour\cite{richardson2020generalised,richardson2020charge,schmalstieg2018full}.
% Because of this and also the fact that the SPM we use is
% typically considered valid for C-rates not exceeding $1$C, we will not
% consider the case with discharging performed at $2$C in this study.

%
%\begin{figure}[h!]
%\includegraphics[scale=0.45]{Ferran_bar_june29.eps}
%\caption{Breakdown of the averaged anode and cathode overpotentials, $\eta_a$
%  and $\eta_c$, and the drop across the electrolyte for the experiments described in Section
%  \ref{sec:datamoderate}\cite{chen2020development}. Since it is quite large, the share
%  of the total voltage corresponding to the equilibrium potential
%  ($V_c - V_a$) is not shown to help elucidate the other potential
%  drops. \label{Ferranpie}}
%\end{figure}

\begin{table}[t!]
\begin{tabular}{c c c c c c c c}
Charge rate & $V_c - V_a$ & $\eta_a$ & $\eta_c$& Electrolyte \\ \hline
 C/$2$ & $      97.2504$ &   $ 1.5916$ &   $ 0.1907$ & $ 0.9674$  \\
 $1$C &   $    94.8758$ &   $2.6239$ &   $ 0.4245$ &  $  2.0758$  \\
%  $2$C &   $    93.9959$ &   $2.6554$ &   $ 0.5261$ &  $ 2.8226$\\ 
\hline
 Ratio & & & & \\ \hline
$1$C:C/$2$ & $   0.9756$ & $1.6486$ &   $ 2.2262$ &  $  2.1459$   \\
% $2$C:$1$C  & $   0.9907$ & $1.0120$ &   $ 1.2394$ &  $  1.3597$  \\
\end{tabular}
\caption{Percentage breakdowns of the time-averaged potential drops
  corresponding to the main components of the cell used in the
  experiments described in Section
  \ref{sec:datamoderate}\cite{chen2020development}.  The results were
  obtained by solving the complete DFN model
  using the material parameters listed in Tables \ref{tab:Ferranparams1}
  and \ref{tab:Ferranparams2} in Supporting Information (Appendix \ref{sec:expferranpara}). 
\label{Ferrantable} }
\end{table}

%\FloatBarrier

We remark that some experiments have made use of thin
  electrodes to ensure that the potential drop in the electrolyte is
  negligible\cite{wu2012high,nguyen2022mathematical}.  In our case,
we emphasize that in both the slow and moderate charge/discharge rate
data we see a non-negligible potential drop in the electrolyte, which
is on the order of the contributions from overpotentials $\eta_a$ and
$\eta_c$, cf.~Tables \ref{Kevintable} and \ref{Ferrantable}.  Thus, an
SPM with a simple ohmic model accounting for the electrolyte as
discussed in the next section will be suitable for our purposes.
%Thus,
%since the standard SPM assumes there is no potential drop in the
%electrolyte\cite{marquis2019asymptotic}, this model will need to be
%augmented to account for this effect, as discussed in the next
%section.

%%%%%%%%%%%%%%%%%%%%%%%%%%%%%%%%%%%%%%%%%%%
%%%%%%%%%%%%%%%%%%%%%%%%%%%%%%%%%%%%%%%%%%%
%Math model

%%%%%%%%%%%%%%%%%%%%%%%%%%%%%%%%%%%%%%%%%%%
\section{Mathematical Model}
\label{mathmodel}

In this section, we first introduce the SPM framework, then formulate
our inverse problem, describe its numerical solution, and finally
discuss the sensitivity analysis. The SPM framework relies on
  several simplifying assumptions used to describe the construction of
  the Li-ion battery. Namely, the microstructure of the cell in each
  electrode is represented in terms of a representative spherical
  particle, the electrodes are spatially homogeneous and the
  intercalation reaction is uniform across the
  electrode\cite{Doyle01061993,fuller1994simulation,nt04}. In
  practice, these assumptions may not reflect the true behaviour of
  the cell.  For example, during the manufacturing process, the active
  particles are usually covered with binder containing carbon and
  therefore active material may not be in a direct contact with the
  electrolyte\cite{mistry2018secondary,mistry2021quantifying}.
  Electrodes are typically produced with spatial inhomogeneities such
  that not all particles undergo intercalation at the same
  rate\cite{mistry2020fingerprinting,hein2020influence}. In addition,
  the electrode particles are not strictly spherical, which can lead
  to preferential intercalation in some parts of the
  electrode\cite{ferraro2020electrode,mistry2022asphericity}.  We note
  here that some of these limitations can be circumvented by employing
  idealized nonporous geometries\cite{verbrugge1996modeling,johnson2022unconventional,amin2015characterization,amin2016characterization,amin2008ionic}.
  However, the aforementioned assumptions are generally invoked when
  modelling cells under low-power operation (generally up to a C-rate
  of
  $1$C)\cite{santhanagopalan2006review,richardson2020generalised,richardson2020charge,schmalstieg2018full,escalante2021uncertainty}.
  We add that in our study the standard SPM is augmented to include an
  Ohmic potential drop in the electrolyte and features
  concentration-dependent diffusivity describing transport in the active
  particles.  As indicated in Section \ref{expsec}, the experimental
  data utilized in this work is within the realm of validity for this
  model.
  
\subsection{Single Particle Model (SPM)}
\label{sec:SPM}

A standard description of the transport of Lithium in battery
electrodes usually relies on the DFN model
\cite{Doyle01061993,fuller1994simulation,nt04}.  However, this model
is computationally expensive and involves a large number of material
parameters, which makes it difficult to calibrate. Both of these
issues are partially alleviated by various simplified models such as
the SPM, which assumes that the electrode particles are uniformly
distributed spheres and that electrochemical reactions are uniform
throughout the electrode\cite{richardson2020generalised}. These
assumptions reduce the model to a transport problem in a single
representative particle.
%In general, the assumptions underlying SPMs are valid when the
%potential drop across the electrolyte is much smaller than the
%potential in the particle. 
The SPM can be derived as a certain asymptotic limit of the DFN model
and the price to be paid for its simplicity is that it is valid for
small C-rates only --- it is usually assumed that it does not provide
an accurate description beyond
1C\cite{santhanagopalan2006review,richardson2020generalised,richardson2020charge,schmalstieg2018full,escalante2021uncertainty}.
In our study, a suitably modified SPM is applied to describe transport
both in the anode and the cathode.

In the SPM, the Lithium transport in the solid state is described
by a nonlinear diffusion equation written in spherical
coordinates\cite{Doyle01061993,fuller1994simulation,nt04}
\begin{equation}
\frac{\partial c_{{\rm{s}},i}}{\partial t} = \frac{1}{r^2}
\frac{\partial}{\partial r} \left[D_{\rm{s}} (c_{{\rm{s}},i}) r^2 \frac{\partial
    c_{{\rm{s}},i}}{\partial r} \right] \qquad \textrm{in}  \ (0,{R_i}) \times (0,t_f],
\label{eq:SPM}
\end{equation}
where $r$ measures the radial distance from the center of the particle
outward to the particle's surface at ${R_i}$, $D_{\rm{s}}{(c_{{\rm{s}},i})}$ is the
(concentration-dependent) diffusivity in the electrodes, $t$ is time
ranging from $0$ to the final time $t_f$, and $i={a,c}$ corresponds to
the anode and cathode, respectively.  The accompanying boundary
conditions are
\begin{alignat}{2}
\frac{\partial c_{{\rm{s}},i}}{\partial r} \bigg|_{r=0} &=0 & \qquad & \textrm{in}  \ (0,t_f],\label{BCzero}\\
-D_{\rm{s}} (c_{{\rm{s}},i})\frac{\partial c_{{\rm{s}},i}}{\partial r} \bigg|_{r={R_i}} &=
\frac{j_i}{F} & &  \textrm{in} \  (0,t_f],\label{jintBC}
\end{alignat}
where $F$ is the Faraday constant, and the initial condition is 
\begin{equation}
  c_{{\rm{s}},i} |_{t=0} = c_{s,i;0},
  \label{eq:SPMic}
\end{equation}
where $c_{s,i;0}$ is the initial concentration of Lithium in the
solid.  Given that transport in the electrolyte is {not
  explicitly modelled} in the SPM, the current density $j_i$ in the
boundary condition \eqref{jintBC} can be deduced directly from the
total current $I(t)$ applied to the cell. { The
  conservation of charge at the anode and cathode
  implies
\begin{equation}\label{charge_con_anode}
	4\pi R_i^2 j_i = \pm \frac{I(t)}{n_i},
\end{equation}
in which the total number of particles in the electrode is given
by $n_i = A_i \, L_i \, \epsilon_i / (4/3 \pi R^3_i)$, where
$A_i$ is the cross-sectional area of the electrode, $L_i$ is the
thickness of the electrode, and $\epsilon_i$ is the volume fraction in
the electrode. {The right-hand side is positive for $i={a}$ and
  negative for $i={c}$.}  Finally, the total current to the cell is
$I(t)$, with $I(t) < 0$ for the anode during
  charging, $I(t) > 0$ for the cathode during
  charging, and vice versa during discharging. Relation
\eqref{charge_con_anode} ensures that there is enough
(de)intercalation reaction to meet the current demand. We emphasize
that, while the Butler-Volmer relation \eqref{eq:BV1} is not used in
the solution of the solution of the SPM, it is required in order to
determine the corresponding overpotential $\eta$ which is done by
post-processing the solution of the SPM. This property will be the
basis for the inverse problem introduced in the next section.
% A summary of the equations constituting the SPM is
% provided in Table \ref{tab:SPM}.
% \begin{table}[h!]
% \begin{tabular}{l l}\hline
% Governing PDE &  $\frac{\partial c_{s}}{\partial t} = \frac{1}{r^2}
%                 \frac{\partial}{\partial r} \left(D_{\rm{s}} (c_{\rm{s}}) r^2
%                 \frac{\partial c_{s}}{\partial r} \right) \qquad
%                 \textrm{in} \  (0,{\revtt{R_i}}) \times (0,t_f] $\\
%   Electrode Boundary Conditions
%  & $\frac{\partial c_{s}}{\partial r} \bigg|_{r=0} =0 \qquad   \textrm{in} \ (0,t_f]$ \\
% & $-D_{\rm{s}} (c_{\rm{s}})\frac{\partial c_{s}}{\partial r} \bigg|_{r={\revtt{R_i}}} =   \frac{j}{F} \quad   \textrm{in} \ (0,t_f]$ \\
% Electrode Initial Condition & $c_{s} |_{t=0} = c_{s,0} $\\
% % Butler-Volmer equation & $j = i_0 \left[ e^{\frac{\alpha_a F}{RT}  \eta} - e^{\frac{-\alpha_c F}{RT}  \eta} \right] $ \\
% % & $i_0 = Fk_0 c_{\rm{e}}^{\alpha_a} \left( c_{\rm{s,max}} - c_{s} \right)^{\alpha_a} c_{s}^{\alpha_c}$ \\
% % & $\eta= \phi_{\rm{s}} - \phi_{\rm{e}} -  U_{\rm{eq}}(c_{s})$ \\ \hline
% \end{tabular}
% \caption{Summary of the SPM. \label{tab:SPM}}
% \end{table}

Since in our experimental data the potential drop in the the
electrolyte is of approximately the same magnitude as the
overpotentials (cf.~Tables \ref{Kevintable} and \ref{Ferrantable}), we
augment the expression for the total potential drop predicted by SPM
with a term representing the potential drop in the electrolyte which
is assumed to be ohmic \cite{marquis2019asymptotic} and proportional to
the applied current $I$ via a constant resistivity $ R_{el}$. This
assumption is supported by the data in Tables \ref{Kevintable} and
\ref{Ferrantable} which indicate that the potential drop in the
electrolyte does increase approximately linearly with the applied
current.  The total potential drop predicted by the model thus takes
the form
\begin{equation}
V_{cell} = {U_{\rm{eq}}^c}(c_{{\rm{s}},c} \big|_{r=R_c}) - {U_{\rm{eq}}^a}(c_{{\rm{s}},a} \big|_{r=R_a}) +  \eta_{c} - \eta_{a} + R_{el} I,
\label{eq:Vcell}
\end{equation} 
where the additional subscripts $a$ and $c$ denote anode and cathode,
respectively. {Relation \eqref{eq:Vcell} will serve as
  the basis for the inverse modelling introduced in Section
  \ref{sec:inverse}.  We note that the SPM is usually defined through
  \eqref{eq:SPM}--\eqref{charge_con_anode} and will refer to the model
  resulting from the inclusion of relation \eqref{eq:Vcell} as an
  ``augmented'' SPM.}

%%%%%%%%%%%%%%%%%%%%%%%%%%%%%%%%%%%%%%%%%%%
\subsection{Inverse Problems}
\label{sec:inverse}

Solution of the SPM given above yields concentration profiles
$c_{{\rm{s}},i}(t)$ as functions of time $t$. Within {the SPM
  \eqref{eq:SPM}--\eqref{charge_con_anode}}, the concentration
$c_{{\rm{s}},i}(t)$ does not depend on the constitutive relation
\eqref{eq:BV}, however, the corresponding overpotential does. The idea
behind the inverse problems is thus to infer the optimal form of the
factor $\iota(c_{{\rm{s}},i})$ by minimizing the discrepancy between the
overpotential predicted by \eqref{eq:BV} and the overpotential deduced
from the experimental data.
The total potential measured experimentally as discussed in Section
\ref{expsec} will be denoted $\overline{V}_{cell}(t)$.

{ In our first, simpler, inverse problem we assume the
  constitutive relation is given by the Butler-Volmer equation
  \eqref{eq:BV1} and calibrate the scalar parameters $\alpha_{{a},m}$ and
  $k_0$ parameterizing this relation for both the anode and the cathode
  as well as the electrolyte resistivity $R_{el}$ in \eqref{eq:Vcell}.
  Since the total potential is dominated by contributions from the
  equilibrium potentials, cf.~relation \eqref{eq:Vcell} and Tables
  \ref{Kevintable}--\ref{Ferrantable}, we consider the following error
  functional}
\begin{equation}
\J_1 = \frac{1}{2}\int\limits_0^{t_f}\left[ \Delta\eta(t) -
  \Delta\overline{\eta}(t) \right]^2  \, dt,
\label{eq:J1}
\end{equation}
where $\Delta\eta(t) = \eta_c(t) -
\eta_a(t){+R_{el}I(t)}$ and $\Delta\overline{\eta}(t) =
\overline{V}_{cell}(t) - \left[{U_{\rm{eq}}^c}(c_{{\rm{s}},c} \big|_{r=R_c}) -
  {U_{\rm{eq}}^a}(c_{{\rm{s}},a} \big|_{r=R_a}) \right]$ {in which
  $\eta_a(t)$ and $\eta_c(t)$ are obtained by inverting relation
  \eqref{eq:BV0} defined on the anode and the cathode (each of these
  relations depends on the parameters $\alpha_{\rm{a},m}$ and $k_0$), whereas
  $c_{{\rm{s}},a}$ and $c_{{\rm{s}},c}$ are the solutions of system
  \eqref{eq:SPM}--\eqref{charge_con_anode} obtained at the two
  electrodes (additional information how $\J_1$ evaluated is provided
  in the next section). The error functional \eqref{eq:J1}} is a
least-square measure of the discrepancy between the portion of the
total potential directly depending on the constitutive relation
describing ion intercalation, and is therefore numerically better
conditioned than a least-square measure of the difference $V_{cell}(t)
- \overline{V}_{cell}(t)$. This leads to the finite-dimensional
optimization problem
\begin{equation*}
\textrm{P1:} \qquad  [\widehat{k}_{0,-},\alphaanhat,\widehat{k}_{0,+},\alphaaphat,\widehat{R}_{el}  ] =  \underset{[k_{0,-},\alphaan,k_{0,+},\alphaap,R_{el}] \in \mathbb{R}^5}{\textrm{argmin}} ~ \mathcal{J}_1(k_{0,-},\alphaan,k_{0,+},\alphaap,R_{el}), 
\end{equation*}
where hats ($\widehat{\cdot}$) denote optimal values of the material
parameters and subscripts $-$ and $+$ on $k_0$ and $\alpha_{{a},m}$ denote
quantities defined on the anode and the cathode, respectively.

{In contrast to Problem P1 where we seek optimal
  parameters in relations \eqref{eq:BV1} defined both on the anode and
  the cathode, in our main inverse problem, Problem P2, we will focus
  on inferring the optimal form of the factor $\iota(c_{{\rm{s}},a})$ in the
  constitutive relation \eqref{eq:BV} on the anode only while assuming
  that on the cathode the constitutive relation has the standard form
  \eqref{eq:BV1} with parameters $\widehat{k}_{0,-}$ and $\alphaanhat$
  obtained by solving Problem P1.  Thus, the second error functional
  is defined to measure the error in the overpotential on the anode
  as}
\begin{subequations}
  \label{eq:J2}
  \begin{align}
\J_2 & = \frac{1}{2}\int\limits_0^{t_f}\left[ \eta_a(t) -
       \overline{\eta}^{P2}(t) \right]^2  \, dt, \qquad \textrm{where}\\
 \eta_a & = \psi^{-1}\left(\frac{I}{F \iota(c_{{\rm{s}},a})  }      \right), \label{eq:etaa} \\
\overline{\eta}^{P2}(t) &=-\left[ \overline{V}_{cell}(t) - \left( {U_{\rm{eq}}^c}(c_{{\rm{s}},c} \big|_{r=R_c}) - {U_{\rm{eq}}^a}(c_{{\rm{s}},a} \big|_{r=R_a})  + \eta^{P1}_c + \widehat{R}_{el}I  \right) \right],  \label{eq:etaP2} 
\end{align}
\end{subequations} {in which the superscript ``P1'' denotes quantities
  computed using information from the solution of Problem P1} and,
with a slight abuse of notation, $\psi^{-1}(\cdot)$ denotes the
function $\Delta\phi = \Delta\phi(c_{{\rm{s}},i})$ which is the solution of
the equation $I / (F \iota(c_{{\rm{s}},a})) -
\psi(\Delta\phi(c_{{\rm{s}},a}),c_{{\rm{s}},a}) =0$ for $c_{{\rm{s}},a} \in
[0,c_{{\rm{s}},a,{\rm{max}}}]$, cf.~\eqref{eq:BV} (the existence of such a function
is guaranteed by the implicit function theorem).  This notation
emphasizes the dependence of the expression in \eqref{eq:etaa} on the
factor $\iota(c_{{\rm{s}},a})$ in the constitutive relation \eqref{eq:BV}
describing ion intercalation in the anode. As stipulated by
assumptions A1--A3 in Section \ref{sec:BV}, see also Supporting
Information (Appendix \ref{sec:appBC}), we must require the function
$\iota(c_{{\rm{s}},a})$ to vanish at both vanishing and maximum
concentrations, i.e., for $c_{{\rm{s}},a} = 0$ and for $c_{{\rm{s}},a} =
c_{{\rm{s}},a,{\rm{max}}}$. We also need the function $\iota(c_{{\rm{s}},a})$ to be
continuous with square-integrable (distributional) derivatives, such
that it will belong to the Sobolev space\cite{af05}
$H_0^1([0,c_{{\rm{s}},a,{\rm{max}}}])$ (the subscript ``0'' indicates that functions
belonging to this space vanish at the boundary). The optimal
functional form of the exchange current $\iota(c_{{\rm{s}},a})$ {on the
  anode} can then be obtained as a solution of the optimization
problem
\begin{equation*}
\textrm{P2:} \qquad [\widehat{\iota}(c_{{\rm{s}},a})] =  \underset{\iota(c_{{\rm{s}},a}) \in {H}_0^1([0,c_{{\rm{s}},a,{\rm{max}}}])}{\textrm{argmin}} ~ \mathcal{J}_2(\iota(c_{{\rm{s}},a})),
\end{equation*}
where the corresponding constitutive relation \eqref{eq:BV} takes the
optimal form
$\widehat{j}(c_{{\rm{s}},a})=
\widehat{\iota}(c_{{\rm{s}},a})\psi(\Delta \phi,
c_{{\rm{s}},a})$. \revtt{We note that the framework outlined above
  could also be utilized to reconstruct the optimal form of the
  exchange current on the cathode.  In this case, the cost functional
  \eqref{eq:J2} would include $\eta_c(t)$, equation \eqref{eq:etaa}
  would involve $\eta_c$ as a function of $\iota(c_{{\rm{s}},c})$
  whereas equation \eqref{eq:etaP2} would use $\eta_a^{P1}$ (with a
  correction to the sign on the right-hand side, cf.~equation
  \eqref{eq:Vcell}).  In such case the optimal form of the
  constitutive relation would be
  $\widehat{j}(c_{{\rm{s}},c})=
  \widehat{\iota}(c_{{\rm{s}},c})\psi(\Delta \phi, c_{{\rm{s}},c})$,
  cf.~\eqref{eq:BV}.  Moreover, one could solve these two problems
  iteratively making it possible to refine the reconstructions of the
  exchange current on both electrodes.}  Our approach to solving
Problems P1 and P2 is described next.

%%%%%%%%%%%%%%%%%%%%%%%%%%%%%%%%%%%%%%%%%%%
\subsection{Computational Approach}
\label{sec:comput}

Evaluation of the error functionals \eqref{eq:J1} and \eqref{eq:J2}
requires solution of the SPM, cf.~\eqref{eq:SPM}--\eqref{eq:Vcell},
which is done using a standard finite element method
\cite{escalante2021uncertainty, korotkin2021dandeliion}. Problem P1
represents unconstrained minimization over $\mathbb{R}^5$ and is
solved numerically in a straightforward manner using the function {\tt
  fminunc} in MATLAB. {The key step is evaluation of
  the error functional \eqref{eq:J1} and the procedure is outlined in
  Algorithm \ref{alg:P1}.}

\begin{algorithm}[H]
\caption{Evaluation of the error functional $\J_1$ in the solution of Problem P1 \\
\noindent\makebox[\linewidth]{\rule{\textwidth}{0.4pt}}
{
{\bf{Input:}} \\
$\overline{V}_{\textrm{cell}}(t)$ --- experimental voltage as a function of time\\
$k^{\omicron}_{0,-}$, $k^{\omicron}_{0,+}$ --- initial guesses for reaction rate for anode and cathode\\
$\alpha^{\omicron}_{a,-}$, $\alpha^{\omicron}_{a,+}$ --- initial guesses for transfer coefficients for anode and cathode\\
$R^{\omicron}_{el}$ --- initial guess for electrolyte resistance\\
{\bf{Output:}}   $\J_1( {k}_{0,-}, {k}_{0,+}, {\alpha}_{a,-}, {\alpha}_{a,+}, {R}_{el})$
}}
{
\begin{algorithmic}
\State Solve the SPM equations \eqref{eq:SPM}-\eqref{charge_con_anode} for $c_{{\rm{s}},a}$ and $c_{{\rm{s}},c}$\\
Compute $\Delta\bar{\eta}(t) = -\left[ \overline{V}_{cell}(t) -
  \left( {U_{\rm{eq}}^c}(c_{{\rm{s}},c} \big|_{r=R_c}) - {U_{\rm{eq}}^a}(c_{{\rm{s}},a} \big|_{r=R_a}) \right) \right]$ \\
Compute  ${\eta}_a$ and ${\eta}_c$ by inverting relations \ref{eq:BV0} and \ref{charge_con_anode} with unknowns  ${k}_{0,-}$, ${k}_{0,+}$,${\alpha}_{a,-}$, and ${\alpha}_{a,+}$ \\
Compute $\Delta{\eta}(t) = {\eta}_c - {\eta}_a + {R}_{el}I(t)$\\
Determine $\J_1( {k}_{0,-}, {k}_{0,+}, {\alpha}_{a,-}, {\alpha}_{a,+}, {R}_{el})$ by evaluating the integral in \eqref{eq:J1} 
\end{algorithmic}
}
\label{alg:P1}
\end{algorithm}

Problem P2 represents minimization over an
infinite-dimensional function space $H^1_0$ and therefore requires a
specialized approach. We compute the optimal exchange current using
the discrete gradient flow as
$\widehat{\iota}(c_{{\rm{s}},i}) = \lim_{n \rightarrow \infty} \iota^{(n)}(c_{{\rm{s}},i})$
where the approximations $\iota^{(n)}(c_{{\rm{s}},i})$ are computed iteratively
by
\begin{equation}
\begin{aligned}
\iota^{(n+1)}(c_{{\rm{s}},i}) & = \iota^{(n)}(c_{{\rm{s}},i}) - \tau^{(n)} \nabla
\mathcal{J}_2(\iota^{(n)}(c_{{\rm{s}},i})), \\
\iota^{(0)}(c_{{\rm{s}},i}) & = \iota_0(c_{{\rm{s}},i}),
\end{aligned}
\label{graddescent}
\end{equation}
in which $\iota_0(c_{{\rm{s}},i})$ is a suitable initial guess,
$\nabla \mathcal{J}_2(\iota^{(n)}(c_{{\rm{s}},i}))$ is the Sobolev
gradient of the error functional \eqref{eq:J2} at the $n$th iteration,
and $\tau^{(n)}$ is the step size along the gradient direction. The
initial guess $\iota_0(c_{{\rm{s}},i})$ is chosen to vanish at
$c_{{\rm{s}},i} = 0$ and at
$c_{{\rm{s}},i} = c_{{\rm{s}},i,{\rm{max}}}$, so that the same
limiting behaviour will be inherited by the optimal solution
$\widehat{\iota}(c_{{\rm{s}},i})$, provided the gradients
$\nabla\J_2(c_{{\rm{s}},i})$ also vanish at $c_{{\rm{s}},i} = 0$ and
$c_{{\rm{s}},i} = c_{{\rm{s}},i,{\rm{max}}}$, cf.~Supporting
Information (Appendix \ref{sec:appBC}). In practice, the initial guess
$\iota_0(c_{{\rm{s}},i})$ is taken as the factor $i_0$ in the
Butler-Volmer relation \eqref{eq:i0} with optimal parameters found by
solving Problem P1 which, by construction, ensures the correct
behaviour at $c_{{\rm{s}},i} =
c_0,c_{{\rm{s}},i,{\rm{max}}}$. Computation of the Sobolev gradient
$\nabla \mathcal{J}_2(\iota(c_{{\rm{s}},i}))$, which is a key element
of approach \eqref{graddescent}, is described in Supporting
Information (Appendix \ref{sec:grad}) and together with the properties
of the initial guess $\iota_0(c_{{\rm{s}},i})$ ensures the required
smoothness of the optimal exchange current
$\widehat{\iota}(c_{{\rm{s}},i})$ and its correct behaviour at
$c_{{\rm{s}},i} = c_0,c_{{\rm{s}},i,{\rm{max}}}$. The optimal step
size in \eqref{graddescent} is found by solving the line-minimization
problem
\begin{equation}
\tau^{(n)} = \underset{\tau > 0}{\operatorname{argmin}} ~ \mathcal{J}_2(\iota(c_{{\rm{s}},i})-\tau \nabla^{{H}^1}\mathcal{J}_2(\iota(c_{{\rm{s}},i}))),
  \label{eq:taun}
\end{equation}
which is done conveniently using Brent's algorithm\cite{pftv86}.
Expression \eqref{graddescent} is evaluated using the collocation
approach on the domain $[0,c_{{\rm{s}},i,{\rm{max}}}]$ discretized with a uniform
grid based on $N =500$ grid points. We also considered finer meshes,
but they did not produce appreciable differences in the obtained
results, hence the aforementioned resolution was used as a trade-off
between accuracy and the computational cost. When determining the
Sobolev gradients, {one must specify the value of the
  Sobolev parameter $0 < \ell < \infty$ which appears in the
  definition of the inner product \eqref{eq:ipH1} and determines the
  relative smoothness of the gradients. More specifically, $\ell$ may
  be regarded as a characteristic, or cut-off, value of the
  concentration below which all information in the gradient $\nabla
  \mathcal{J}_2(\iota(c_{{\rm{s}},i}))$ is smoothed out.}  At the validation
stage, values of $\ell$ spanning several orders of magnitude were
considered to determine the value resulting in the fastest rate of
convergence of iterations \eqref{graddescent}.  {Based
  on these tests, we chose} $\ell = 3000000$ and $\ell = 100000$ for
the case of slow charge and moderate discharge rates, respectively.
The steps leading to the solution of Problem P2 are summarized as
Algorithm \ref{alg:P2}.

The computational approach was systematically validated by solving
Problems P1 and P2 with ``manufactured'' measurement data. More
specifically, we assumed a certain form of the constitutive relation
(\eqref{eq:BV1} with some values of $\alpha_{{a},m}$ and $k_0$ for
Problem P1 and \eqref{eq:BV} with some function
$\iota(c_{{\rm{s}},i})$ for Problem P2) which was used to generate
measurements. The true constitutive relations were then reconstructed
using Algorithm \ref{alg:P2}. Excellent reconstruction accuracy was
achieved with errors essentially at the level of errors due to
numerical discretization. We add that Problems P1 and P2 are
non-convex and we have found some evidence for the presence of
nonunique (local) minimizers when different initial guesses are
  used in Algorithms \ref{alg:P1} and \ref{alg:P2}. However, in all
instances these local minimizers involved physically inconsistent
material properties (e.g., with incorrect signs), which made it easy
to eliminate them.  Analysis of the sensitivity of solutions to
Problems P1 and P2 to perturbations of measurement data is discussed
in the next section.

\begin{algorithm}[H]
\caption{Solution of Problem P2 using discrete gradient flow \eqref{graddescent}\\
\noindent\makebox[\linewidth]{\rule{\textwidth}{0.4pt}}
{\bf{Input:}} \\
$\epsilon$ --- tolerance\\
$\ell$ --- Sobolev parameter\\
$\overline{V}_{\textrm{cell}}(t)$ --- experimental voltage as a function of time\\
$k^{\omicron}_{0,-}$, $k^{\omicron}_{0,+}$ --- initial guesses for reaction rate for anode and cathode\\
$\alpha^{\omicron}_{a,-}$, $\alpha^{\omicron}_{a,+}$ --- initial guesses for transfer coefficients for anode and cathode\\
$R^{\omicron}_{el}$ --- initial guess for electrolyte resistance\\
{\bf{Output:}}    $\widehat{k}_{0,-}$, $\widehat{k}_{0,+}$,$\widehat{\alpha}_{a,-}$, $\widehat{\alpha}_{a,+}$, $\widehat{R}_{el}$, and $\widehat{\iota}(c_{{\rm{s}},i})$
}\label{alg:cap}
\begin{algorithmic}
\State Solve Problem P1  using  {\tt fminunc}\\
Compute $\overline{\eta}(t) \gets -\left[ \overline{V}_{cell}(t) -
  \left( {U_{\rm{eq}}^c}(c_{{\rm{s}},c} \big|_{r=R_c}) - {U_{\rm{eq}}^a}(c_{{\rm{s}},a} \big|_{r=R_a})
    + \eta^{P1}_c + \widehat{R}_{el}I \right) \right]$ \\
Set $\iota_0(c_{{\rm{s}},i})  = F\widehat{k}_{0,-} c_{\rm{e}}^{\widehat{\alpha}_{a,-}} \left( c_{{\rm{s}},i,{\rm{max}}} - c_{{\rm{s}},i} \right)^{\widehat{\alpha}_{a,-}} c_{{\rm{s}},i}^{(1-\widehat{\alpha}_{a,-})}$
\Repeat \\
\hspace{\algorithmicindent}Evaluate $\nabla^{L^2} \mathcal{J}(\iota(c_{{\rm{s}},i})))$\\
\hspace{\algorithmicindent}Compute $\nabla^{H^1} \mathcal{J}(\iota(c_{{\rm{s}},i})))$ given $\nabla^{L^2} \mathcal{J}(\iota(c_{{\rm{s}},i})))$\\
\hspace{\algorithmicindent}Perform a Polak-Ribiere conjugate gradient update  $\nabla^{H^1} \mathcal{J}(\iota(c_{{\rm{s}},i})))$\\
\hspace{\algorithmicindent}Find the step size $\tau^{(n)}$
by solving the line-search problem \eqref{eq:taun} with Brent's method \\
\hspace{\algorithmicindent}$\iota^{(n+1)}(c_{{\rm{s}},i}) \gets \iota^{(n)}(c_{{\rm{s}},i}) - \tau^{(n)} \nabla  \mathcal{J}(\iota^{(n)}(c_{{\rm{s}},i})) $ 
\State \Until{$ \frac{|\mathcal{J}(\iota^{(n)}(c_{{\rm{s}},i})))| - |\mathcal{J}(\iota^{(n-1)}(c_{{\rm{s}},i})))|}{|\mathcal{J}(\iota^{(n)}(c_{{\rm{s}},i})))|} < \epsilon $}
\end{algorithmic}
\label{alg:P2}
\end{algorithm}

%%%%%%%%%%%%%%%%%%%%%%%%%%%%%%%%%%%%%%%%%%%
\subsection{Sensitivity Analysis}
\label{sec:sensitivity}

Inverse problems are generally ill-posed and variations in the
experimental data can result in significant shifts in the
reconstructed solutions \cite{Tarantola2005}. To explore the effects
of small perturbations in the experimental data on the solution of
Problems P1 and P2, a Monte-Carlo analysis is developed where the
measurement data is perturbed as 
\begin{equation}
\widehat{V}_{cell}(t) = V_{cell}(t) + \sum\limits_{k=1}^3 \frac{a_k}{k^2}
\sin \left(\frac{2 \pi k t}{t_f} \right)+ \sum\limits_{k=1}^3
\frac{b_k}{k^2} \cos \left(\frac{2 \pi k t}{t_f} \right), \qquad t
\in [0,t_f],
\label{eq:VcPert}
\end{equation}
in which $a_k$ and $b_k$, $k=1,2,3$, are normally distributed random
variables with zero mean and a prescribed standard deviation. Problems
P1 and P2 with measurement data perturbed as in \eqref{eq:VcPert} are
then solved $M$ times ($M \gg 1$), each time using a different sample
of the random variables $a_k$ and $b_k$. This allows us to compute the
statistics of the reconstruction errors and of the optimal
constitutive relations $\widehat{\iota}(c_{{\rm{s}},a})$ corresponding to
perturbed measurements which provide information about their
sensitivity to noise.

\begin{figure}
   \mbox{
     \subfigure[]{\includegraphics[width=0.5\textwidth]{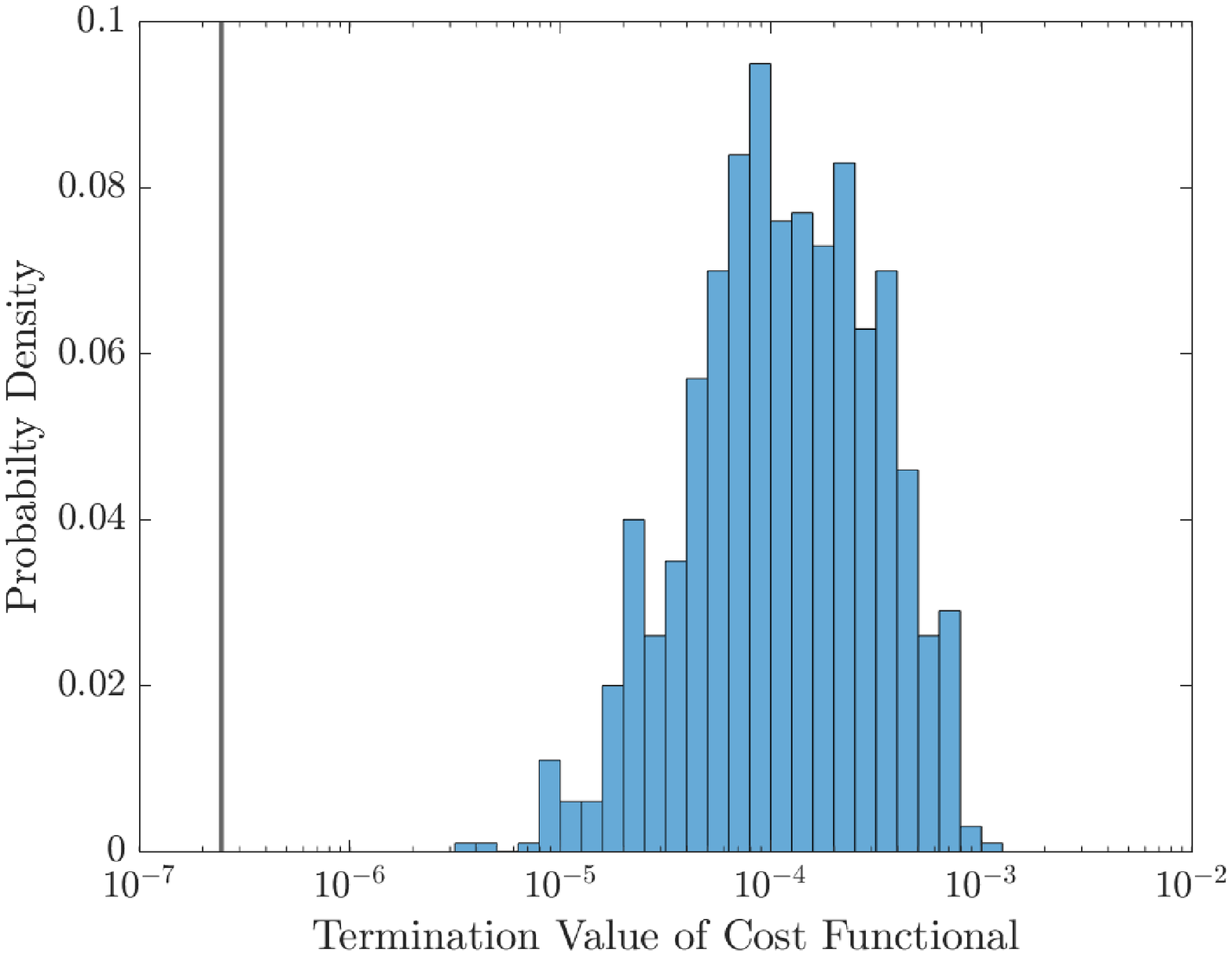}}\quad
     \subfigure[]{\includegraphics[width=0.5\textwidth]{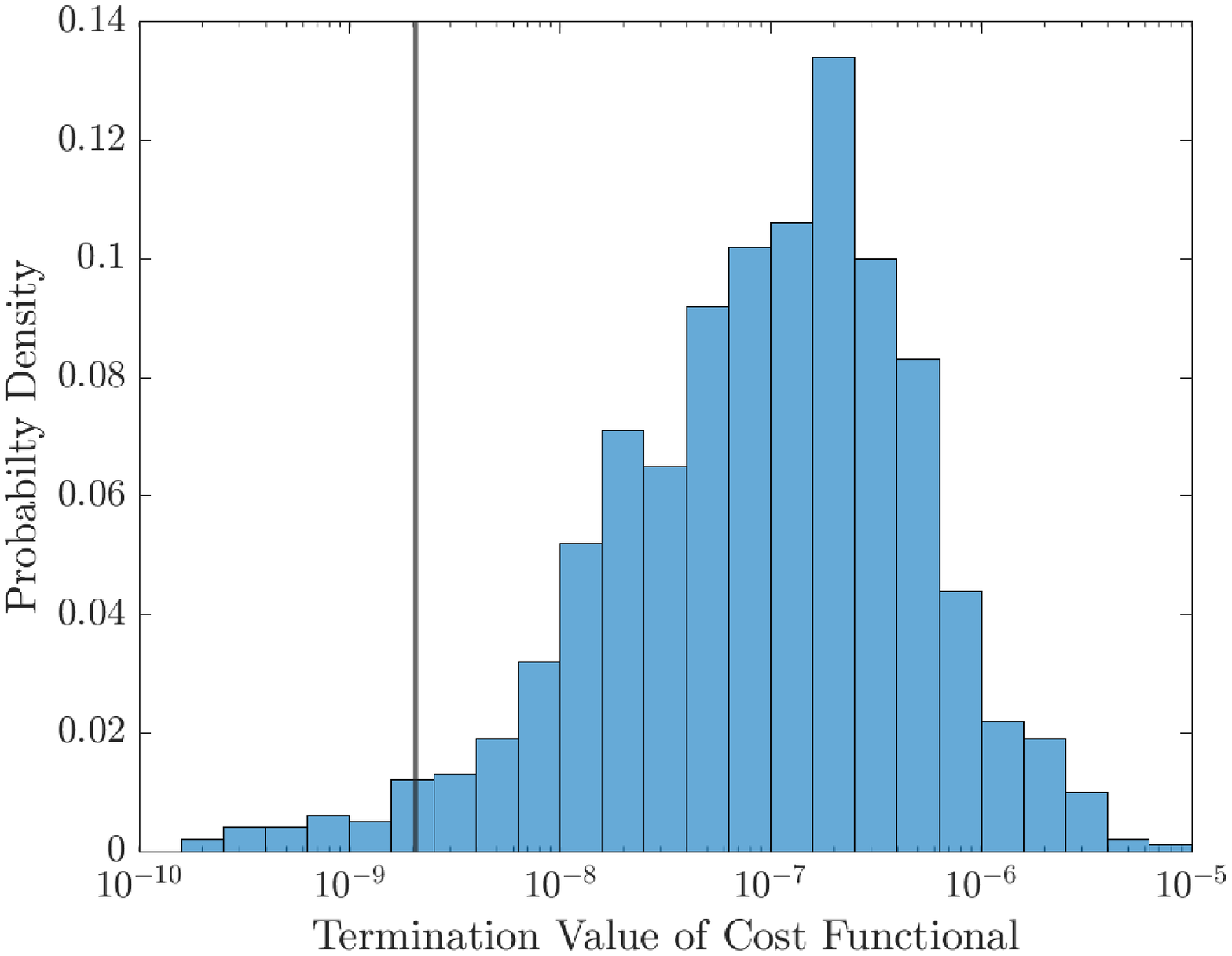}}}
   \caption{PDFs of the error functionals (a) $\J_1$ and (b) $\J_2$
     obtained in Monte-Carlo sensitivity analysis. The vertical lines
     represent the values of $\J_1$ and $\J_2$ obtained in solutions
     of Problems P1 and P2 with no noise.}
   \label{MClogcost}
\end{figure}

Our sensitivity analysis is then performed based on the manufactured
measurement data at 1C (which was also used to validate Algorithm
\ref{alg:P2}) and using $M = 3000$ normally distributed noise samples
with zero mean and $0.0025$ standard deviation. Using the empirical
$3$-sigma rule\cite{devore2011probability}, the relative variation of
the measurement data due to noise can be estimated as $6 \times 0.0025
/ 1.3 \approx 0.012$ or $1.2\%$, where $1.3$ in the denominator is the
approximate total voltage drop during the (dis)charge cycle.  The
effect of this noise is thus comparable to the potential drop in the
electrolyte, cf.~Tables \ref{Kevintable} and \ref{Ferrantable}, which
is accounted for in the voltage \eqref{eq:Vcell} predicted by the
augmented SPM. The probability density functions (PDFs) of the error
functionals \eqref{eq:J1} and \eqref{eq:J2} obtained by solving
Problems P1 and P2 with measurement data perturbed as described above
are shown in Figure \ref{MClogcost}. The corresponding {\em mean}
inferred exchange currents are shown in Figure \ref{MCExchangeCurrent}
together with their 95\% confidence intervals and the standard
Butler-Volmer relation \eqref{eq:BV0} with some
{reference} parameter values that were used to
``manufacture'' the measurements. We observe that optimal
reconstructions obtained by solving both Problem P1 and P2 capture the
general functional form of the true constitutive relation quite well,
but the magnitude is off by a bit more than 10\% even though the noise
amplitude is rather small. It ought to be emphasized however that such
level of uncertainty is not uncommon in parametrization of
electrochemical
systems\cite{escalante2020discerning,escalante2021uncertainty}.  The
corresponding distributions of optimal anode overpotentials are shown
in Figure \ref{MCAnode} where deviations from the original data are
rather small.  We also mention that additional Monte-Carlo simulations
were performed using the random coefficients $a_k$ and $b_k$ in
\eqref{eq:VcPert} with several decreasing standard deviations and in
this limit of vanishing noise amplitude, the deterministic solution is
recovered, confirming the consistency of the calculations.
{These observations allow us to conclude that a modest
  amount of noise in the measurement data has some quantitative and
  little qualitative effect on the obtained reconstructions.}

\begin{figure}[h!]
\mbox{
\subfigure[]{\includegraphics[width=0.5\textwidth]{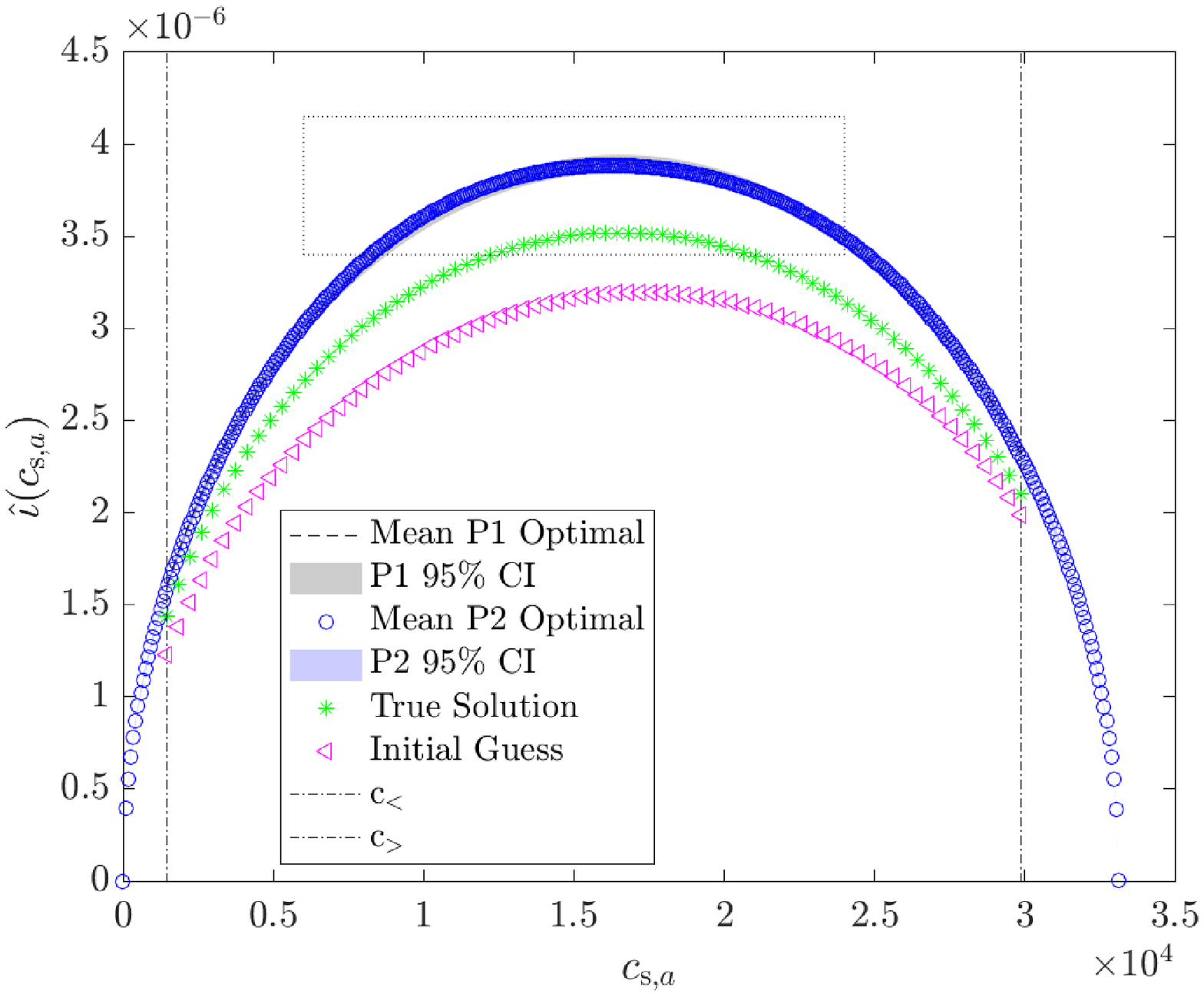}}
\subfigure[]{\includegraphics[width=0.5\textwidth]{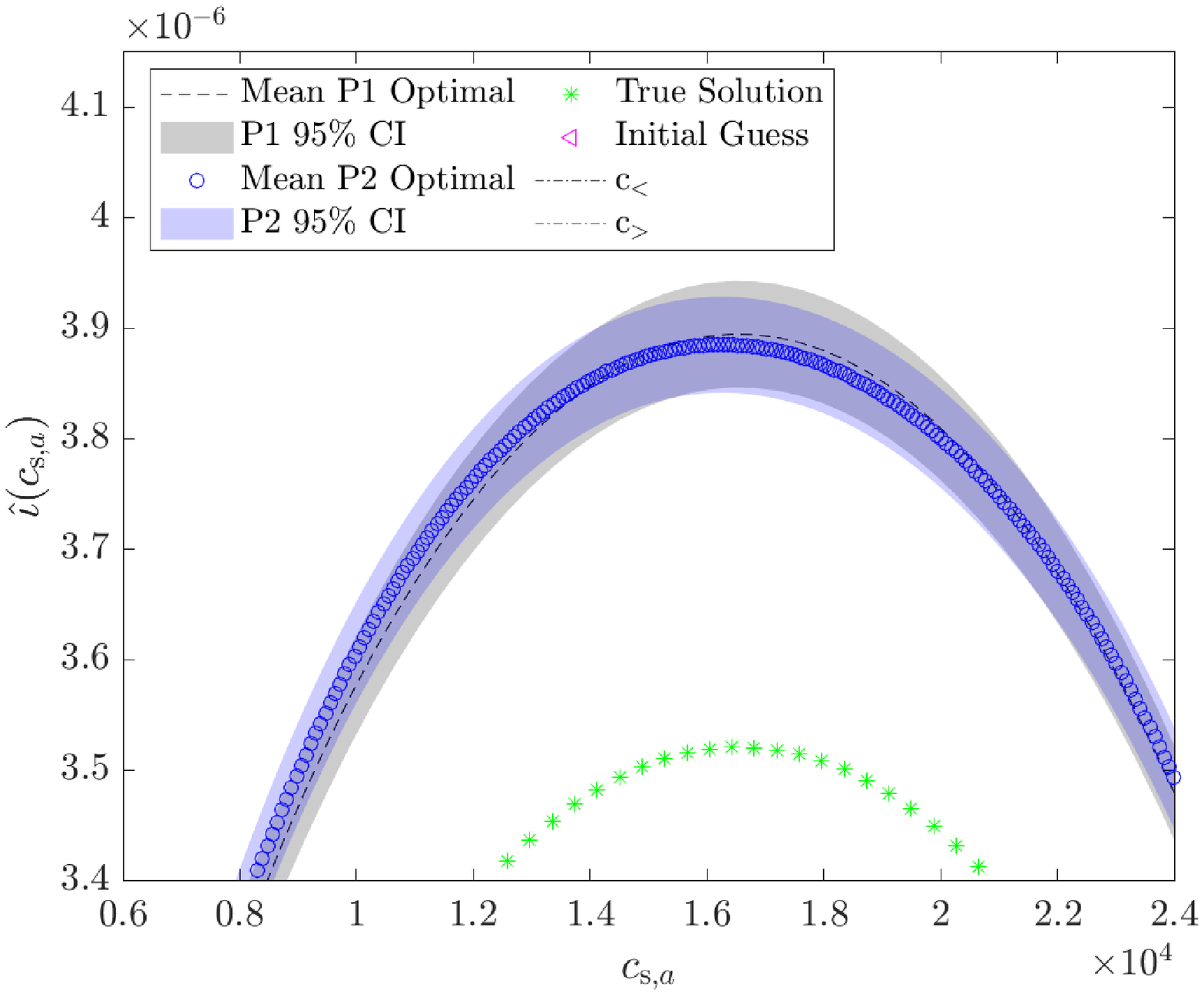}}}
\caption{Optimal reconstructions of the exchange current (black dashed
  lines) $i_0(c_{\rm{s}})$ and (blue circles) $\widehat{\iota}(c_{{\rm{s}},a})$
  obtained by solving Problems P1 and P2 with measurement data
  perturbed as in \eqref{eq:VcPert} {for (a) the entire
    concentration range and (b) intermediate concentrations indicated by the dotted box in (a)}.  The
  shaded bands indicate the $95\%$ confidence intervals whereas the
  vertical lines represent the bounds $c_<$ and $c_>$ of the
  identifiability interval $\I$. The true exchange current used to
  manufacture the measurements is shown in green.
  \label{MCExchangeCurrent}}
  \end{figure}

\begin{figure}[h!]
\mbox{
\subfigure[]{\includegraphics[width=0.5\textwidth]{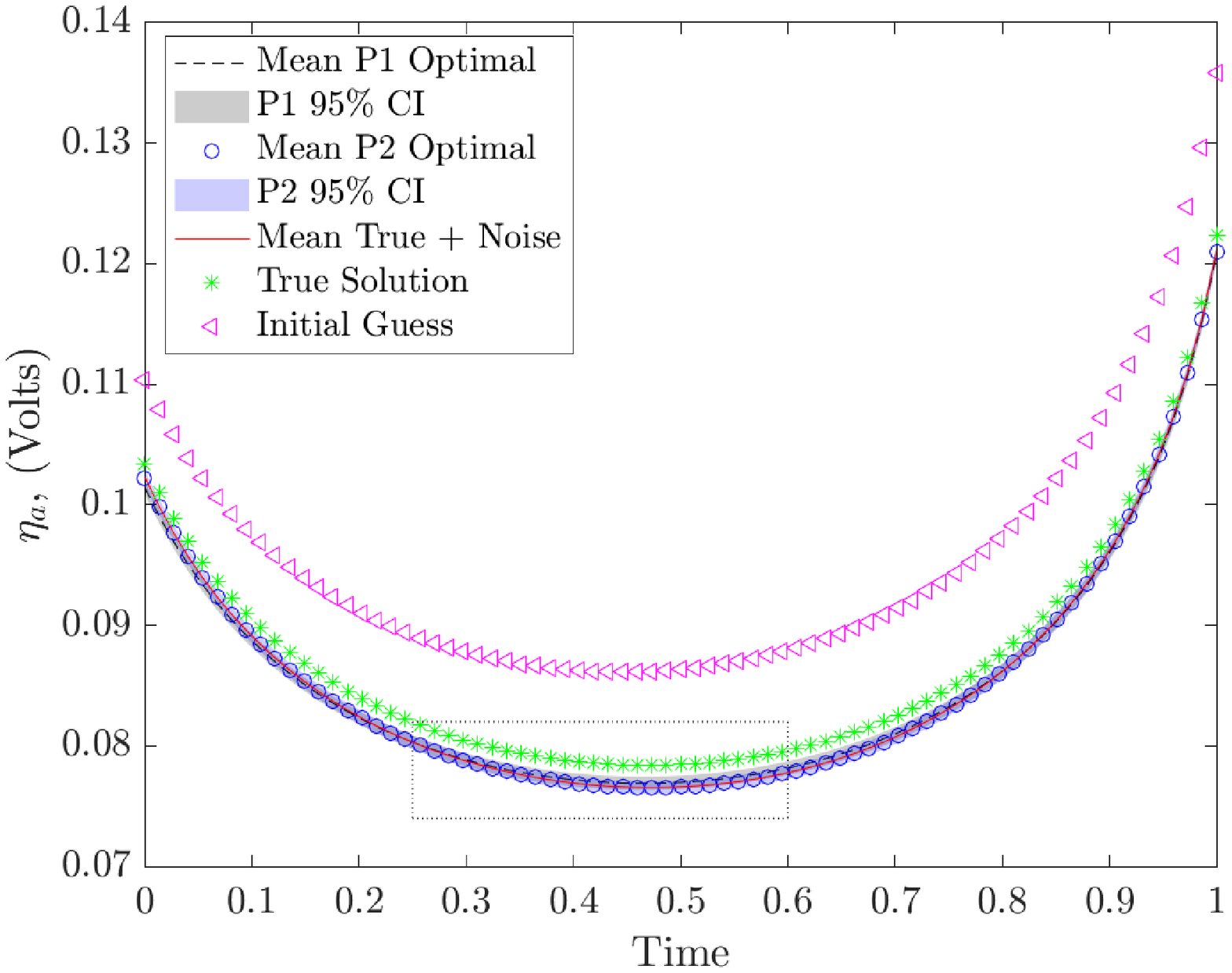}}
\subfigure[]{\includegraphics[width=0.5\textwidth]{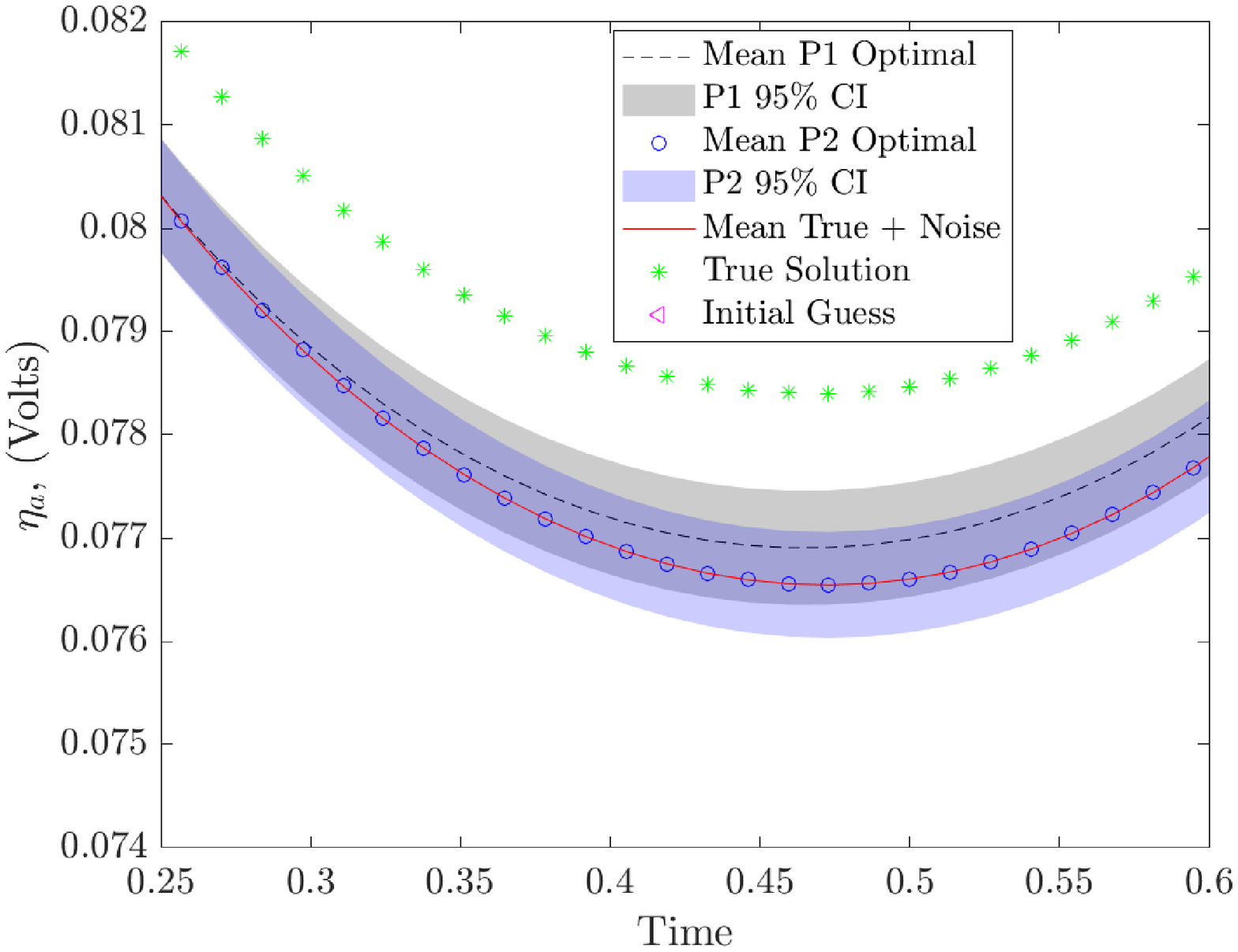}}}
\caption{Anode overpotentials corresponding to the optimal exchange
  currents shown in Figure \ref{MCExchangeCurrent} for (a) the entire
  normalized time window and (b) magnification around intermediate
  times indicated by the dotted box in (a). The line types and colour-coding are the same as in Figure
  \ref{MCExchangeCurrent} with an added red solid line to represent
  the average value of the overpotential when the total potential is
  perturbed with noise.}
  \label{MCAnode}
\end{figure}

\FloatBarrier

% To remove any discrepancies due to experimental measurement error in
% either the voltage measurements or parameter estimates, the initial
% experimental voltage, $V_{cell} $, is calculated through manufacturing
% data with parameters given by the moderate charge dataset
% \cite{chen2020development}.  Iteratively solving the proposed
% numerical framework for multiple simulations will provide insight into
% the sensitivity of the optimal reconstruction of the exchange current
% to noise in the experimental measurements. Here, it is important to
% note that the total voltage is varied as this is the directly
% measurable quantity in experiments.  However, the curve used as the
% target function in the cost functional is the proportion related to
% the anode overpotential.  So while a $1\%$ standard deviation is
% fairly small compared to the total voltage of the cell, this results
% in a large variation in the overpotential used for the optimization
% algorithm.

%%%%%%%%%%%%%%%%%%%%%%%%%%%%%%%%%%%%%%%%%%%
%%%%%%%%%%%%%%%%%%%%%%%%%%%%%%%%%%%%%%%%%%%
%Results \& Discussion

%%%%%%%%%%%%%%%%%%%%%%%%%%%%%%%%%%%%%%%%%%%
\section{Results   \label{sec:results}}

We compare the optimal constitutive relations reconstructed by solving
Problems P1 and P2 using data for slow charge and moderate discharge
rates against the standard form of the Butler-Volmer relation
\eqref{eq:BV0} for a graphite electrode. In the latter, we set
$\alpha_{{a},m} = \alpha_{{c},m} = 1/2$, which is a standard choice for the
transfer coefficients\cite{nt04,bard2001electrochemical}, and in the
problem with slow charge rates we set $k_0 = 1.8742 \times 10^{-12}$
which is taken from a similar set-up of a graphite anode with an
NMC$622$ cathode\cite{Ecker01012015}, whereas in the problem with
moderate charge rates we set $k_0 = 6.72 \times 10^{-12}$ which was
measured experimentally when collecting the data for a graphite anode
with an NMC$811$ cathode \cite{chen2020development}. Hereafter we will
refer to these values of $\alpha_{{a},m}$, $\alpha_{{c},m}$ and $k_0$ as
``nominal''.

%%%%%%%%%%%%%%%%%%%%%%%%%%%%%%%%%%%%%%%%%%%
\subsection{Slow Charge Rates}
\label{sec:Kevinresults}

We first consider solutions of Problem P1 where scalar parameters are
calibrated in expression \eqref{eq:i0} before analyzing the results
obtained by solving Problem P2 in which the exchange current
$\iota(c_{{\rm{s}},a})$ is sought in a fairly general form. In Figure
\ref{fig:Kevincost}, we show the values of the error functional
\eqref{eq:J1} corresponding to solutions of Problem P1 and see that
compared to the values of $\J_1$ obtained using the standard
Butler-Volmer relation \eqref{eq:BV0}, they are reduced only mildly,
by less than one order of magnitude for both C-rates.
{The solution to Problem P1 suggests that a small
  deviation from the standard functional form of the exchange current
  density already provides a relatively good fit.}

\begin{figure}[h!]
\includegraphics[width=0.5\textwidth]{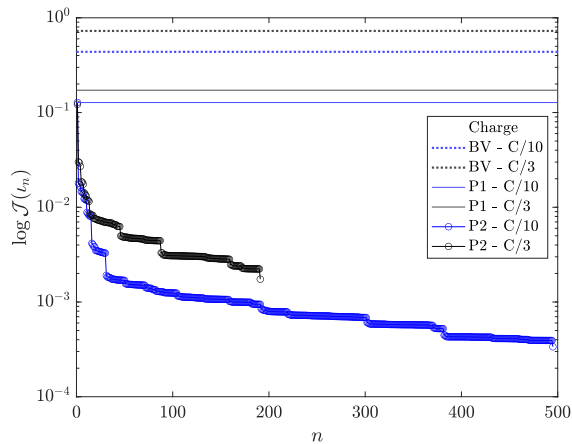}
\caption{Cost functional $\J_2$ as function of iterations in the
  solution of Problem P2 using Algorithm \ref{alg:P2} at the C-rate 
  of C$/10$ (blue lines with circles) and C$/3$ (black lines with
  circles). The horizontal lines represent the values of the error
  functional $\J_1$ corresponding to the 
  standard Butler-Volmer equation \eqref{eq:BV0}  (dotted lines) and
  the terminal values obtained in the solutions of Problem P1 
  at C-rates of C$/10$ (blue solid lines) and C$/3$ (black solid lines).
  \label{fig:Kevincost}}
\end{figure}

\begin{table}[h!]
\begin{tabular}{l l l}
Charge Rate & Parameter & Value\\ \hline
C$/10$  & anode reaction rate constant, $\widehat{k}_{0,-}$ & $1.076 \times 10^{-12}$\\
	& cathode reaction rate constant, $\widehat{k}_{0,+}$ & $4.416 \times 10^{-12}$ \\
	& anode anodic transfer coefficient, $\alphaanhat$ & $ 0.5667$\\ 
	& cathodic anodic transfer coefficient, $\alphaaphat$ & $0.4948$\\ 
	& electrolyte resistance, $\widehat{R}_{el}$ & $0.5813$\\ \hline
C$/3$   & anode reaction rate constant, $\widehat{k}_{0,-}$ & $2.428 \times 10^{-12}$ \\
	& cathode reaction rate constant, $\widehat{k}_{0,+}$ & $4.93 \times 10^{-12}$ \\
	& anode anodic transfer coefficient, $\alphaanhat$ & $0.4999$ \\ 
	& cathodic anodic transfer coefficient, $\alphaaphat$ & $  0.4959$ \\ 
	& electrolyte resistance, $\widehat{R}_{el}$ & $ 0.5625$\\
\end{tabular}
\caption{Material parameters obtained by solving Problem P1. \label{tab:KevinP1}}
\end{table}
The material parameters inferred by solving Problem P1 are summarized
in Table
\ref{tab:KevinP1}.  %where we see that the values of the anodic and cathodic transfer coefficients $\alpha_a$ and $\alpha_c$ do not differ much from the nominal value of 1/2.
We see that the optimal parameter values obtained at the two C-rates
are quite similar and from Figure \ref{fig:Kevini0}, we see that
the dependence on $c_{{\rm{s}},a}$ is similar to the standard Butler-Volmer
relation, but with a reduction in magnitude due to values of
$\widehat{k}_0$ being smaller by two orders of magnitude.

%The forms of the corresponding exchange currents obtained on
%the anode are shown in Figures \ref{fig:Kevini0}a,b where we see that
%their dependence on $c_s$ is quite similar to the reference case, but
%the magnitude is much smaller due to the value of $\widehat{k}_0$ being reduced
%by two orders of magnitude relative to the nominal value of $k_0$.

\begin{figure}
   \mbox{
     \subfigure[]{\includegraphics[width=0.5\textwidth]{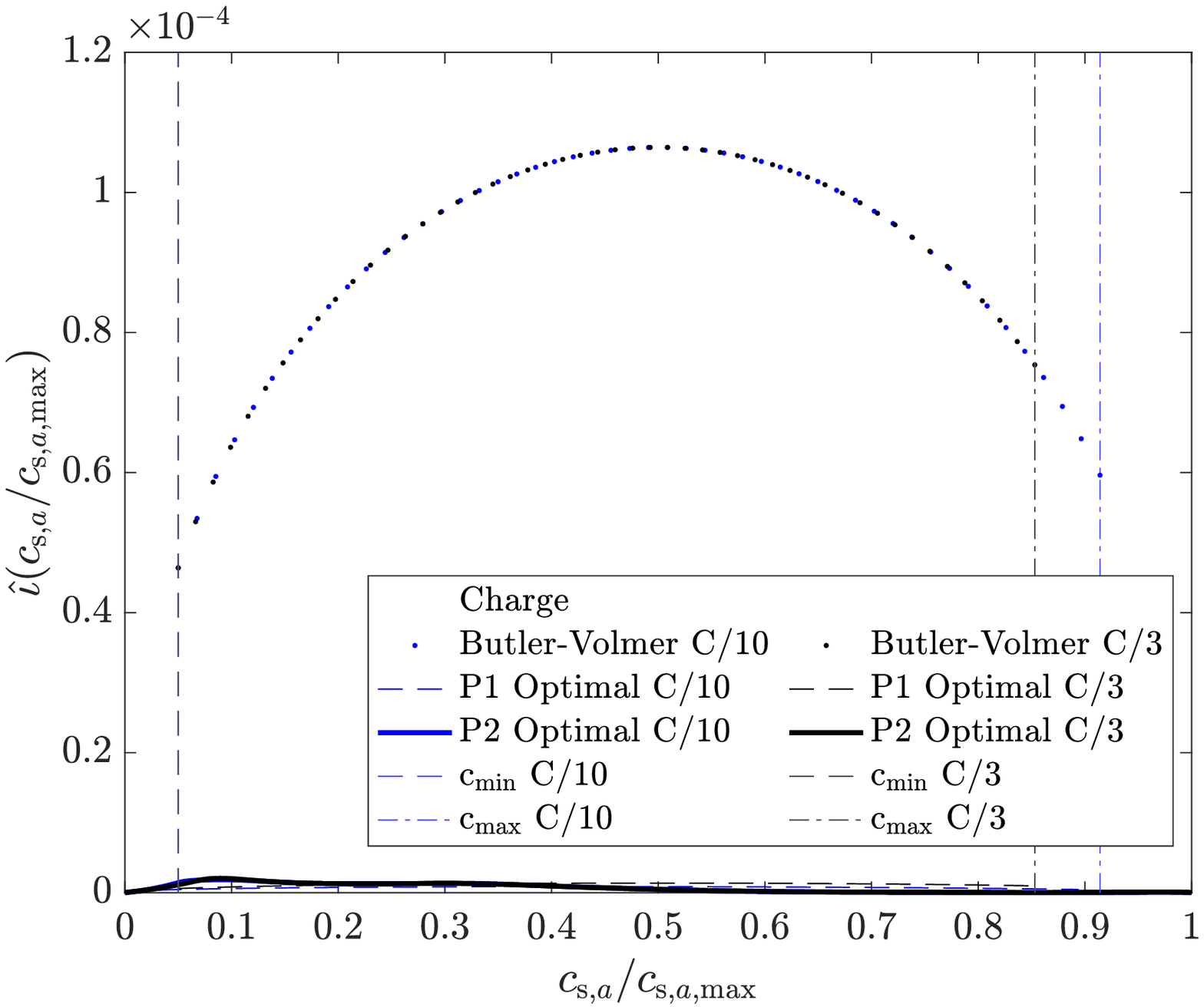}}
     \subfigure[]{\includegraphics[width=0.5\textwidth]{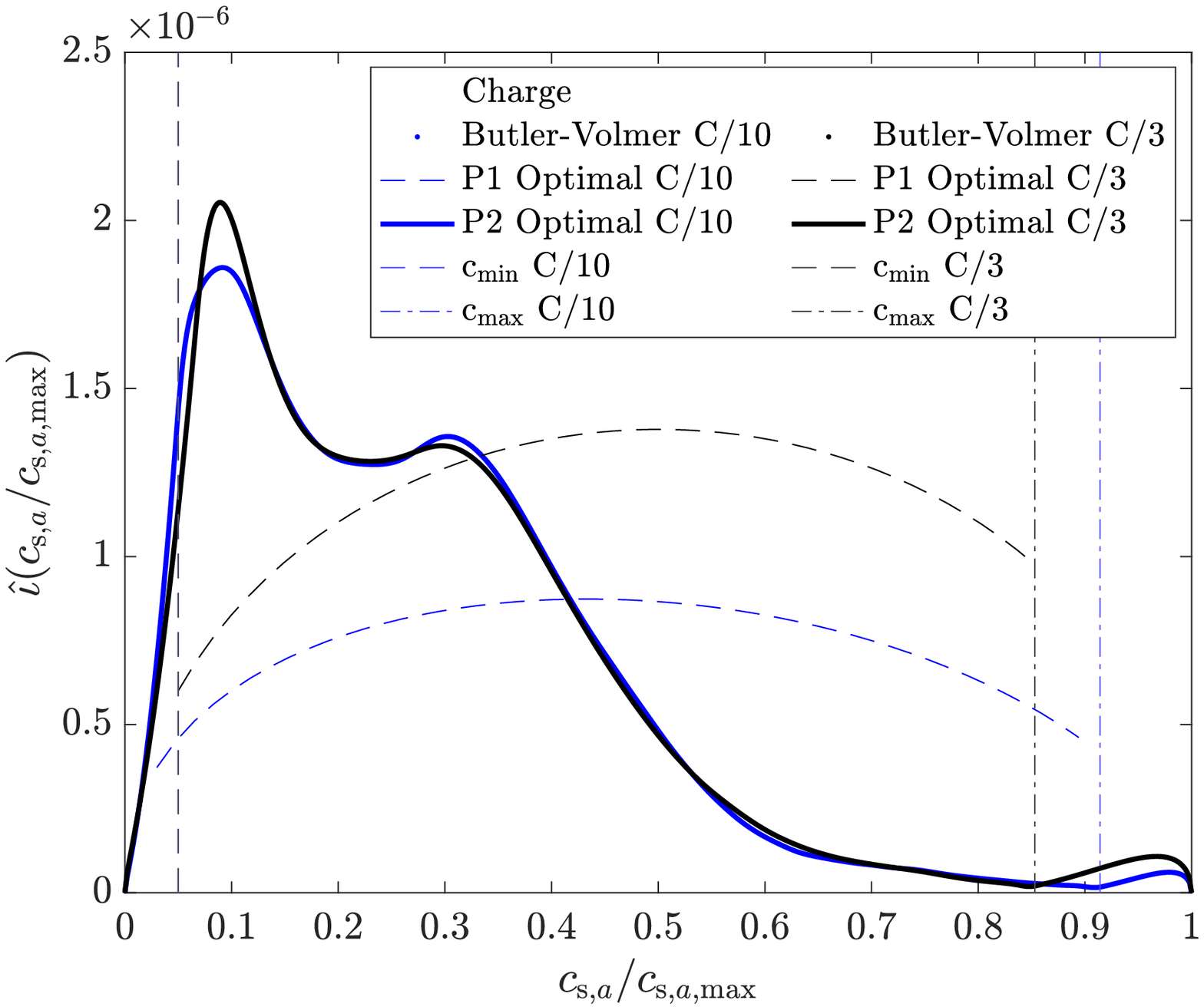}}}
   \caption{Optimal reconstructions of the exchange current obtained
     by solving Problem P1 (dashed lines) and Problem P2 (solid lines)
     as function of the normalized concentration $c_{\rm{s}} / c_{\rm{s,max}}$ for
     (a) the entire concentration range and (b) small values of the
     exchange current. Panel (a) also shows the standard Butler-Volmer
     relation \eqref{eq:BV0} with the nominal parameter values (dotted lines), 
     whereas the vertical lines
     represent the bounds of the identifiability intervals at the two
     C-rates.
     \label{fig:Kevini0}}
\end{figure}

The reconstructed overpotentials and the total cell potentials are
compared as functions of time to the experimental data in Figure
\ref{fig:KevinPot}. While the anode overpotential corresponding to
the Butler-Volmer relation with parameters calibrated by solving
Problem P1, cf.~Table \ref{tab:KevinP1}, has values coinciding with a
subset of the values observed in the actual measurements, its
dependence on time $t$ is incorrect. The same behaviour is also
evident, though less pronounced due to the presence of large
equilibrium potentials, in Figure \ref{fig:KevinPot}b where we show
the total potentials.

We now move on to discuss the results obtained by solving Problem P2.
In Figure \ref{fig:Kevincost}, we see that after a few hundred
iterations in Algorithm \ref{alg:P2}, the value of the error
functional $\J_2$ is reduced by several orders of magnitude. The
obtained optimal forms $\widehat{\iota}(c_{{\rm{s}},a})$ of the exchange
current shown in Figure \ref{fig:Kevini0} {are
  strikingly similar for both C-rates. The optimal solutions }are
skewed towards smaller concentrations where larger exchange currents
are predicted, unlike solutions of Problem P1. In addition, they
feature localized peaks at concentration values near the smallest
resolved values. As was the case with solutions of Problem P1, the
optimal functions $\widehat{\iota}(c_{{\rm{s}},a})$ inferred at the two
C-rates are rather similar, except that the peak at low concentrations
is steeper at the higher C-rate. Figure \ref{fig:KevinPot} shows
that, as expected from the small errors evident in Figure
\ref{fig:Kevincost}, both the overpotentials and the total potentials
predicted by the constitutive relation \eqref{eq:BV} with the optimal
exchange currents $\widehat{\iota}(c_{{\rm{s}},a})$ shown in Figure
\ref{fig:KevinPot} provide a better fit to the measured data.

\begin{figure}
  \mbox{
    \subfigure[]{\includegraphics[width=0.5\textwidth]{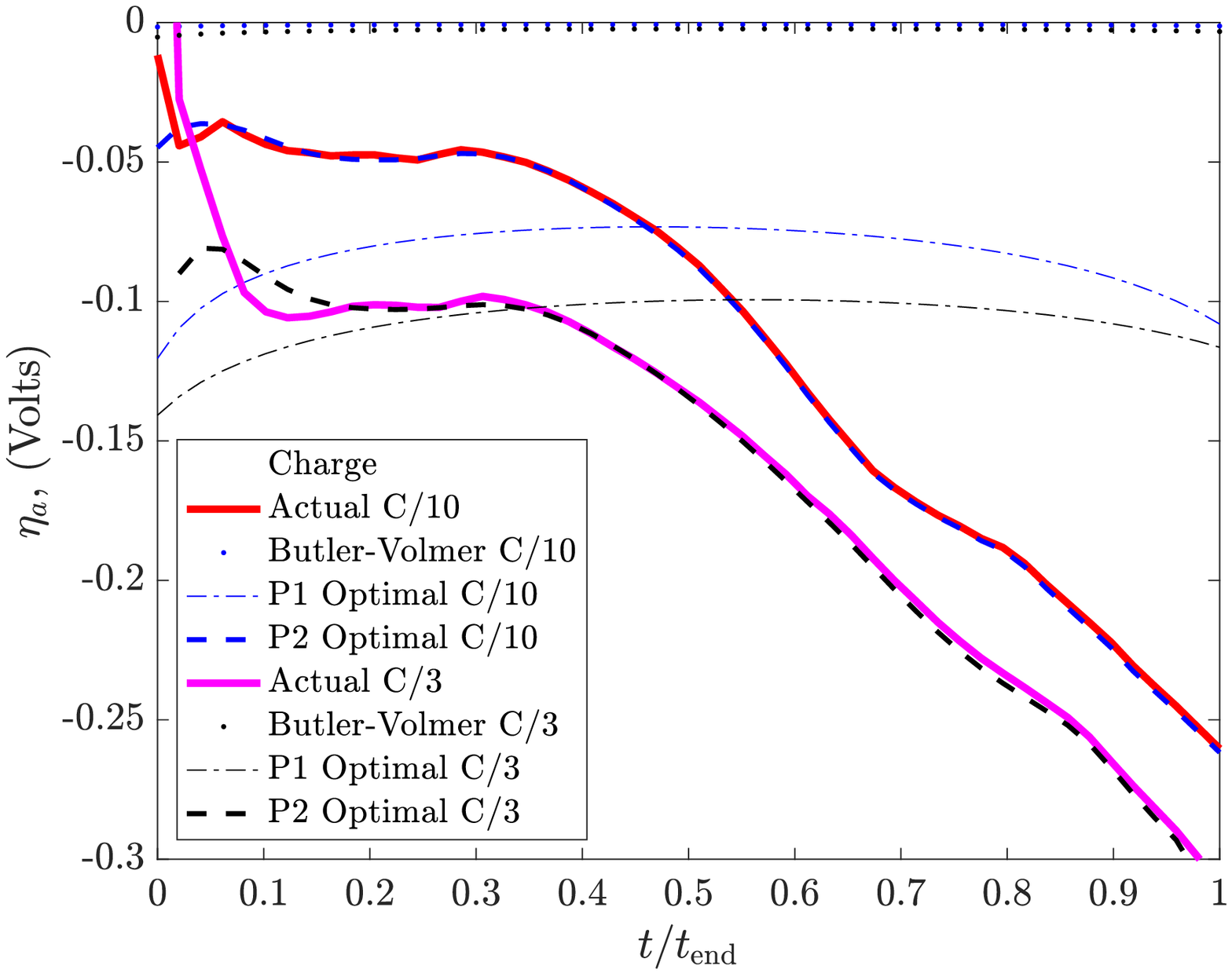}}
    \subfigure[]{\includegraphics[width=0.5\textwidth]{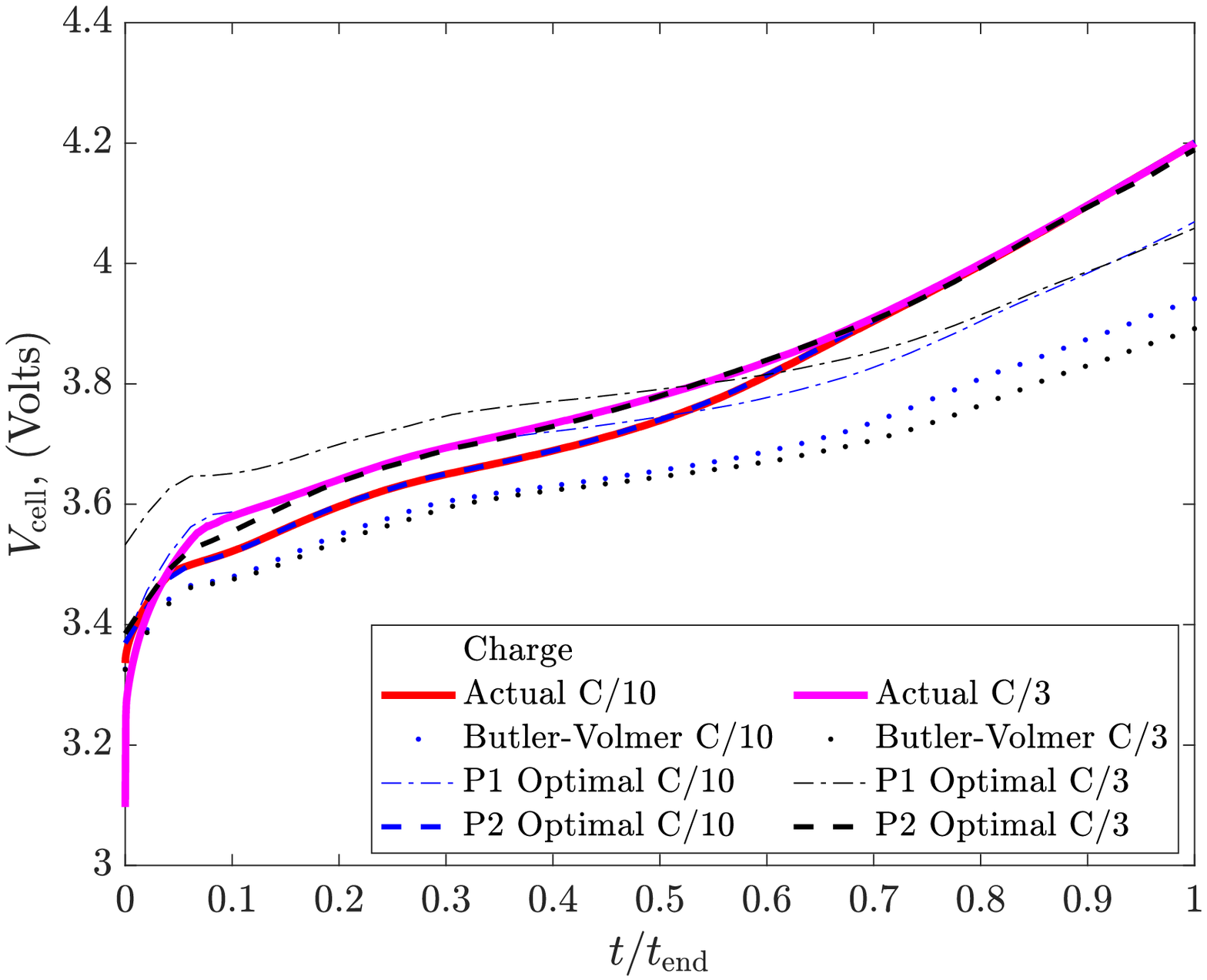}}}
  \caption{(a) Anode overpotentials and (b) total potentials predicted
    by the constitutive relation \eqref{eq:BV} using optimal
    reconstructions of the exchange current obtained by solving
    Problem P1 (dot-dashed lines) and Problem P2 (dashed lines) as
    function of the normalized time $t/t_{tot}$. Solid lines and
    dotted line represent the experimental measurements and
    predictions obtained using the standard Butler-Volmer relation
    \eqref{eq:BV0} with the nominal parameter values, respectively.
\label{fig:KevinPot}}
\end{figure}

\FloatBarrier

% \textcolor{blue}{Jamie: I'm struggling to rationalise Fig8a: P1 optimal undershoots the actual
% for early times and overshoots for later times. The current is determined, thus, to increase
% eta we decrease the exchange current and vice versa. This seems to be the opposite of what is
% happening. Probably worth talking about this as it might
% be worth giving the interpretation of this, once we have it\\
% Lindsey: I think this may also go along with why we see the spike in the P2 optimal curves, where we have a spike at early times.  Since the P1 form is ``locked" in to a semi-circular shape, there is not a lot of flexibility for fminunc to try to fit the total voltage, and we end up with somewhat of an average of the $\eta$ curve. }

%%%%%%%%%%%%%%%%%%%%%%%%%%%%%%%%%%%%%%%%%%%
 
\subsection{Moderate Discharge Rates}
\label{sec:Ferranresults}
Overall, the results obtained by solving Problems P1 and P2 using the
data for moderate discharge rates are qualitatively similar to the
results presented in the previous section. In our analysis below, we
will therefore focus on highlighting the differences.

\begin{table}[h!]
\begin{tabular}{l l l}
Charge Rate & Parameter & Value\\ \hline
C/2, discharge  & anode reaction rate constant, $k_{0,-}$ & $3.165 \times 10^{-12}$\\
	& cathode reaction rate constant, $k_{0,+}$ & $1.138 \times 10^{-12}$ \\
	& anode anodic transfer coefficient, $\alphaan$ & $ 0.5001$\\ 
	& cathodic anodic Transfer Coefficient, $\alpha_{c,a}$ & $0.4999$\\ 
	& electrolyte resistance, $R_{el}$ & $0.0010$\\ \hline
1C, discharge  & anode reaction rate constant, $k_{0,-}$ & $2.414 \times 10^{-12}$ \\
	& cathode reaction rate constant, $k_{0,+}$ & $0.867 \times 10^{-12}$ \\
	& anode anodic transfer coefficient, $\alphaan$ & $0.5001$ \\ 
	& cathodic anodic transfer coefficient, $\alpha_{c,a}$ & $  0.4999$ \\ 
	& electrolyte resistance, $R_{el}$ & $0.0010$\\
\end{tabular}
\caption{Material parameters obtained by solving Problem P1. \label{tab:FerranP1}}
\end{table}

The optimal parameters inferred by solving Problem P1 and collected in
Table \ref{tab:FerranP1} are in the same range as those obtained at
low charging rates, cf.~Table \ref{tab:KevinP1}, and also show little
variability between the two C-rates considered. %In particular, the
%anodic and cathodic transfer coefficients are both close to 1/2.
The values of the error functional \eqref{eq:J1} shown in Figure
\ref{fig:Ferrancost} reveal a similar improvement resulting from
solving Problem P1 as for the slow discharge rates, cf.~Figure
\ref{fig:Kevincost}. However, while in solutions of Problem P2 the
reduction of the least-square error $\J_2$ is still substantial, the
errors saturate at the level of $\mathcal{O}(10^{-2}-10^{-3})$.

The optimal form of expression \eqref{eq:i0} inferred by solving
Problem P1 and shown in Figure \ref{fig:Ferrani0} is again similar to
the nominal form, but has a smaller amplitude, roughly by a factor of
2. On the other hand, the optimal forms of the exchange current
$\widehat{\iota}(c_{{\rm{s}},a})$ inferred by solving Problem P2 reveal
significant differences with respect to the corresponding functions
obtained at low charging rates, cf.~Figure \ref{fig:Kevini0}. Most
importantly, in the present case large values of the exchange current
$\widehat{\iota}(c_{{\rm{s}},a})$ occur at large concentrations close to
$c_{{\rm{s}},a,{\rm{max}}}$.  The data in Figure \ref{fig:FerranPot} confirm the
small errors evident in Figure \ref{fig:Ferrancost} are achieved with
total potentials closely matching the measurement data.  We note that
by reducing the value of parameter $\ell$ in \eqref{eq:ipH1} it is in
fact possible to obtain overpotentials and total potentials that are
indistinguishable (within graphical resolution) from the measured
data. However, the price to be paid for that would be increasingly
irregular forms of the optimal exchange current $\widehat{\iota}(c_{{\rm{s}},a})$.

%The effect of parameter $\ell$ on the reconstructed exchange
%currents $\widehat{\iota}(c_{\rm{s}})$ is explored in Appendix
%\ref{smallell_results}.

%and computed from
%the optimal exchange currents $\widehat{\iota}(c_{\rm{s}})$ shown in Figures
%\ref{fig:KevinPot}a,b produce total potentials that more closely align with the measured data.  We note that it is possible to provide overpotentials and total potentials that are indistinguishable (within graphical resolution) from the measured data by lowering the $\ell$ parameter.  These results are showcased in Appendix \ref{smallell_results}.

\begin{figure}[t!]
\includegraphics[width=0.5\textwidth]{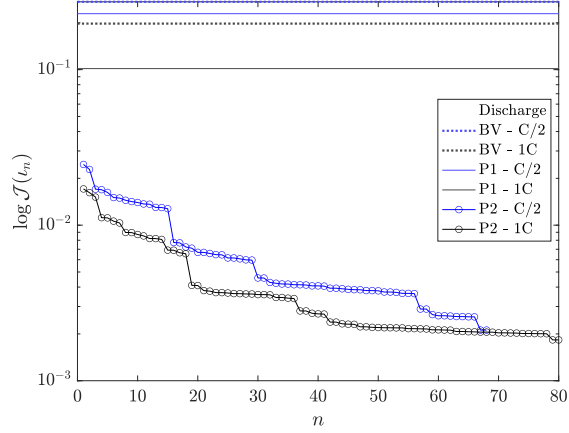}
\caption{Cost functional $\J_2$ as function of iterations in the
  solution of Problem P2 using Algorithm \ref{alg:P2} at the C-rate 
  of C/2 (blue lines with circles) and 1C (black lines with
  circles). The horizontal lines represent the values of the error
  functional $\J_1$ corresponding to the 
  standard Butler-Volmer equation \eqref{eq:BV0}  (dotted lines) and
  the terminal values obtained in the solutions of Problem P1 
  at C-rates of C/2 (blue solid lines) and 1C (black solid lines).
  \label{fig:Ferrancost}}
\end{figure}

\begin{figure}
   \includegraphics[width=0.5\textwidth]{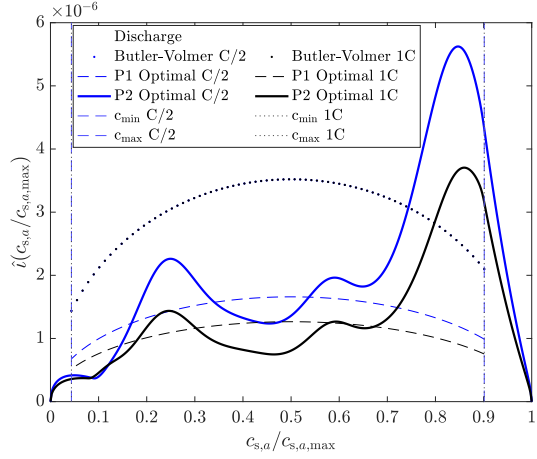}
   \caption{Optimal reconstructions of the exchange current obtained
     by solving Problem P1 (dashed lines) and Problem P2 (solid
     lines) as function of the normalized concentration     
     $c_{\rm{s}} / c_{\rm{s,max}}$ for the entire concentration range. The
     standard Butler-Volmer relation \eqref{eq:BV0} with the nominal
     parameter values is shown using a dotted line whereas the vertical
     lines represent the bounds of the identifiability intervals at
     the two C-rates.
     \label{fig:Ferrani0}}
\end{figure}

\begin{figure}
  \mbox{
    \subfigure[]{\includegraphics[width=0.5\textwidth]{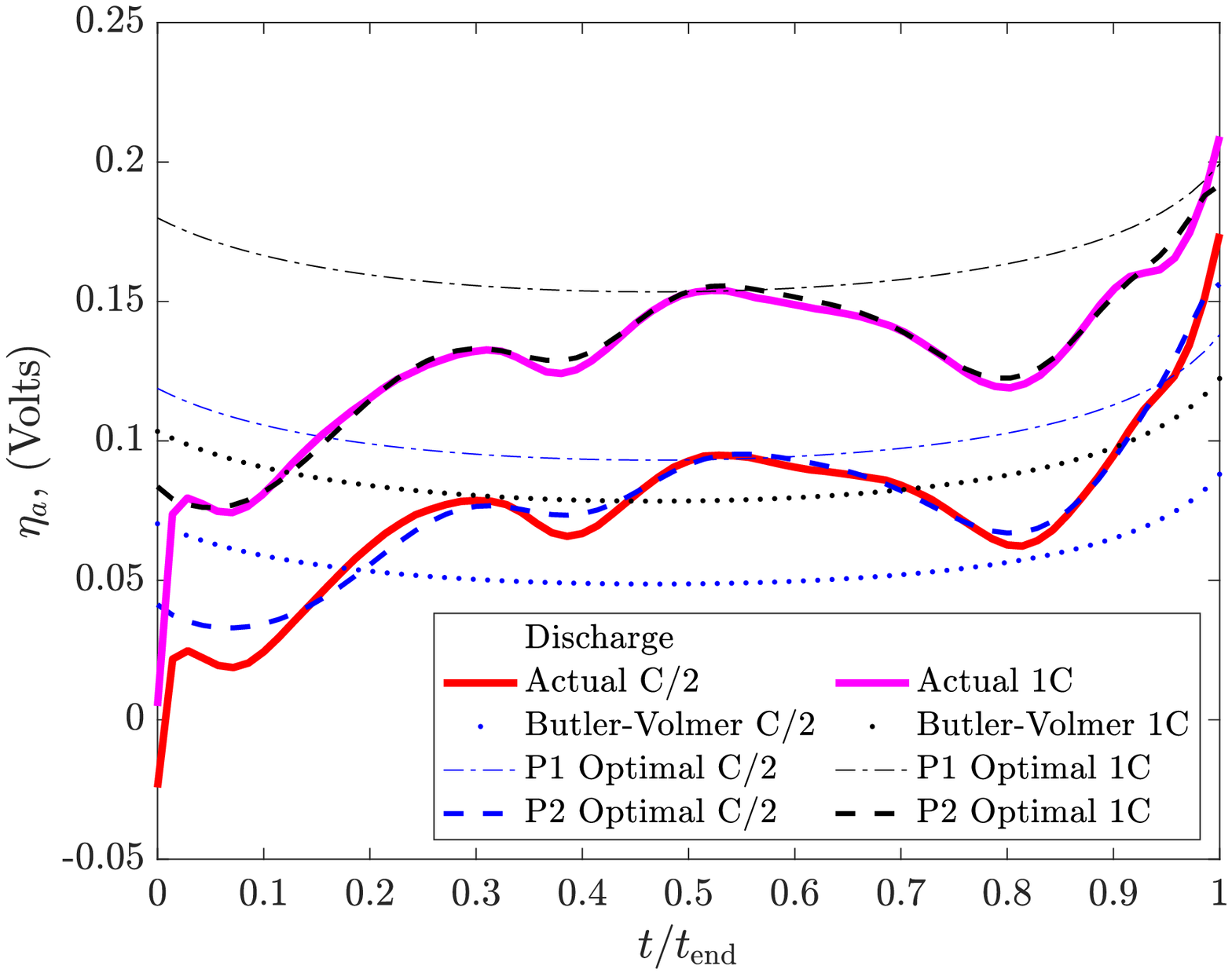}}
    \subfigure[]{\includegraphics[width=0.5\textwidth]{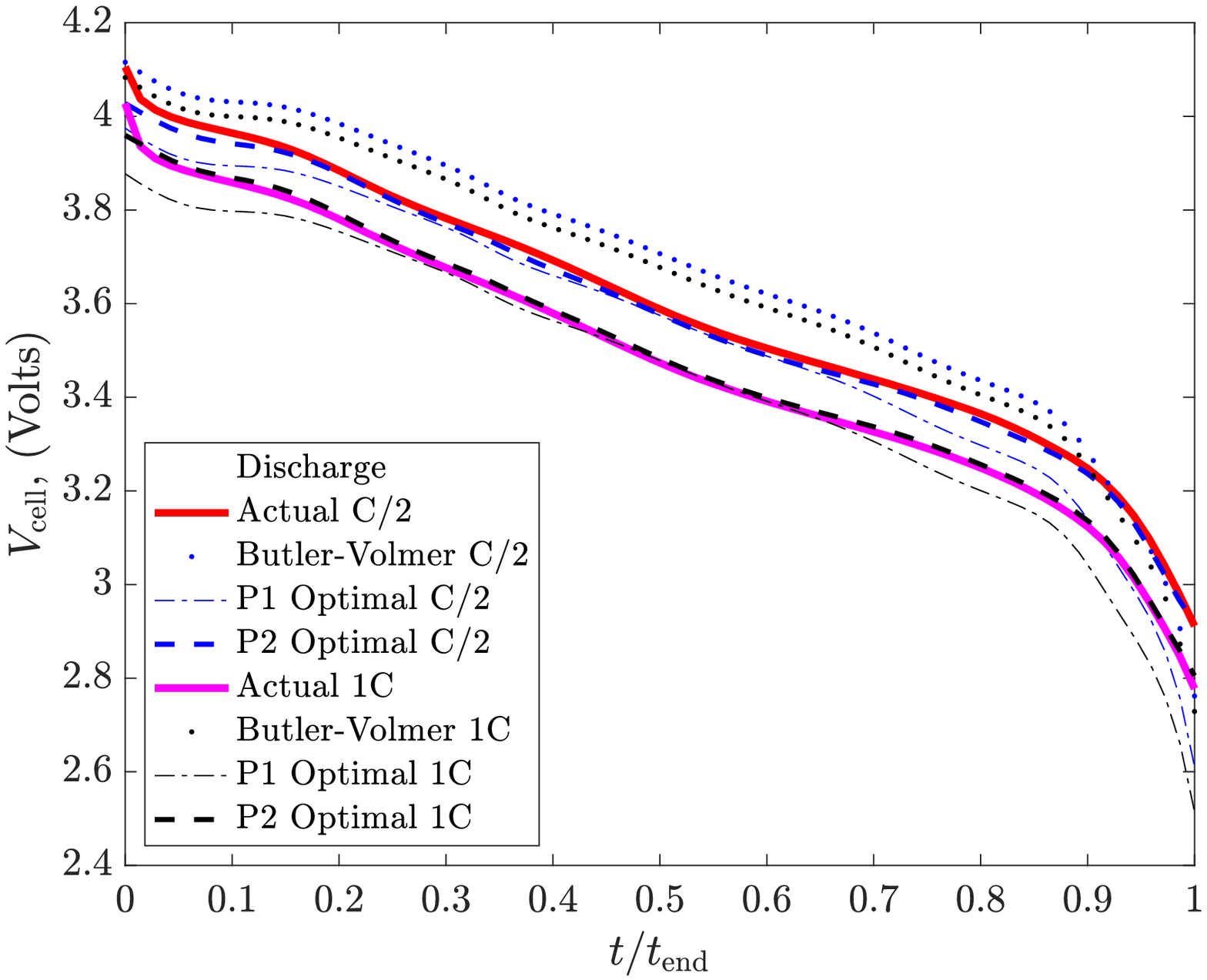}}}
  \caption{(a) Anode overpotentials and (b) total potentials predicted
    by the constitutive relation \eqref{eq:BV} using optimal
    reconstructions of the exchange current obtained by solving
    Problem P1 (dot-dashed lines) and Problem P2 (dashed lines) as
    function of the normalized time $t/t_{tot}$. Solid lines and
    dotted line represent the experimental measurements and
    predictions obtained using the standard Butler-Volmer relation
    \eqref{eq:BV0} with the nominal parameter values, respectively.
\label{fig:FerranPot}}
\end{figure}

\begin{figure}
\includegraphics[width=0.6\textwidth]{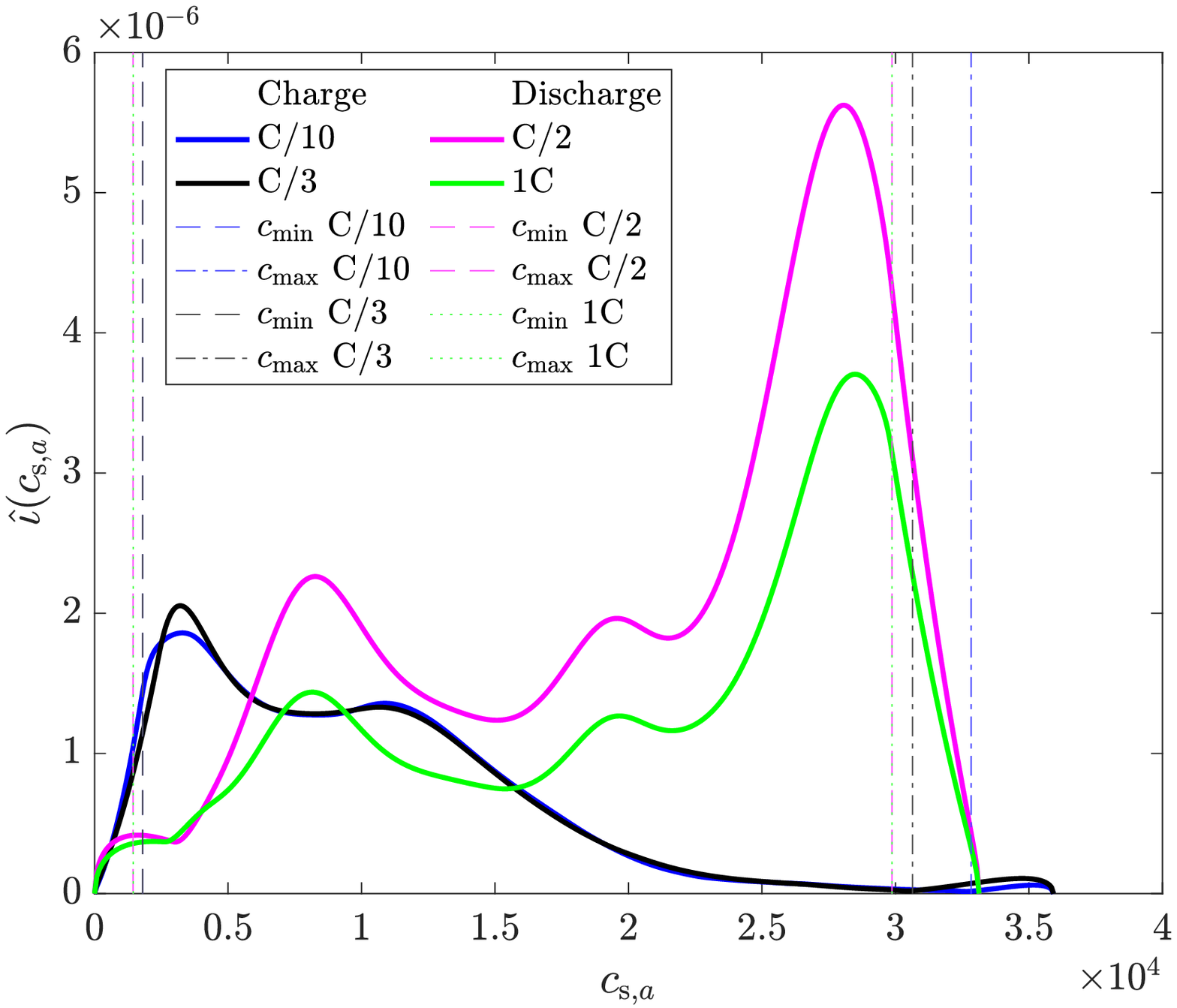}
\caption{Comparison of optimal reconstructions $\widehat{\iota}(c_{\rm{s}})$
  of the exchange current obtained by solving Problem P2 using all
  four datasets considered.}
\label{fig:iota_all}
\end{figure}

Finally, we attempt to extract some universal features of the optimal
exchange currents determined using measurements acquired at the
different charge/discharge rates and, to this end, in Figure
\ref{fig:iota_all} we compare the optimal reconstructions
$\widehat{\iota}(c_{{\rm{s}},a})$ obtained by solving Problem P2 for all
considered datasets. We see that there is a range of concentrations
$0.25\times 10^4 \lessapprox c_{{\rm{s}},a} \lessapprox 2\times 10^4$ where the
inferred exchange currents are qualitatively similar revealing some
possibly universal features which include a steep increase of
$\widehat{\iota}(c_{{\rm{s}},a})$ at lower concentrations $c_{{\rm{s}},a}$ followed by a
gentle and nonmonotonic decrease at higher concentrations. Outside
that range the optimal reconstructions $\widehat{\iota}(c_{{\rm{s}},a})$ exhibit
distinct behaviour specific to each dataset.

\FloatBarrier

%%%%%%%%%%%%%%%%%%%%%%%%%%%%%%%%%%%%%%%%%%%
%%%%%%%%%%%%%%%%%%%%%%%%%%%%%%%%%%%%%%%%%%%
%Concluding Remarks

%%%%%%%%%%%%%%%%%%%%%%%%%%%%%%%%%%%%%%%%%%%
\section{Discussion and Concluding Remarks \label{conclusion}}

In this work, we have developed an inverse modelling approach to infer
an optimal form of the concentration-dependent exchange current
$\widehat{\iota}(c_{{\rm{s}},a})$ appearing in a generalization
\eqref{eq:BV} of the Butler-Volmer relation describing intercalation
of Li ions into an active material. While many inverse problems have
already been solved in the area of
electrochemistry\cite{bvp11,sethurajan2015accurate,doi:10.1002/jcc.25759,bukshtynov2013889,escalante2020discerning,escalante2021uncertainty},
this is to the best of our knowledge the first attempt to infer
constitutive relations describing an interfacial phenomenon. Although
the proposed approach relies on many simplifications, most
importantly, on the use of the SPM to describe the transport of Li, it
is shown to be computationally robust and thermodynamically
consistent. In particular, the model equipped with the inferred
optimal exchange currents $\widehat{\iota}(c_{{\rm{s}},a})$ predicts
very accurately the time evolution of the cell voltage which
  strongly suggests that ansatz \eqref{eq:BV} and assumptions A$1$--A$3$
  are physically consistent (we note that such a good match with
  experimental data need not occur if the model used in the inverse
  problem does not accurately describe the physics of the
  problem\cite{escalante2020discerning}).  Interestingly, by
exploiting the coupling between different scales in the model, our
approach allows us to reconstruct a constitutive relation defined on
the microscale based on macroscale measurements. The
  variational formulation of the inverse problem involves averaging
  of the reconstruction errors over time and concentrations such that
  the vanishing of the exchange current for small and large
  concentrations, cf.~Assumptions A$2$ and A$3$, does not cause problems.

We emphasize that the observations made below are predicated on the
accuracy of the material properties which were assumed known in the
solution of Problems P1 and P2, cf.~Supporting Information (Appendix
\ref{exppara}).  While a significant effort was made to determine
these parameters accurately, their inherent uncertainties may have
some effect on the obtained results. In particular, one material
property known to be difficult to determine with high accuracy is the
diffusivity $D_{\rm{s}}(c_{\rm{s}})$ appearing in \eqref{eq:SPM}
\cite{park2010review}. To probe the effect of its uncertainty, we have
solved Problems P1 and P2 in which this diffusivity was allowed to
vary over several orders of magnitude. While quantitative details of
the reconstructed constitutive relations were different, the key
qualitative features of the optimal exchange currents as discussed
below were essentially unchanged. That being said, the present study
represents only a proof of the concept and more work needs to be done
to understand the extent to which our observations about optimal forms
of the exchange current could generalize to other battery materials.

The optimal exchange currents $\widehat{\iota}(c_{{\rm{s}},a})$
reconstructed by solving Problem P2 based on datasets corresponding to
different charging/discharging rates show a qualitatively similar
dependence on $c_{{\rm{s}},a}$ for intermediate concentrations
$0.25\times 10^4 \lessapprox c_{{\rm{s}},a} \lessapprox 2\times 10^4$,
cf.~Figure \ref{fig:iota_all}. This behaviour is different from the
form of the exchange current predicted by the standard Butler-Volmer
relation \eqref{eq:i0}, even after its parameters are calibrated by
solving Problem P1. Outside that range, the optimal exchange currents
exhibit non-universal features specific to individual cases. In
particular, the optimal exchange currents inferred from data
representing charging and discharging feature sharp peaks at,
respectively, low and high concentrations, cf.~Figures
\ref{fig:Kevini0} and \ref{fig:Ferrani0}, indicating possible
hysteresis effects in the constitutive relations.  \revtt{In addition,
  the results shown in Figure \ref{fig:iota_all} indicate that the
  exchange current density tends to be higher during discharging which
  is consistent with the observations made in earlier studies
  \cite{jow2018factors} where this effect was partly attributed to a
  higher de-solvation energy barrier.}

The sharp peaks characterizing the optimal exchange current at low
{concentrations when charging} and at high concentrations {when
  discharging} suggest that the electrode kinetics are faster at the
beginning of the cycle. This might be due to some interfacial
phenomena not accounted for in our model such as, e.g., capacitance
effects and variation in the activity
  coefficient\cite{christensen2006stress}.  It is known that the
structure of the double layer and ion adsorption can affect electrode
kinetics\cite{bard2001electrochemical}.  These phenomena have been
implicated in causing anomalies in rate constants and the exchange
current, as well as variations in current-potential
plots\cite{bard2001electrochemical}.  The behaviour observed at early
times in the charge and discharge cycles could be compensating for
residual effects produced by the double layer at the electrode
interface.  In this context, modelling efforts have produced
corrections to the Butler-Volmer equation, such as the Frumkin
correction which allows the rate constant to be dependent on
potential, in order to take into account the effects of the electric
double layer and ion adsorption in electrode
kinetics\cite{bard2001electrochemical}. On the other hand, a
  thermodynamically consistent derivation of the exchange current
  density shows that its functional form is dependent on the rate
  constant, concentration, and the activity coefficient
  \cite{christensen2006stress}. At the same time, the activity of the
  relevant species is often assumed to be $1$, which may not be
  reflective of the dynamics in the system and could lead to its
  non-ideal behaviour\cite{bard2001electrochemical}.

Another possible reason for the presence of large exchange currents at
low and high concentrations could be due to the system moving away
from equilibrium at the beginning of the cycle protocol when there is
a uniform distribution of Li ions in the electrolyte throughout the
cell\cite{bard2001electrochemical,nt04}.  When charging begins, there
is a richness of Li near the particle and as the protocol continues,
Li moves in the electrolyte towards the anode.  The spike observed at
early times of the charging protocol could possibly be attributed to
the richness of Li near the electrode, which could contribute to a
larger exchange current and, consequently, a larger current density at
the interface.  In the case of a discharge protocol, there is a
richness of Li ions in the anode near the surface and as the cycle
continues, Li ions must diffuse from deeper in the electrode to the
surface. {This is not accounted for in the SPM because
  the concentration $c_{\rm{e}}$ in the electrolyte is assumed
  constant,~cf.~Section \ref{sec:SPM}.}

The standard formulation of the Butler-Volmer model is based on a
simple, single step reaction and as such does not allow for formation
of short-lived intermediate products. This is a key assumption which
may not be valid in experimental settings; for example, Li-ion
batteries often suffer from Li-ion plating, which is a result of fast
charging and Li metal plating out onto the surface of the
electrode\cite{o2021lithium}.  While this plated material
{may} absorb into the electrode after a certain
relaxation time, the side reaction is not captured by the
Butler-Volmer kinetic relation
{\cite{zinth2014lithium,uhlmann2015situ,wandt2018quantitative}}.
Some modelling efforts have attempted to extend the Butler-Volmer to
account for such
effects\cite{yang2017,yang2018look,o2020physical,sahu2023continuum}.

It is also possible that with peaks at low and high Li concentrations
the optimal exchange currents $\widehat{\iota}(c_{{\rm{s}},a})$ may be
compensating for certain effects other than interfacial phenomena
which are not accounted for in the model. One such possibility has to
do with the nature of Li transport in the active material, which under
certain circumstances is known to lead to phase transformations and in
turn can significantly affect the
diffusivity\cite{levi2003noteworthy}.  The spike in the reconstructed
exchange current is located at approximately $12$--$15\%$ capacity,
cf.~Figure \ref{fig:Kevini0}, which could be a result of a rapid
increase in the diffusion directly after the phase transition in
graphite.  When comparing the plots of ${U_{\rm{eq}}^a}$ and $\frac{d {U_{\rm{eq}}^a}}{d
  (c_{{\rm{s}},a}/c_{{\rm{s}},a,{\rm{max}}})}$ vs. $c_{{\rm{s}},a}/c_{{\rm{s}},a,{\rm{max}}}$ (the latter not shown for
brevity), there does appear to be agreement between the minima/maxima
observed in the optimal exchange current and the remnants of these
phase transitions captured by the equilibrium potential, ${U_{\rm{eq}}^a}$.
The reconstructed constitutive relations could be therefore
compensating for an inadequate transport model used for graphite.  It
is known that diffusion models are difficult to quantify
experimentally and that the physics of transport in the solid is
complicated by interfacial phenomena\cite{escalante2020discerning}.
The literature often reports use of different models of transport in
the solid phase based on Fickian diffusion, non-linear diffusion and
the Cahn-Hilliard phase separation \cite{escalante2020discerning}.
However, the anomalous behaviour of optimal exchange currents at low
and high concentrations persisted even when a strongly non-monotonic
diffusivities $D(c_{{\rm{s}},i})$, similar to the properties measured or inferred
for graphite\cite{levi2003noteworthy,escalante2020discerning} were
used in the transport equation \eqref{eq:SPM}. 

More work based on additional datasets is thus needed in order to
understand the precise origins of the universal features in the
optimal exchange currents $\widehat{\iota}(c_{{\rm{s}},i})$ inferred based on
different datasets.  In particular, it would be beneficial to
understand what these universal features can tell us about the
validity of the different assumptions used to derive the standard
Butler-Volmer relation \eqref{eq:BV0}, so that these assumptions could
be suitably refined to produce more accurate constitutive models.
{Even though we have given proof of concept for the
  approach and its application to interfacial phenomena in
  electrochemistry, the question of whether it can be used to learn a
  general improvement to the widely-used ``standard'' exchange current
  density in the Butler-Volmer relation \eqref{eq:BV0} remains open.
  Next steps towards answering this question might include the
  application of the method using the full DFN framework, thereby
  generalizing the physics and allowing us to examine higher C-rates.
  Or, arguably better, one would like to apply this approach to
  experimental data specifically designed for our purposes, e.g., this
  might involve a non-porous electrode such that there are fewer
  device and model parameters, thereby spotlighting the dependence on
  the interfacial constitutive relation. In the best case, it may then
  be possible to learn something about the fundamental
  electrochemistry occurring at material interfaces.}

As is evident from the discussion in Section \ref{sec:sensitivity}, a
key limiting factor of our approach is its sensitivity to noise. The
main issue is that the useful information about the interfacial
reaction kinetics is encoded in the overpotential $\eta$ which at low
C-rates represents only a small fraction of the measured cell voltage,
cf.~Tables \ref{Kevintable}--\ref{Ferrantable}. On the other hand, at
higher C-rates where this proportion increases, the SPM loses its
validity. A preferred solution would thus be to formulate the inverse
problem based on the full DFN model (or one of its reductions, but
more complete than the SPM), as this would make it possible to use
data obtained at higher C-rates where the overpotential $\eta$
represents a larger proportion of the measured voltage. Such an
inverse problem will be mathematically more complicated since, unlike
the case of the SPM, the solutions of the DFN model depend on the form
of the constitutive relation describing ion intercalation.
Another limitation of the proposed approach is that the optimal
  forms of the constitutive relation depend on the C-rate, cf.~Figures
  \ref{fig:Ferrani0} and \ref{fig:iota_all}. While this is not
  uncommon when an inverse problem is solved using multiple
  experimental datasets, this complicates application of the method in
  practice when one needs to model battery cells at different C-rates.
  A possible solution is to infer an ``average'' optimal exchange
  current by simultaneously fitting voltage curves corresponding to
  several C-rates. When extending the proposed analysis to
  alternative materials, considerations such as multiphase
  intercalation\cite{verbrugge2022cation,baker2019multi} and mixed
  conduction effects\cite{johnson2022unconventional} would need to be
  taken into account. The former would necessitate redefining
  diffusivity in the bulk whereas the latter would require
  simultaneous solution of the diffusion and conduction problems at
  the particle level. We plan to pursue such extensions of the
present approach in the near future.

In principle, the framework developed in the present investigation
could be used to study other interfacial phenomena, where experimental
data can be leveraged to learn the optimal forms of the relevant
constitutive relations.  Some interesting problems of this type
include the reconstruction of constitutive relations governing the
Hodgkin-Huxley dynamics of ion transfer through a cell membrane
\cite{hodgkin1952quantitative}, the Shockley-Read-Hall recombination
of electrons in a semiconductor
\cite{shockley1952statistics,hall1952electron}, or the charge density
of an adsorbed layer formed between a charged interface and an
electrolyte \cite{bousse1983influence}.

%%%%%%%%%%%%%%%%%%%%%%%%%%%%%%%%%%%%%%%%%%%
\begin{acknowledgement}
  LD, KJS, GRG and BP were supported by a Collaborative Research \&
  Development grant \# CRD494074-16 from Natural Sciences \&
  Engineering Research Council of Canada. {JF and SS
    were supported by the Faraday Institution, UK (grant number
    FIRG003).}

\end{acknowledgement}

% \begin{suppinfo}
%   The Supporting Information discusses constraints on the behavior of
%   the exchange current $\iota(c_{\rm{s}})$ for small and large concentrations
%   (Appendix \ref{sec:appBC}), describes the computation of the
%   gradient $\nabla \mathcal{J}_2(\iota(c_{{\rm{s}},i}))$ of the objective
%   functional, cf.~\eqref{graddescent}, (Appendix \ref{sec:grad}) and
%   summarizes experimental parameters (Appendix \ref{exppara}).
% \end{suppinfo}

%%%%%%%%%%%%%%%%%%%%%%%%%%%%%%%%%%%%%%%%%%%
%%%%%%%%%%%%%%%%%%%%%%%%%%%%%%%%%%%%%%%%%%%
%Appendices

%%%%%%%%%%%%%%%%%%%%%%%%%%%%%%%%%%%%%%%%%%%
\appendix
\section{Limiting Behaviour of Constitutive
  Relationships at Small and Large Concentrations \label{sec:appBC}}

We derive conditions that must be satisfied by the function
$\iota(c_s)$ for $c_s \rightarrow 0,c_{s,max}$ in order for the
constitutive relation \eqref{eq:BV} to be physically consistent. We
note that $c_s = \iota^{-1}(\Delta \phi)$ relates the potential
$\Delta \phi$ to the concentration $c_s$ and the range of
$\iota^{-1}(\Delta \phi)$ is $0 \leq \iota^{-1}(\Delta \phi) \leq
c_{s,max}$. The equilibrium potential can be separated into a bounded
portion $u_{eq}$ and an unbounded part given by
\cite{richardson2020charge}
\begin{equation}
  {U_{eq}^i} (\Delta \phi) = \frac{RT}{F(\alpha_a + \alpha_c)} \log \left( \frac{k_c c_e (c_{s,max} - \iota^{-1}(\Delta \phi) )}{k_a  \iota^{-1}(\Delta \phi)} \right) + u_{eq}. \label{boundedUeq}
\end{equation}
Plugging this expression  into \eqref{eq:BV1}, we obtain
\begin{equation}
j =F\iota(c_s) \left[e^{\frac{\alpha_a F}{RT} \left( \Delta \phi-u_{eq}\right)}\left( \frac{k_c c_e (c_{s,max} - c_s)}{k_ac_s} \right)^{-\frac{\alpha_a}{\alpha_a + \alpha_c}}- e^{\frac{\alpha_c F}{RT} \left(\Delta \phi -u_{eq}\right)}  \left( \frac{k_c c_e (c_{s,max} - c_s)}{k_ac_s} \right)^{\frac{\alpha_c}{\alpha_a + \alpha_c}} \right],  \label{fullBVBC}
\end{equation}
which should vanish when $c_s \rightarrow 0$ as the electrode cannot
be depleted at vanishing concentrations, and when
$c_s \rightarrow c_{s,max}$ as the electrode cannot continue to be
filled once the maximum concentration is reached. In order to
overcome the divergence of the factors in the parentheses in these
limits, we must require the function $\iota(c_s)$ to vanish
sufficiently rapidly as $c_s \rightarrow 0,c_{s,max}$. More
specifically, expression \eqref{fullBVBC} will vanish in these limits
provided
\begin{equation}
\iota(c_s) = \begin{cases}
  \mathcal{O}\left[\left( \frac{k_ac_s} {k_c c_e (c_{s,max} -
        c_s)}\right)^{\omega}\right],  & \omega >
  \frac{\alpha_c}{\alpha_a + \alpha_c}, \quad  \textrm{as}  \ c_s \to 0, \\
   \mathcal{O}\left[ \left( \frac {k_c c_e (c_{s,max} -
         c_s)}{k_ac_s}\right)^{\gamma} \right],  & \gamma >
   \frac{\alpha_a}{\alpha_a + \alpha_c}, \quad \textrm{as} \ c_s \to c_{s,max} \\
\end{cases}.
  \label{eq:iotaBC}
\end{equation}
Condition \eqref{eq:iotaBC} must be incorporated into the formulation
of Problem P2.

%%%%%%%%%%%%%%%%%%%%%%%%%%%%%%%%%%%%%%%%%%%
\section{Computation of Gradients \label{sec:grad}}

We begin by computing the G\^ateaux (directional) derivative of the
error functional \eqref{eq:J2} defined as
$\mathcal{J}_2^{\prime}(\iota(c_s); \iota^{\prime}(c_s)) = \lim
\limits_{\epsilon \to 0}
\epsilon^{-1}\left[\mathcal{J}_2(\iota(c_s)+\epsilon
  \iota^{\prime}(c_s))- \mathcal{J}_2(\iota(c_s))\right]$, where
$\iota'(c_s)$ is an arbitrary perturbation of $\iota(c_s)$, and obtain
\begin{equation}
\mathcal{J}_2^{\prime}(\iota(c_s); \iota^{\prime}(c_s)) = \int\limits_0^{t_f} \eta^{\prime}(\iota(c_s(t));\iota^{\prime}(c_s))\left[
 \eta(\iota(c_s(t)))- \overline{\eta}(\iota(c_s(t))) \right]  \, dt.
\label{eq:dJ2}
\end{equation}
The Riesz representation theorem then asserts that the G\^ateaux
differential \eqref{eq:dJ2} can be expressed as
\begin{equation}
\mathcal{J}_2^{\prime}(\iota(c_s);\iota^{\prime}(c_s))   = \left\langle
  \nabla^{L^2} \mathcal{J}, \iota^{\prime} \right\rangle_{L^2} = 
\int_{0}^{c_{s,max}} \nabla^{L^2} \J_2(c_s) \cdot \iota^{\prime}(c_s)
\, dc,
\label{eq:Riesz}
\end{equation}
where $\nabla^{L^2} \J_2(c_s)$ is the gradient of the error functional
computed with respect to the $L^2$ inner product
$\langle\cdot,\cdot\rangle_{L^2}$. To transform \eqref{eq:dJ2} into a
form consistent with \eqref{eq:Riesz}, where $\iota'(c_s)$ appears as
a factor and integration is performed with respect to $c_s$, we define
$\iota_\epsilon(c_s) = \iota(c_s) + \epsilon \iota^{\prime}(c_s)
+\mathcal{O}(\epsilon^2)$ so that the perturbed Butler-Volmer relation
is $j_\epsilon = \iota(c_s) \psi(\Delta \phi, c_s)+ \epsilon
\iota^{\prime}(c_s) \psi(\Delta \phi, c_s)+\mathcal{O}(\epsilon^2)$.
Then, the corresponding cell overpotential, $\eta$, is
\begin{align}
\eta& = \psi^{-1}\left( \frac{I}{A F  \left[ \iota(c_s) + \epsilon
      \iota^{\prime}(c_s) + \mathcal{O}(\epsilon^2)\right]} \right) \nonumber \\
        &=  \psi^{-1}\left( \frac{I}{A F  \iota(c_s)}
          \right)+
          \epsilon \underbrace{\left(\left[\psi^{-1}\left( \frac{I}{A F
          \iota(c_s)}\right)\right]^{\prime}
          \frac{- I}{(A F ) (\iota(c_s))^2}
          \right)\iota^{\prime}(c_s)}_{\eta^{\prime}(\iota(c_s(t));\iota^{\prime}(c_s))} + \mathcal{O}(\epsilon^2). \label{Vcellexpand}
\end{align}
Using \eqref{Vcellexpand} in \eqref{eq:dJ2}, we obtain
\begin{align}
 \mathcal{J}_2^{\prime}(\iota(c_s); \iota^{\prime}(c_s)) & =
  \int\limits_0^{t_f} \frac{ \frac{- I\iota^{\prime}(c_s(t)) }{(A F )
  (\iota(c_s(t)))^2}  }{\psi^{\prime} \left( \frac{I}{AF
  \iota(c_s(t))}\right)} \left[ \eta(\iota(c_s(t)) )- \overline{\eta}(\iota(c_s(t)))\right] \, dt \nonumber \\
 & = \int\limits_{0}^{c_{s,max}} \frac{1}{\psi^{\prime} \left(
   \frac{I}{AF  \iota(c_s)}\right)} \frac{- I\iota^{\prime}(c_s)}{(A
   F) (\iota(c_s))^2}\left( \eta(\iota(c_s) )-
   \overline{\eta}(\iota(c_s))\right) \, \left| \frac{d t}{dc_s}
   \right| \, dc_s, \label{eq:dJ3}
\end{align}
where the integration variable has been changed from $t$ to $c_s$
using the relation $c_s = c_s(t)$ defined by the numerical solution of
the SPM. We note that \eqref{eq:dJ3} is now in a form consistent with
the Riesz representation \eqref{eq:Riesz}, which allows us to define
the $L^2$ gradient as
 \begin{equation}
 \nabla^{L^2} \mathcal{J}_2 =  \left(\frac{1}{\psi^{\prime} \left(
       \frac{I}{AF \iota(c_s)}\right)} \frac{- I}{(A F )
     (\iota(c_s))^2}  \right) \left(\eta(\iota(c_s) )-
  \overline{\eta}(\iota(c_s))\right) \, \left| \frac{d t}{dc_s}
 \right|, \qquad c_s \in [0,c_{s,max}].
 \label{eq:gradL2}
\end{equation}

We observe that while expression \eqref{eq:gradL2} is defined for $c_s
\in [0,c_{s,max}]$, it will be nonzero only for $c_s \in [c_<,c_>] =
\I$, where $c_< = \inf_{t \in [0,t_f]} c_s(t)$, $c_> = \sup_{t \in
  [0,t_f]} c_s(t)$ and $\I$ is referred to as the {\em identifiability
  interval}. In other words, the sensitivity information can be
obtained only for concentration values in the identifiability
interval, which is the range of concentrations attained in a given
experiment. In addition, expression \eqref{eq:gradL2} can produce
discontinuous functions of $c_s$ which, when used in
\eqref{graddescent}, would not produce constitutive relations with the
required regularity, cf.~Assumption A1 in Section \ref{sec:BV}.  In
the light of expression \eqref{graddescent}, the regularity of the
optimal constitutive relation $\widehat{\iota}(c_s)$ is determined by
the smoothness of the gradients $\nabla\J_2(c_s)$.

Gradients with the required regularity are constructed using the Riesz
identity \eqref{eq:Riesz} extended to the Sobolev space
$H^1([0,c_{s,max}])$ of functions with square-integrable
distributional derivatives, namely,
 \begin{equation}
\mathcal{J}_2^{\prime}(\iota(c_s);\iota^{\prime}(c_s))   
= \left\langle \nabla^{L^2} \mathcal{J}_2, \iota^{\prime} \right\rangle_{L^2}   = \Big\langle \nabla \mathcal{J}_2, \iota^{\prime} \Big\rangle_{H^1},
\label{eq:RieszH1}
\end{equation}
which makes it possible to identify the G\^ateaux differential
\eqref{eq:dJ3} with the $H^1$ inner product defined as
\begin{equation}
\Big\langle \nabla \mathcal{J}_2, \iota^{\prime} \Big\rangle_{H^1}
= \int\limits_{0}^{c_{s,max}} \left[ \nabla
  \mathcal{J}_2 \, \iota^{\prime} + \ell^2 \frac{d\left(\nabla
      \mathcal{J}_2\right)}{dc_s} \frac{d \iota^{\prime} }{dc_s} \right]
\, dc_s,
\label{eq:ipH1}
\end{equation}
where $\ell \in \mathbb{R}$ is a parameter with the meaning of a
characteristic concentration. We note that as long as $0 < \ell <
\infty$, inner products \eqref{eq:ipH1} with different values of
$\ell$ are equivalent in the sense of norm equivalence. However, the
choice of the value of parameter $\ell$ is important in computations
as it controls the degree of regularization in the solution of Problem
P2.  Performing integration by parts with respect to $c_s$ on the
second term in \eqref{eq:ipH1}, we obtain
\begin{equation}
\Big\langle \nabla \mathcal{J}_2, \iota^{\prime}
\Big\rangle_{H^1} =  \int\limits_{0}^{c_{s,max}} \left( Id -
  \ell^2 \frac{d^2}{dc_s^2} \right) \nabla \mathcal{J}_2\,
\iota^{\prime} \,dc_s -  \ell^2\frac{d\left(\nabla
    \mathcal{J}_2\right)}{dc_s} \iota^{\prime}
\Bigg|_{0}^{c_{s,max}}.
\label{eq:dJ4}
\end{equation}
Noting that the perturbation of the exchange current $\iota'(c_s)$
(i.e., the ``test'' function) is arbitrary and the Sobolev gradient
$\nabla\J_2(c_s)$ needs to vanish at the endpoints, such that
$\iota'(0) = \iota'(c_{s,max}) = 0$ (cf.~Assumptions A2--A3 in
Section \ref{sec:BV} and Appendix \ref{sec:appBC}), relation
\eqref{eq:dJ4} implies
\begin{subequations}
  \label{eq:gradH1}
\begin{alignat}{2}
\left( Id - \ell^2 \frac{d^2}{dc_s^2} \right) \nabla
  \mathcal{J}_2 &=  \nabla^{L^2}\mathcal\J_2 & \qquad & \textrm{for} \ c_s
  \in (0,c_{s,max}),   \label{eq:gradH1eq} \\
  \nabla\J_2 &= 0 & \qquad & \textrm{at} \ c_s = 0,c_{s,max}, \label{eq:gradH1bc}
\end{alignat}
\end{subequations}
which shows that the Sobolev gradient can be obtained by solving a
boundary-value problem with the $L^2$ gradient \eqref{eq:gradL2}
appearing as a source term. We add that this formalism allows us to
propagate the sensitivity information outside the identifiability
interval $\I$ such that the Sobolev gradient $\nabla\J_2(c_s)$ is a
continuous functions not identically equal to zero for
$c_s \in [0,c_{s,max}] \: \backslash \: \I$. It is
known\cite{ProtasBewleyHagen2004} that extraction of gradients in
spaces of smoother functions such as $H^1$ can be interpreted as
low-pass filtering of the $L^2$ gradients with parameter $\ell$ acting
as the cut-off length-scale of the filter. The value of $\ell$ can
significantly affect the rate of convergence of the iterative
procedure \eqref{graddescent}.

%%%%%%%%%%%%%%%%%%%%%%%%%%%%%%%%%%%%%%%%%%%
\section{Experimental Parameters \label{exppara}}

%%%%%%%%%%%%%%%%%%%%%%%%%%%%%%%%%%%%%%%%%%%
\subsection{Slow charge rates \label{sec:expkevinpara}}

\begin{table}[H]
\hspace*{-1.0cm}
\begin{tabular}{l l l l}
Parameter & Units & Anode & Cathode\\ \hline
electrode thickness, $L$, & $\mu$m & $50$ & $50$ \\
electrode particle radius, $R_p$ &  $\mu$m & $9$ & $6.5$\\
electrode cross-section area, $A$ & m$^2$ & $0.000551$ & $0.000551$\\
volume fraction of electrolyte, $\epsilon_l$ & unitless & $0.1$ & $0.1$ \\
Brunauer-Emmett-Teller surface area, $b_0$ & m$^{-1}$ & $3.0 \times 10^5$ & $4.1538 \times 10^5$ \\
conductivity in solid, $\sigma_0$ & Sm$^{-1}$ & $14$ & $68.1$ \\
permeability factor of electrode, $\mathcal{B}_0$ & unitless & $0.0493$ & $0.0515$ \\
reaction rate constant, $k_0$ & m$^{5/2}$s$^{-1}$mol$^{-1/2}$ & $1.8742 \times 10^{-10}$ & $4.9375\times 10^{-11}$ \\
maximum concentration of Lithium in solid, $c_{s,max}$ & mol~m$^{-3}$ & $35920$ & $42580$\\ 
initial concentration of Lithium in electrode, $c_{0}$ & mol~m$^{-3}$ & $1000$ & $1000$ \\ 
{activation energy}, $E_i$ & J mol$^{-1}$ & $30300$ & $80600$\\ \hline
\end{tabular}
\caption{Electrode specific experimental parameters for slow charge
  rate experiments\cite{sanders2021transient}.}
\label{tab:kevinparams1}
\end{table}

\begin{table}[h!]
\begin{tabular}{l l l}
Parameter & Units  & Value\\ \hline
temperature, $T$ & K & $295.15$\\ 
{relative temperature}, $\hat{T}$ & K & $296$ \\  
{universal gas constant}, $R$ &J mol$^{-1}$ K$^{-1}$ & $8.3144$  \\
C$/10$ current, $I_0$ & A & $2.29 \times 10^{-3}$ \\
C$/3$ current, $I_0$  & A & $0.0076$\\
%Contact Resistance, $R_c$  & $\Omega$ & $0$\\
transference number, $t^+$ & unitless & $0.26$
\end{tabular}
\caption{Global experimental parameters for slow charge rate
  experiments\cite{sanders2021transient}.}
\label{tab:kevinparams2}
\end{table}

{\hspace*{-\parindent}Diffusivities in the electrodes
  are assumed to depend on Arrhenius temperature and are given by
\begin{align*}
D_{s}(c_{s,i},T) &= D_{s}(c_{s,i})e^{\left( \frac{E_{i}}{R\hat{T}} - \frac{E_{i}}{RT}\right)},\\
D_{s}(c_{s,a}) & = 8.4\times 10^{-9}e^{-11.3\left(\frac{c_{s,a}}{c_{s,a,max}}\right)}+8.2\times 10^{-12},\\
D_{s}(c_{s,c}) & = 3.7\times 10^{-13} - 3.4\times 10^{-13}e^{-12\left(\frac{c_{s,c}}{c_{s,c,max}}-0.62 \right)}
\end{align*}
with units of m$^2$/s.}

%%%%%%%%%%%%%%%%%%%%%%%%%%%%%%%%%%%%%%%%%%%
\subsection{Moderate discharge rates \label{sec:expferranpara}}

\begin{table}[H]
\hspace*{-1.0cm}
\begin{tabular}{l l l l}
Parameter & Units & Anode & Cathode\\ \hline
electrode thickness, $L$, & $\mu$m & $85.2$ & $75.6$ \\
electrode particle radius, $R_p$ &  $\mu$m & $5.86$ & $5.22$\\
electrode cross-section area, $A$ & m$^2$ & $0.10465$ & $0.1027$\\
volume fraction of electrolyte, $\epsilon_l$ & unitless & $0.25$ & $0.335$ \\
Brunauer-Emmett-Teller surface area, $b_0$ & m$^{-1}$ & $ 3.8396 \times 10^5$ & $3.8218 \times 10^5$ \\
conductivity in solid, $\sigma_0$ & Sm$^{-1}$ & $215$ & $0.18$ \\
permeability factor of electrode, $\mathcal{B}_0$ & unitless & $0.0177$ & $0.0701$ \\
reaction rate Constant, $k_0$ & m$^{5/2}$s$^{-1}$mol$^{-1/2}$ & $0.0672 \times 10^{-10}$ & $0.3545\times 10^{-10}$ \\
maximum concentration of Lithium in solid, $c_{s,max}$ & mol~m$^{-3}$ & $33133$ & $63104$\\ 
initial concentration of Lithium in electrode, $c_{0}$ & mol~m$^{-3}$ & $1000$ & $1000$ \\ \hline
\end{tabular}
\caption{Electrode specific experimental parameters for moderate discharge
  rate experiments\cite{chen2020development}.}
\label{tab:Ferranparams1}
\end{table}

\begin{table}[h!]
\begin{tabular}{l l l}
Parameter & Units  & Value\\ \hline
temperature, $T$ & K & $298.15$\\ 
C$/2$ current, $I_0$ & A & $2.5$ \\
$1$C current, $I_0$  & A & $5$\\
%Contact Resistance, $R_c$  & $\Omega$ & $0$\\
Transference Number, $t^+$ & unitless & $0.259$\\
\end{tabular}
\caption{Global experimental parameters for moderate discharge rate
  experiments\cite{chen2020development}.}
\label{tab:Ferranparams2}
\end{table}

{\hspace*{-\parindent}Transport in the electrodes is
  assumed to be governed by Fickian diffusion with
\begin{align*}
D_{s}(c_{s,a}) & = 3.3 \times 10^{-10},\\
D_{s}(c_{s,c}) & = 4 \times 10^{-11}
\end{align*}
and with units of m$^2$/s.}

%%%%%%%%%%%%%%%%%%%%%%%%%%%%%%%%%%%%%%%%%%%
\section{Checklist To Report Theoretical Battery Studies}

Here, we complete the checklist for the minimal information set \cite{mistry2021minimal}.

\begin{enumerate}
\item {\em Have you provided all assumptions, theory, governing equations,
  initial and boundary conditions, material properties (e.g.,
  open-circuit potential) with appropriate precision and literature
  sources, constant states (e.g., temperature), etc.? }

Remarks: Yes, all elements of the model have been stated in the main text and the supplemental information. 

\item {\em If the calculations have a probabilistic component (e.g.,
    Monte Carlo, initial configuration in Molecular Dynamics, etc.),
    did you provide statistics (mean, standard deviation, confidence
    interval, etc.) from multiple ($\ge 3$) runs of a representative
    case?}
  
  Remarks: Yes, Monte Carlo simulations were run with $3000$
  samples. The methodology is reported in the main text and confidence
  intervals are included on the relevant figures.

\item  {\em If data-driven calculations are performed (e.g., machine
    learning), did you specify dataset origin, the rationale behind
    choosing it, what information it contains, and the specific
    portion of it being utilized? Have you described the thought
    process for choosing a specific modeling paradigm?}
  
  Remarks: While the methodology used is data driven, it relies on
  calculus of variations rather than machine learning.  The
  experimental data is described in Section $3$ whereas the
  optimization scheme is outlined in Algorithms 1 and 2.

\item {\em Have you discussed all sources of potential uncertainty,
    variability, and errors in the modeling results and their impact
    on quantitative results and qualitative trends? Have you discussed
    the sensitivity of modeling (and numerical) inputs such as
    material properties, time step, domain size, neural network
    architecture, etc. where they are variable or uncertain?}
  
  Remarks: Yes, this discussion is provided in Sections 4.4 and 6 in
  the main text.

\item {\em Have you sufficiently discussed new or not widely familiar
    terminology and descriptors for clarity? Did you use these terms
    in their appropriate context to avoid misinterpretation? Enumerate
    these terms in the “Remarks”.}
  
  Remarks: Yes, the terminology is consistent with the conventions
  used in the research fields.
\end{enumerate}

%%%%%%%%%%%%%%%%%%%%%%%%%%%%%%%%%%%%%%%%%%%
%%%%%%%%%%%%%%%%%%%%%%%%%%%%%%%%%%%%%%%%%%%
%acknowledgments

%%%%%%%%%%%%%%%%%%%%%%%%%%%%%%%%%%%%%%%%%%%
%bibliography
%%%%%%%%%%%%%%%%%%%%%%%%%%%%%%%%%%%%%%%%%%%
%\bibliography{electrochemistry}
%\bibliography{../Bib/electrochemistry,../Bib/allPROTAS}

\begin{mcitethebibliography}{74}
\providecommand*\natexlab[1]{#1}
\providecommand*\mciteSetBstSublistMode[1]{}
\providecommand*\mciteSetBstMaxWidthForm[2]{}
\providecommand*\mciteBstWouldAddEndPuncttrue
  {\def\EndOfBibitem{\unskip.}}
\providecommand*\mciteBstWouldAddEndPunctfalse
  {\let\EndOfBibitem\relax}
\providecommand*\mciteSetBstMidEndSepPunct[3]{}
\providecommand*\mciteSetBstSublistLabelBeginEnd[3]{}
\providecommand*\EndOfBibitem{}
\mciteSetBstSublistMode{f}
\mciteSetBstMaxWidthForm{subitem}{(\alph{mcitesubitemcount})}
\mciteSetBstSublistLabelBeginEnd
  {\mcitemaxwidthsubitemform\space}
  {\relax}
  {\relax}

\bibitem[Wang \latin{et~al.}(2022)Wang, O'Kane, Planella, Le~Houx, O'Regan,
  Zyskin, Edge, Monroe, Cooper, Howey, Kendrick, and Foster]{wang2022review}
Wang,~A.; O'Kane,~S.~E.; Planella,~F.~B.; Le~Houx,~J.; O'Regan,~K.; Zyskin,~M.;
  Edge,~J.~S.; Monroe,~C.; Cooper,~S.; Howey,~D.~A. \latin{et~al.}  Review of
  parameterisation and a novel database (LiionDB) for continuum Li-ion battery
  models. \emph{Progress in Energy} \textbf{2022}, \emph{4}, 032004\relax
\mciteBstWouldAddEndPuncttrue
\mciteSetBstMidEndSepPunct{\mcitedefaultmidpunct}
{\mcitedefaultendpunct}{\mcitedefaultseppunct}\relax
\EndOfBibitem
\bibitem[Lv \latin{et~al.}(2022)Lv, Zhou, Zhong, Yan, Srinivasan, Seh, Liu,
  Pan, Li, Wen, and Yan]{lv2022machine}
Lv,~C.; Zhou,~X.; Zhong,~L.; Yan,~C.; Srinivasan,~M.; Seh,~Z.~W.; Liu,~C.;
  Pan,~H.; Li,~S.; Wen,~Y. \latin{et~al.}  Machine learning: an advanced
  platform for materials development and state prediction in lithium-ion
  batteries. \emph{Advanced Materials} \textbf{2022}, \emph{34}, 2101474\relax
\mciteBstWouldAddEndPuncttrue
\mciteSetBstMidEndSepPunct{\mcitedefaultmidpunct}
{\mcitedefaultendpunct}{\mcitedefaultseppunct}\relax
\EndOfBibitem
\bibitem[Houchins and Viswanathan(2020)Houchins, and
  Viswanathan]{houchins2020accurate}
Houchins,~G.; Viswanathan,~V. An accurate machine-learning calculator for
  optimization of Li-ion battery cathodes. \emph{The Journal of Chemical
  Physics} \textbf{2020}, \emph{153}, 054124\relax
\mciteBstWouldAddEndPuncttrue
\mciteSetBstMidEndSepPunct{\mcitedefaultmidpunct}
{\mcitedefaultendpunct}{\mcitedefaultseppunct}\relax
\EndOfBibitem
\bibitem[Chemali \latin{et~al.}(2018)Chemali, Kollmeyer, Preindl, and
  Emadi]{chemali2018state}
Chemali,~E.; Kollmeyer,~P.~J.; Preindl,~M.; Emadi,~A. State-of-charge
  estimation of Li-ion batteries using deep neural networks: A machine learning
  approach. \emph{Journal of Power Sources} \textbf{2018}, \emph{400},
  242--255\relax
\mciteBstWouldAddEndPuncttrue
\mciteSetBstMidEndSepPunct{\mcitedefaultmidpunct}
{\mcitedefaultendpunct}{\mcitedefaultseppunct}\relax
\EndOfBibitem
\bibitem[Klett \latin{et~al.}(2012)Klett, Giesecke, Nyman, Hallberg,
  Lindstr{\"o}m, Lindbergh, and Furo]{kgnhllf12}
Klett,~M.; Giesecke,~M.; Nyman,~A.; Hallberg,~F.; Lindstr{\"o}m,~R.~W.;
  Lindbergh,~G.; Furo,~I. {Quantifying Mass Transport during Polarization in a
  Li Ion Battery Electrolyte by in Situ Li NMR Imaging}. \emph{J. Am. Chem.
  Soc.} \textbf{2012}, \emph{134}, 14654--14657\relax
\mciteBstWouldAddEndPuncttrue
\mciteSetBstMidEndSepPunct{\mcitedefaultmidpunct}
{\mcitedefaultendpunct}{\mcitedefaultseppunct}\relax
\EndOfBibitem
\bibitem[Sethurajan \latin{et~al.}(2015)Sethurajan, Krachkovskiy, Halalay,
  Goward, and Protas]{sethurajan2015accurate}
Sethurajan,~A.~K.; Krachkovskiy,~S.~A.; Halalay,~I.~C.; Goward,~G.~R.;
  Protas,~B. Accurate characterization of ion transport properties in binary
  symmetric electrolytes using in situ NMR imaging and inverse modeling.
  \emph{The Journal of Physical Chemistry B} \textbf{2015}, \emph{119},
  12238--12248\relax
\mciteBstWouldAddEndPuncttrue
\mciteSetBstMidEndSepPunct{\mcitedefaultmidpunct}
{\mcitedefaultendpunct}{\mcitedefaultseppunct}\relax
\EndOfBibitem
\bibitem[Richardson \latin{et~al.}(2018)Richardson, Foster, Sethurajan,
  Krachkovskiy, Halalay, Goward, and Protas]{richardson2018effect}
Richardson,~G.; Foster,~J.~M.; Sethurajan,~A.~K.; Krachkovskiy,~S.~A.;
  Halalay,~I.~C.; Goward,~G.~R.; Protas,~B. The effect of ionic aggregates on
  the transport of charged species in lithium electrolyte solutions.
  \emph{Journal of The Electrochemical Society} \textbf{2018}, \emph{165},
  H561--H567\relax
\mciteBstWouldAddEndPuncttrue
\mciteSetBstMidEndSepPunct{\mcitedefaultmidpunct}
{\mcitedefaultendpunct}{\mcitedefaultseppunct}\relax
\EndOfBibitem
\bibitem[Sethurajan \latin{et~al.}(2019)Sethurajan, Foster, Richardson,
  Krachkovskiy, Bazak, Goward, and Protas]{sethurajan2019dendrites}
Sethurajan,~A.~K.; Foster,~J.~M.; Richardson,~G.; Krachkovskiy,~S.~A.;
  Bazak,~J.~D.; Goward,~G.~R.; Protas,~B. Incorporating Dendrite Growth into
  Continuum Models of Electrolytes: Insights from NMR Measurements and Inverse
  Modeling. \emph{Journal of The Electrochemical Society} \textbf{2019},
  \emph{166}, A1591--A1602\relax
\mciteBstWouldAddEndPuncttrue
\mciteSetBstMidEndSepPunct{\mcitedefaultmidpunct}
{\mcitedefaultendpunct}{\mcitedefaultseppunct}\relax
\EndOfBibitem
\bibitem[Escalante \latin{et~al.}(2020)Escalante, Ko, Foster, Krachkovskiy,
  Goward, and Protas]{escalante2020discerning}
Escalante,~J.~M.; Ko,~W.; Foster,~J.~M.; Krachkovskiy,~S.; Goward,~G.;
  Protas,~B. Discerning models of phase transformations in porous graphite
  electrodes: Insights from inverse modelling based on MRI measurements.
  \emph{Electrochimica Acta} \textbf{2020}, \emph{349}, 136290\relax
\mciteBstWouldAddEndPuncttrue
\mciteSetBstMidEndSepPunct{\mcitedefaultmidpunct}
{\mcitedefaultendpunct}{\mcitedefaultseppunct}\relax
\EndOfBibitem
\bibitem[Escalante \latin{et~al.}(2021)Escalante, Sahu, Foster, and
  Protas]{escalante2021uncertainty}
Escalante,~J.~M.; Sahu,~S.; Foster,~J.~M.; Protas,~B. On Uncertainty
  Quantification in the Parametrization of Newman-type Models of Lithium-ion
  Batteries. \emph{Journal of The Electrochemical Society} \textbf{2021},
  \emph{168}, 110519\relax
\mciteBstWouldAddEndPuncttrue
\mciteSetBstMidEndSepPunct{\mcitedefaultmidpunct}
{\mcitedefaultendpunct}{\mcitedefaultseppunct}\relax
\EndOfBibitem
\bibitem[Chen \latin{et~al.}(2019)Chen, Liang, Yang, and Li]{chen2019review}
Chen,~W.; Liang,~J.; Yang,~Z.; Li,~G. A review of lithium-ion battery for
  electric vehicle applications and beyond. \emph{Energy Procedia}
  \textbf{2019}, \emph{158}, 4363--4368\relax
\mciteBstWouldAddEndPuncttrue
\mciteSetBstMidEndSepPunct{\mcitedefaultmidpunct}
{\mcitedefaultendpunct}{\mcitedefaultseppunct}\relax
\EndOfBibitem
\bibitem[Blomgren(2016)]{blomgren2016development}
Blomgren,~G.~E. The development and future of lithium ion batteries.
  \emph{Journal of The Electrochemical Society} \textbf{2016}, \emph{164},
  A5019\relax
\mciteBstWouldAddEndPuncttrue
\mciteSetBstMidEndSepPunct{\mcitedefaultmidpunct}
{\mcitedefaultendpunct}{\mcitedefaultseppunct}\relax
\EndOfBibitem
\bibitem[Newman and Thomas-Alyea(2004)Newman, and Thomas-Alyea]{nt04}
Newman,~J.; Thomas-Alyea,~K.~E. \emph{{Electrochemical Systems}}; John Wiley
  and Sons, 2004\relax
\mciteBstWouldAddEndPuncttrue
\mciteSetBstMidEndSepPunct{\mcitedefaultmidpunct}
{\mcitedefaultendpunct}{\mcitedefaultseppunct}\relax
\EndOfBibitem
\bibitem[Hamelers \latin{et~al.}(2011)Hamelers, Ter~Heijne, Stein, Rozendal,
  and Buisman]{hamelers2011butler}
Hamelers,~H.~V.; Ter~Heijne,~A.; Stein,~N.; Rozendal,~R.~A.; Buisman,~C.~J.
  Butler--Volmer--Monod model for describing bio-anode polarization curves.
  \emph{Bioresource technology} \textbf{2011}, \emph{102}, 381--387\relax
\mciteBstWouldAddEndPuncttrue
\mciteSetBstMidEndSepPunct{\mcitedefaultmidpunct}
{\mcitedefaultendpunct}{\mcitedefaultseppunct}\relax
\EndOfBibitem
\bibitem[Compton and Banks(2011)Compton, and Banks]{compton2011voltammetry}
Compton,~R.~G.; Banks,~C.~E. \emph{Understanding voltammetry, 2nd ed.};
  Imperial College Press London, 2011\relax
\mciteBstWouldAddEndPuncttrue
\mciteSetBstMidEndSepPunct{\mcitedefaultmidpunct}
{\mcitedefaultendpunct}{\mcitedefaultseppunct}\relax
\EndOfBibitem
\bibitem[Oyarzun \latin{et~al.}(2021)Oyarzun, Zhan, Hawks, Cer{\'o}n, Kuo,
  Loeb, Aydin, Pham, Stadermann, and Campbell]{oyarzun2021unraveling}
Oyarzun,~D.~I.; Zhan,~C.; Hawks,~S.~A.; Cer{\'o}n,~M.~R.; Kuo,~H.~A.;
  Loeb,~C.~K.; Aydin,~F.; Pham,~T.~A.; Stadermann,~M.; Campbell,~P.~G.
  Unraveling the Ion Adsorption Kinetics in Microporous Carbon Electrodes: A
  Multiscale Quantum-Continuum Simulation and Experimental Approach. \emph{ACS
  Applied Materials \& Interfaces} \textbf{2021}, \emph{13}, 23567--23574\relax
\mciteBstWouldAddEndPuncttrue
\mciteSetBstMidEndSepPunct{\mcitedefaultmidpunct}
{\mcitedefaultendpunct}{\mcitedefaultseppunct}\relax
\EndOfBibitem
\bibitem[Hodgkin and Huxley(1952)Hodgkin, and Huxley]{hodgkin1952quantitative}
Hodgkin,~A.~L.; Huxley,~A.~F. A quantitative description of membrane current
  and its application to conduction and excitation in nerve. \emph{The Journal
  of physiology} \textbf{1952}, \emph{117}, 500\relax
\mciteBstWouldAddEndPuncttrue
\mciteSetBstMidEndSepPunct{\mcitedefaultmidpunct}
{\mcitedefaultendpunct}{\mcitedefaultseppunct}\relax
\EndOfBibitem
\bibitem[Banks and Kunisch(2012)Banks, and Kunisch]{banks2012estimation}
Banks,~H.~T.; Kunisch,~K. \emph{Estimation techniques for distributed parameter
  systems}; Springer Science \& Business Media, 2012\relax
\mciteBstWouldAddEndPuncttrue
\mciteSetBstMidEndSepPunct{\mcitedefaultmidpunct}
{\mcitedefaultendpunct}{\mcitedefaultseppunct}\relax
\EndOfBibitem
\bibitem[Tarantola(2005)]{Tarantola2005}
Tarantola,~A. \emph{Inverse Problem Theory and Methods for Model Parameter
  Estimation}; Society for Industrial and Applied Mathematics, 2005\relax
\mciteBstWouldAddEndPuncttrue
\mciteSetBstMidEndSepPunct{\mcitedefaultmidpunct}
{\mcitedefaultendpunct}{\mcitedefaultseppunct}\relax
\EndOfBibitem
\bibitem[Isakov(1998)]{i98}
Isakov,~V. \emph{Inverse Problems for Partial Differential Equations}; Applied
  Mathematical Sciences 127; Springer, 1998\relax
\mciteBstWouldAddEndPuncttrue
\mciteSetBstMidEndSepPunct{\mcitedefaultmidpunct}
{\mcitedefaultendpunct}{\mcitedefaultseppunct}\relax
\EndOfBibitem
\bibitem[Bukshtynov \latin{et~al.}(2011)Bukshtynov, Volkov, and Protas]{bvp11}
Bukshtynov,~V.; Volkov,~O.; Protas,~B. {On Optimal Reconstruction of
  Constitutive Relations}. \emph{Physica D: Nonlinear Phenomena} \textbf{2011},
  \emph{240}, 1228--1244\relax
\mciteBstWouldAddEndPuncttrue
\mciteSetBstMidEndSepPunct{\mcitedefaultmidpunct}
{\mcitedefaultendpunct}{\mcitedefaultseppunct}\relax
\EndOfBibitem
\bibitem[Sethurajan \latin{et~al.}(2019)Sethurajan, Krachkovskiy, Goward, and
  Protas]{doi:10.1002/jcc.25759}
Sethurajan,~A.; Krachkovskiy,~S.; Goward,~G.; Protas,~B. Bayesian uncertainty
  quantification in inverse modeling of electrochemical systems. \emph{Journal
  of Computational Chemistry} \textbf{2019}, \emph{40}, 740--752\relax
\mciteBstWouldAddEndPuncttrue
\mciteSetBstMidEndSepPunct{\mcitedefaultmidpunct}
{\mcitedefaultendpunct}{\mcitedefaultseppunct}\relax
\EndOfBibitem
\bibitem[Bukshtynov and Protas(2013)Bukshtynov, and Protas]{bukshtynov2013889}
Bukshtynov,~V.; Protas,~B. Optimal reconstruction of material properties in
  complex multiphysics phenomena. \emph{Journal of Computational Physics}
  \textbf{2013}, \emph{242}, 889 -- 914\relax
\mciteBstWouldAddEndPuncttrue
\mciteSetBstMidEndSepPunct{\mcitedefaultmidpunct}
{\mcitedefaultendpunct}{\mcitedefaultseppunct}\relax
\EndOfBibitem
\bibitem[Dickinson and Wain(2020)Dickinson, and Wain]{dickinson2020butler}
Dickinson,~E.~J.; Wain,~A.~J. The Butler-Volmer equation in electrochemical
  theory: Origins, value, and practical application. \emph{Journal of
  Electroanalytical Chemistry} \textbf{2020}, \emph{872}, 114145\relax
\mciteBstWouldAddEndPuncttrue
\mciteSetBstMidEndSepPunct{\mcitedefaultmidpunct}
{\mcitedefaultendpunct}{\mcitedefaultseppunct}\relax
\EndOfBibitem
\bibitem[Latz and Zausch(2013)Latz, and Zausch]{latz2013thermodynamic}
Latz,~A.; Zausch,~J. Thermodynamic derivation of a Butler--Volmer model for
  intercalation in Li-ion batteries. \emph{Electrochimica Acta} \textbf{2013},
  \emph{110}, 358--362\relax
\mciteBstWouldAddEndPuncttrue
\mciteSetBstMidEndSepPunct{\mcitedefaultmidpunct}
{\mcitedefaultendpunct}{\mcitedefaultseppunct}\relax
\EndOfBibitem
\bibitem[Bard and Faulkner(2001)Bard, and Faulkner]{bard2001electrochemical}
Bard,~A.~J.; Faulkner,~L.~R. \emph{Electrochemical methods: Fundamentals and
  applications}; Wiley New York, 2001\relax
\mciteBstWouldAddEndPuncttrue
\mciteSetBstMidEndSepPunct{\mcitedefaultmidpunct}
{\mcitedefaultendpunct}{\mcitedefaultseppunct}\relax
\EndOfBibitem
\bibitem[Fletcher(2009)]{fletcher2009tafel}
Fletcher,~S. Tafel slopes from first principles. \emph{Journal of Solid State
  Electrochemistry} \textbf{2009}, \emph{13}, 537--549\relax
\mciteBstWouldAddEndPuncttrue
\mciteSetBstMidEndSepPunct{\mcitedefaultmidpunct}
{\mcitedefaultendpunct}{\mcitedefaultseppunct}\relax
\EndOfBibitem
\bibitem[Richardson \latin{et~al.}(2020)Richardson, Foster, Ranom, Please, and
  Ramos]{richardson2020charge}
Richardson,~G.~W.; Foster,~J.~M.; Ranom,~R.; Please,~C.~P.; Ramos,~A.~M. Charge
  transport modelling of Lithium-ion batteries. \emph{European Journal of
  Applied Mathematics} \textbf{2020}, 1--49\relax
\mciteBstWouldAddEndPuncttrue
\mciteSetBstMidEndSepPunct{\mcitedefaultmidpunct}
{\mcitedefaultendpunct}{\mcitedefaultseppunct}\relax
\EndOfBibitem
\bibitem[Sanders \latin{et~al.}(2022)Sanders, Aguilera, Keffer, Balcom,
  Halalay, and Goward]{sanders2021transient}
Sanders,~K.~J.; Aguilera,~A.~R.; Keffer,~J.~R.; Balcom,~B.~J.; Halalay,~I.~C.;
  Goward,~G.~R. Transient lithium metal plating on graphite: Operando $^7${L}i
  nuclear magnetic resonance investigation of a battery cell using a novel {RF}
  probe. \emph{Carbon} \textbf{2022}, 377--385\relax
\mciteBstWouldAddEndPuncttrue
\mciteSetBstMidEndSepPunct{\mcitedefaultmidpunct}
{\mcitedefaultendpunct}{\mcitedefaultseppunct}\relax
\EndOfBibitem
\bibitem[Chen \latin{et~al.}(2020)Chen, Planella, O’regan, Gastol, Widanage,
  and Kendrick]{chen2020development}
Chen,~C.-H.; Planella,~F.~B.; O’regan,~K.; Gastol,~D.; Widanage,~W.~D.;
  Kendrick,~E. Development of experimental techniques for parameterization of
  multi-scale lithium-ion battery models. \emph{Journal of The Electrochemical
  Society} \textbf{2020}, \emph{167}, 080534\relax
\mciteBstWouldAddEndPuncttrue
\mciteSetBstMidEndSepPunct{\mcitedefaultmidpunct}
{\mcitedefaultendpunct}{\mcitedefaultseppunct}\relax
\EndOfBibitem
\bibitem[Richardson \latin{et~al.}(2020)Richardson, Korotkin, Ranom, Castle,
  and Foster]{richardson2020generalised}
Richardson,~G.; Korotkin,~I.; Ranom,~R.; Castle,~M.; Foster,~J. Generalised
  single particle models for high-rate operation of graded lithium-ion
  electrodes: systematic derivation and validation. \emph{Electrochimica Acta}
  \textbf{2020}, \emph{339}, 135862\relax
\mciteBstWouldAddEndPuncttrue
\mciteSetBstMidEndSepPunct{\mcitedefaultmidpunct}
{\mcitedefaultendpunct}{\mcitedefaultseppunct}\relax
\EndOfBibitem
\bibitem[Korotkin \latin{et~al.}(2021)Korotkin, Sahu, O’Kane, Richardson, and
  Foster]{korotkin2021dandeliion}
Korotkin,~I.; Sahu,~S.; O’Kane,~S.~E.; Richardson,~G.; Foster,~J.~M.
  DandeLiion v1: An extremely fast solver for the Newman model of lithium-ion
  battery (dis) charge. \emph{Journal of The Electrochemical Society}
  \textbf{2021}, \emph{168}, 060544\relax
\mciteBstWouldAddEndPuncttrue
\mciteSetBstMidEndSepPunct{\mcitedefaultmidpunct}
{\mcitedefaultendpunct}{\mcitedefaultseppunct}\relax
\EndOfBibitem
\bibitem[Wu \latin{et~al.}(2012)Wu, Zhang, Song, Shukla, Liu, Battaglia, and
  Srinivasan]{wu2012high}
Wu,~S.-L.; Zhang,~W.; Song,~X.; Shukla,~A.~K.; Liu,~G.; Battaglia,~V.;
  Srinivasan,~V. High rate capability of Li (Ni1/3Mn1/3Co1/3) O2 electrode for
  Li-ion batteries. \emph{Journal of The Electrochemical Society}
  \textbf{2012}, \emph{159}, A438\relax
\mciteBstWouldAddEndPuncttrue
\mciteSetBstMidEndSepPunct{\mcitedefaultmidpunct}
{\mcitedefaultendpunct}{\mcitedefaultseppunct}\relax
\EndOfBibitem
\bibitem[Nguyen \latin{et~al.}(2022)Nguyen, Delobel, Berthe, Fleutot,
  Demorti{\`e}re, and Delacourt]{nguyen2022mathematical}
Nguyen,~T.-T.; Delobel,~B.; Berthe,~M.; Fleutot,~B.; Demorti{\`e}re,~A.;
  Delacourt,~C. Mathematical Modeling of Energy-Dense NMC Electrodes: I.
  Determination of Input Parameters. \emph{Journal of The Electrochemical
  Society} \textbf{2022}, \emph{169}, 040546\relax
\mciteBstWouldAddEndPuncttrue
\mciteSetBstMidEndSepPunct{\mcitedefaultmidpunct}
{\mcitedefaultendpunct}{\mcitedefaultseppunct}\relax
\EndOfBibitem
\bibitem[Doyle \latin{et~al.}(1993)Doyle, Fuller, and Newman]{Doyle01061993}
Doyle,~M.; Fuller,~T.~F.; Newman,~J. Modeling of Galvanostatic Charge and
  Discharge of the Lithium/Polymer/Insertion Cell. \emph{Journal of The
  Electrochemical Society} \textbf{1993}, \emph{140}, 1526--1533\relax
\mciteBstWouldAddEndPuncttrue
\mciteSetBstMidEndSepPunct{\mcitedefaultmidpunct}
{\mcitedefaultendpunct}{\mcitedefaultseppunct}\relax
\EndOfBibitem
\bibitem[Fuller \latin{et~al.}(1994)Fuller, Doyle, and
  Newman]{fuller1994simulation}
Fuller,~T.~F.; Doyle,~M.; Newman,~J. Simulation and optimization of the dual
  lithium ion insertion cell. \emph{Journal of the Electrochemical Society}
  \textbf{1994}, \emph{141}, 1\relax
\mciteBstWouldAddEndPuncttrue
\mciteSetBstMidEndSepPunct{\mcitedefaultmidpunct}
{\mcitedefaultendpunct}{\mcitedefaultseppunct}\relax
\EndOfBibitem
\bibitem[Mistry \latin{et~al.}(2018)Mistry, Smith, and
  Mukherjee]{mistry2018secondary}
Mistry,~A.~N.; Smith,~K.; Mukherjee,~P.~P. Secondary-phase stochastics in
  lithium-ion battery electrodes. \emph{ACS applied materials \& interfaces}
  \textbf{2018}, \emph{10}, 6317--6326\relax
\mciteBstWouldAddEndPuncttrue
\mciteSetBstMidEndSepPunct{\mcitedefaultmidpunct}
{\mcitedefaultendpunct}{\mcitedefaultseppunct}\relax
\EndOfBibitem
\bibitem[Mistry \latin{et~al.}(2021)Mistry, Trask, Dunlop, Jeka, Polzin,
  Mukherjee, and Srinivasan]{mistry2021quantifying}
Mistry,~A.; Trask,~S.; Dunlop,~A.; Jeka,~G.; Polzin,~B.; Mukherjee,~P.~P.;
  Srinivasan,~V. Quantifying negative effects of carbon-binder networks from
  electrochemical performance of porous li-ion electrodes. \emph{Journal of The
  Electrochemical Society} \textbf{2021}, \emph{168}, 070536\relax
\mciteBstWouldAddEndPuncttrue
\mciteSetBstMidEndSepPunct{\mcitedefaultmidpunct}
{\mcitedefaultendpunct}{\mcitedefaultseppunct}\relax
\EndOfBibitem
\bibitem[Mistry \latin{et~al.}(2020)Mistry, Usseglio-Viretta, Colclasure,
  Smith, and Mukherjee]{mistry2020fingerprinting}
Mistry,~A.; Usseglio-Viretta,~F.~L.; Colclasure,~A.; Smith,~K.;
  Mukherjee,~P.~P. Fingerprinting redox heterogeneity in electrodes during
  extreme fast charging. \emph{Journal of The Electrochemical Society}
  \textbf{2020}, \emph{167}, 090542\relax
\mciteBstWouldAddEndPuncttrue
\mciteSetBstMidEndSepPunct{\mcitedefaultmidpunct}
{\mcitedefaultendpunct}{\mcitedefaultseppunct}\relax
\EndOfBibitem
\bibitem[Hein \latin{et~al.}(2020)Hein, Danner, Westhoff, Prifling, Scurtu,
  Kremer, Hoffmann, Hilger, Osenberg, Manke, Wohlfahrt-Mehrens, Schmidt, and
  Latz]{hein2020influence}
Hein,~S.; Danner,~T.; Westhoff,~D.; Prifling,~B.; Scurtu,~R.; Kremer,~L.;
  Hoffmann,~A.; Hilger,~A.; Osenberg,~M.; Manke,~I. \latin{et~al.}  Influence
  of conductive additives and binder on the impedance of lithium-ion battery
  electrodes: effect of morphology. \emph{Journal of The Electrochemical
  Society} \textbf{2020}, \emph{167}, 013546\relax
\mciteBstWouldAddEndPuncttrue
\mciteSetBstMidEndSepPunct{\mcitedefaultmidpunct}
{\mcitedefaultendpunct}{\mcitedefaultseppunct}\relax
\EndOfBibitem
\bibitem[Ferraro \latin{et~al.}(2020)Ferraro, Trembacki, Brunini, Noble, and
  Roberts]{ferraro2020electrode}
Ferraro,~M.~E.; Trembacki,~B.~L.; Brunini,~V.~E.; Noble,~D.~R.; Roberts,~S.~A.
  Electrode mesoscale as a collection of particles: coupled electrochemical and
  mechanical analysis of NMC cathodes. \emph{Journal of The Electrochemical
  Society} \textbf{2020}, \emph{167}, 013543\relax
\mciteBstWouldAddEndPuncttrue
\mciteSetBstMidEndSepPunct{\mcitedefaultmidpunct}
{\mcitedefaultendpunct}{\mcitedefaultseppunct}\relax
\EndOfBibitem
\bibitem[Mistry \latin{et~al.}(2022)Mistry, Heenan, Smith, Shearing, and
  Mukherjee]{mistry2022asphericity}
Mistry,~A.; Heenan,~T.; Smith,~K.; Shearing,~P.; Mukherjee,~P.~P. Asphericity
  can cause nonuniform lithium intercalation in battery active particles.
  \emph{ACS Energy Letters} \textbf{2022}, \emph{7}, 1871--1879\relax
\mciteBstWouldAddEndPuncttrue
\mciteSetBstMidEndSepPunct{\mcitedefaultmidpunct}
{\mcitedefaultendpunct}{\mcitedefaultseppunct}\relax
\EndOfBibitem
\bibitem[Verbrugge and Koch(1996)Verbrugge, and Koch]{verbrugge1996modeling}
Verbrugge,~M.~W.; Koch,~B.~J. Modeling lithium intercalation of single-fiber
  carbon microelectrodes. \emph{Journal of the Electrochemical Society}
  \textbf{1996}, \emph{143}, 600\relax
\mciteBstWouldAddEndPuncttrue
\mciteSetBstMidEndSepPunct{\mcitedefaultmidpunct}
{\mcitedefaultendpunct}{\mcitedefaultseppunct}\relax
\EndOfBibitem
\bibitem[Johnson \latin{et~al.}(2022)Johnson, Mistry, Yin, Murphy, Wolfman,
  Fister, Lapidus, Cabana, Srinivasan, and Ingram]{johnson2022unconventional}
Johnson,~I.~D.; Mistry,~A.~N.; Yin,~L.; Murphy,~M.; Wolfman,~M.; Fister,~T.~T.;
  Lapidus,~S.~H.; Cabana,~J.; Srinivasan,~V.; Ingram,~B.~J. Unconventional
  Charge Transport in MgCr2O4 and Implications for Battery Intercalation Hosts.
  \emph{Journal of the American Chemical Society} \textbf{2022}, \emph{144},
  14121--14131\relax
\mciteBstWouldAddEndPuncttrue
\mciteSetBstMidEndSepPunct{\mcitedefaultmidpunct}
{\mcitedefaultendpunct}{\mcitedefaultseppunct}\relax
\EndOfBibitem
\bibitem[Amin \latin{et~al.}(2015)Amin, Ravnsb{\ae}k, and
  Chiang]{amin2015characterization}
Amin,~R.; Ravnsb{\ae}k,~D.~B.; Chiang,~Y.-M. Characterization of electronic and
  ionic transport in Li1-xNi0. 8Co0. 15Al0. 05O2 (NCA). \emph{Journal of the
  electrochemical society} \textbf{2015}, \emph{162}, A1163\relax
\mciteBstWouldAddEndPuncttrue
\mciteSetBstMidEndSepPunct{\mcitedefaultmidpunct}
{\mcitedefaultendpunct}{\mcitedefaultseppunct}\relax
\EndOfBibitem
\bibitem[Amin and Chiang(2016)Amin, and Chiang]{amin2016characterization}
Amin,~R.; Chiang,~Y.-M. Characterization of electronic and ionic transport in
  Li1-xNi0. 33Mn0. 33Co0. 33O2 (NMC333) and Li1-xNi0. 50Mn0. 20Co0. 30O2
  (NMC523) as a function of Li content. \emph{Journal of The Electrochemical
  Society} \textbf{2016}, \emph{163}, A1512\relax
\mciteBstWouldAddEndPuncttrue
\mciteSetBstMidEndSepPunct{\mcitedefaultmidpunct}
{\mcitedefaultendpunct}{\mcitedefaultseppunct}\relax
\EndOfBibitem
\bibitem[Amin \latin{et~al.}(2008)Amin, Maier, Balaya, Chen, and
  Lin]{amin2008ionic}
Amin,~R.; Maier,~J.; Balaya,~P.; Chen,~D.; Lin,~C. Ionic and electronic
  transport in single crystalline LiFePO4 grown by optical floating zone
  technique. \emph{Solid State Ionics} \textbf{2008}, \emph{179},
  1683--1687\relax
\mciteBstWouldAddEndPuncttrue
\mciteSetBstMidEndSepPunct{\mcitedefaultmidpunct}
{\mcitedefaultendpunct}{\mcitedefaultseppunct}\relax
\EndOfBibitem
\bibitem[Santhanagopalan \latin{et~al.}(2006)Santhanagopalan, Guo, Ramadass,
  and White]{santhanagopalan2006review}
Santhanagopalan,~S.; Guo,~Q.; Ramadass,~P.; White,~R.~E. Review of models for
  predicting the cycling performance of lithium ion batteries. \emph{Journal of
  power sources} \textbf{2006}, \emph{156}, 620--628\relax
\mciteBstWouldAddEndPuncttrue
\mciteSetBstMidEndSepPunct{\mcitedefaultmidpunct}
{\mcitedefaultendpunct}{\mcitedefaultseppunct}\relax
\EndOfBibitem
\bibitem[Schmalstieg \latin{et~al.}(2018)Schmalstieg, Rahe, Ecker, and
  Sauer]{schmalstieg2018full}
Schmalstieg,~J.; Rahe,~C.; Ecker,~M.; Sauer,~D.~U. Full cell parameterization
  of a high-power lithium-ion battery for a physico-chemical model: Part {I}.
  Physical and electrochemical parameters. \emph{Journal of The Electrochemical
  Society} \textbf{2018}, \emph{165}, A3799\relax
\mciteBstWouldAddEndPuncttrue
\mciteSetBstMidEndSepPunct{\mcitedefaultmidpunct}
{\mcitedefaultendpunct}{\mcitedefaultseppunct}\relax
\EndOfBibitem
\bibitem[Marquis \latin{et~al.}(2019)Marquis, Sulzer, Timms, Please, and
  Chapman]{marquis2019asymptotic}
Marquis,~S.~G.; Sulzer,~V.; Timms,~R.; Please,~C.~P.; Chapman,~S.~J. An
  asymptotic derivation of a single particle model with electrolyte.
  \emph{Journal of The Electrochemical Society} \textbf{2019}, \emph{166},
  A3693\relax
\mciteBstWouldAddEndPuncttrue
\mciteSetBstMidEndSepPunct{\mcitedefaultmidpunct}
{\mcitedefaultendpunct}{\mcitedefaultseppunct}\relax
\EndOfBibitem
\bibitem[Adams and Fournier(2005)Adams, and Fournier]{af05}
Adams,~R.~A.; Fournier,~J.~F. \emph{Sobolev Spaces}; Elsevier, 2005\relax
\mciteBstWouldAddEndPuncttrue
\mciteSetBstMidEndSepPunct{\mcitedefaultmidpunct}
{\mcitedefaultendpunct}{\mcitedefaultseppunct}\relax
\EndOfBibitem
\bibitem[Press \latin{et~al.}(1986)Press, Flanner, Teukolsky, and
  Vetterling]{pftv86}
Press,~W.~H.; Flanner,~B.~P.; Teukolsky,~S.~A.; Vetterling,~W.~T.
  \emph{Numerical Recipes: the Art of Scientific Computations}; Cambridge
  University Press, 1986\relax
\mciteBstWouldAddEndPuncttrue
\mciteSetBstMidEndSepPunct{\mcitedefaultmidpunct}
{\mcitedefaultendpunct}{\mcitedefaultseppunct}\relax
\EndOfBibitem
\bibitem[Devore(2011)]{devore2011probability}
Devore,~J.~L. \emph{Probability and Statistics for Engineering and the
  Sciences}; Cengage learning, 2011\relax
\mciteBstWouldAddEndPuncttrue
\mciteSetBstMidEndSepPunct{\mcitedefaultmidpunct}
{\mcitedefaultendpunct}{\mcitedefaultseppunct}\relax
\EndOfBibitem
\bibitem[Ecker \latin{et~al.}(2015)Ecker, Tran, Dechent, Käbitz, Warnecke, and
  Sauer]{Ecker01012015}
Ecker,~M.; Tran,~T. K.~D.; Dechent,~P.; Käbitz,~S.; Warnecke,~A.; Sauer,~D.~U.
  Parameterization of a Physico-Chemical Model of a Lithium-Ion Battery: I.
  Determination of Parameters. \emph{Journal of The Electrochemical Society}
  \textbf{2015}, \emph{162}, A1836--A1848\relax
\mciteBstWouldAddEndPuncttrue
\mciteSetBstMidEndSepPunct{\mcitedefaultmidpunct}
{\mcitedefaultendpunct}{\mcitedefaultseppunct}\relax
\EndOfBibitem
\bibitem[Park \latin{et~al.}(2010)Park, Zhang, Chung, Less, and
  Sastry]{park2010review}
Park,~M.; Zhang,~X.; Chung,~M.; Less,~G.~B.; Sastry,~A.~M. A review of
  conduction phenomena in Li-ion batteries. \emph{Journal of Power Sources}
  \textbf{2010}, \emph{195}, 7904--7929\relax
\mciteBstWouldAddEndPuncttrue
\mciteSetBstMidEndSepPunct{\mcitedefaultmidpunct}
{\mcitedefaultendpunct}{\mcitedefaultseppunct}\relax
\EndOfBibitem
\bibitem[Jow \latin{et~al.}(2018)Jow, Delp, Allen, Jones, and
  Smart]{jow2018factors}
Jow,~T.~R.; Delp,~S.~A.; Allen,~J.~L.; Jones,~J.-P.; Smart,~M.~C. Factors
  limiting Li+ charge transfer kinetics in Li-ion batteries. \emph{Journal of
  the electrochemical society} \textbf{2018}, \emph{165}, A361\relax
\mciteBstWouldAddEndPuncttrue
\mciteSetBstMidEndSepPunct{\mcitedefaultmidpunct}
{\mcitedefaultendpunct}{\mcitedefaultseppunct}\relax
\EndOfBibitem
\bibitem[Christensen and Newman(2006)Christensen, and
  Newman]{christensen2006stress}
Christensen,~J.; Newman,~J. Stress generation and fracture in lithium insertion
  materials. \emph{Journal of Solid State Electrochemistry} \textbf{2006},
  \emph{10}, 293--319\relax
\mciteBstWouldAddEndPuncttrue
\mciteSetBstMidEndSepPunct{\mcitedefaultmidpunct}
{\mcitedefaultendpunct}{\mcitedefaultseppunct}\relax
\EndOfBibitem
\bibitem[O'Kane \latin{et~al.}(2022)O'Kane, Ai, Madabattula, Alonso-Alvarez,
  Timms, Sulzer, Edge, Wu, Offer, and Marinescu]{o2021lithium}
O'Kane,~S.~E.; Ai,~W.; Madabattula,~G.; Alonso-Alvarez,~D.; Timms,~R.;
  Sulzer,~V.; Edge,~J.~S.; Wu,~B.; Offer,~G.~J.; Marinescu,~M. Lithium-ion
  battery degradation: how to model it. \emph{Physical Chemistry Chemical
  Physics} \textbf{2022}, \emph{24}, 7909--7922\relax
\mciteBstWouldAddEndPuncttrue
\mciteSetBstMidEndSepPunct{\mcitedefaultmidpunct}
{\mcitedefaultendpunct}{\mcitedefaultseppunct}\relax
\EndOfBibitem
\bibitem[Zinth \latin{et~al.}(2014)Zinth, Von~L{\"u}ders, Hofmann, Hattendorff,
  Buchberger, Erhard, Rebelo-Kornmeier, Jossen, and Gilles]{zinth2014lithium}
Zinth,~V.; Von~L{\"u}ders,~C.; Hofmann,~M.; Hattendorff,~J.; Buchberger,~I.;
  Erhard,~S.; Rebelo-Kornmeier,~J.; Jossen,~A.; Gilles,~R. Lithium plating in
  lithium-ion batteries at sub-ambient temperatures investigated by in situ
  neutron diffraction. \emph{Journal of Power Sources} \textbf{2014},
  \emph{271}, 152--159\relax
\mciteBstWouldAddEndPuncttrue
\mciteSetBstMidEndSepPunct{\mcitedefaultmidpunct}
{\mcitedefaultendpunct}{\mcitedefaultseppunct}\relax
\EndOfBibitem
\bibitem[Uhlmann \latin{et~al.}(2015)Uhlmann, Illig, Ender, Schuster, and
  Ivers-Tiff{\'e}e]{uhlmann2015situ}
Uhlmann,~C.; Illig,~J.; Ender,~M.; Schuster,~R.; Ivers-Tiff{\'e}e,~E. In situ
  detection of lithium metal plating on graphite in experimental cells.
  \emph{Journal of Power Sources} \textbf{2015}, \emph{279}, 428--438\relax
\mciteBstWouldAddEndPuncttrue
\mciteSetBstMidEndSepPunct{\mcitedefaultmidpunct}
{\mcitedefaultendpunct}{\mcitedefaultseppunct}\relax
\EndOfBibitem
\bibitem[Wandt \latin{et~al.}(2018)Wandt, Jakes, Granwehr, Eichel, and
  Gasteiger]{wandt2018quantitative}
Wandt,~J.; Jakes,~P.; Granwehr,~J.; Eichel,~R.-A.; Gasteiger,~H.~A.
  Quantitative and time-resolved detection of lithium plating on graphite
  anodes in lithium ion batteries. \emph{Materials Today} \textbf{2018},
  \emph{21}, 231--240\relax
\mciteBstWouldAddEndPuncttrue
\mciteSetBstMidEndSepPunct{\mcitedefaultmidpunct}
{\mcitedefaultendpunct}{\mcitedefaultseppunct}\relax
\EndOfBibitem
\bibitem[Yang \latin{et~al.}(2017)Yang, Leng, Zhang, Ge, and Wang]{yang2017}
Yang,~X.-G.; Leng,~Y.; Zhang,~G.; Ge,~S.; Wang,~C.-Y. Modeling of lithium
  plating induced aging of lithium-ion batteries: Transition from linear to
  nonlinear aging. \emph{Journal of Power Sources} \textbf{2017}, \emph{360},
  28--40\relax
\mciteBstWouldAddEndPuncttrue
\mciteSetBstMidEndSepPunct{\mcitedefaultmidpunct}
{\mcitedefaultendpunct}{\mcitedefaultseppunct}\relax
\EndOfBibitem
\bibitem[Yang \latin{et~al.}(2018)Yang, Ge, Liu, Leng, and Wang]{yang2018look}
Yang,~X.-G.; Ge,~S.; Liu,~T.; Leng,~Y.; Wang,~C.-Y. A look into the voltage
  plateau signal for detection and quantification of lithium plating in
  lithium-ion cells. \emph{Journal of Power Sources} \textbf{2018}, \emph{395},
  251--261\relax
\mciteBstWouldAddEndPuncttrue
\mciteSetBstMidEndSepPunct{\mcitedefaultmidpunct}
{\mcitedefaultendpunct}{\mcitedefaultseppunct}\relax
\EndOfBibitem
\bibitem[O’Kane \latin{et~al.}(2020)O’Kane, Campbell, Marzook, Offer, and
  Marinescu]{o2020physical}
O’Kane,~S.~E.; Campbell,~I.~D.; Marzook,~M.~W.; Offer,~G.~J.; Marinescu,~M.
  Physical origin of the differential voltage minimum associated with lithium
  plating in Li-ion batteries. \emph{Journal of The Electrochemical Society}
  \textbf{2020}, \emph{167}, 090540\relax
\mciteBstWouldAddEndPuncttrue
\mciteSetBstMidEndSepPunct{\mcitedefaultmidpunct}
{\mcitedefaultendpunct}{\mcitedefaultseppunct}\relax
\EndOfBibitem
\bibitem[Sahu and Foster(2023)Sahu, and Foster]{sahu2023continuum}
Sahu,~S.; Foster,~J.~M. A continuum model for lithium plating and dendrite
  formation in lithium-ion batteries: Formulation and validation against
  experiment. \emph{Journal of Energy Storage} \textbf{2023}, \emph{60},
  106516\relax
\mciteBstWouldAddEndPuncttrue
\mciteSetBstMidEndSepPunct{\mcitedefaultmidpunct}
{\mcitedefaultendpunct}{\mcitedefaultseppunct}\relax
\EndOfBibitem
\bibitem[Levi \latin{et~al.}(2003)Levi, Wang, Markevich, Aurbach, and
  Chvoj]{levi2003noteworthy}
Levi,~M.; Wang,~C.; Markevich,~E.; Aurbach,~D.; Chvoj,~Z. Noteworthy
  electroanalytical features of the stage 4 to stage 3 phase transition in
  lithiated graphite. \emph{Journal of Solid State Electrochemistry}
  \textbf{2003}, \emph{8}, 40--43\relax
\mciteBstWouldAddEndPuncttrue
\mciteSetBstMidEndSepPunct{\mcitedefaultmidpunct}
{\mcitedefaultendpunct}{\mcitedefaultseppunct}\relax
\EndOfBibitem
\bibitem[Verbrugge \latin{et~al.}(2022)Verbrugge, Baker, Chen, He, and
  Cai]{verbrugge2022cation}
Verbrugge,~M.~W.; Baker,~D.~R.; Chen,~S.; He,~M.; Cai,~M. Cation Mixing and
  Capacity Loss in Li|| Ni0. 6Mn0. 2Co0. 2O2 Cells: Experimental Investigation
  and Application of the Multi-Site, Multi-Reaction Model. \emph{Frontiers in
  Energy Research} \textbf{2022}, \emph{10}, 671\relax
\mciteBstWouldAddEndPuncttrue
\mciteSetBstMidEndSepPunct{\mcitedefaultmidpunct}
{\mcitedefaultendpunct}{\mcitedefaultseppunct}\relax
\EndOfBibitem
\bibitem[Baker \latin{et~al.}(2019)Baker, Verbrugge, and Gu]{baker2019multi}
Baker,~D.~R.; Verbrugge,~M.~W.; Gu,~W. Multi-species, multi-reaction model for
  porous intercalation electrodes: Part II. Model-experiment comparisons for
  linear-sweep voltammetry of spinel lithium manganese oxide electrodes.
  \emph{Journal of The Electrochemical Society} \textbf{2019}, \emph{166},
  A521\relax
\mciteBstWouldAddEndPuncttrue
\mciteSetBstMidEndSepPunct{\mcitedefaultmidpunct}
{\mcitedefaultendpunct}{\mcitedefaultseppunct}\relax
\EndOfBibitem
\bibitem[Shockley and Read~Jr(1952)Shockley, and
  Read~Jr]{shockley1952statistics}
Shockley,~W.; Read~Jr,~W. Statistics of the recombinations of holes and
  electrons. \emph{Physical review} \textbf{1952}, \emph{87}, 835\relax
\mciteBstWouldAddEndPuncttrue
\mciteSetBstMidEndSepPunct{\mcitedefaultmidpunct}
{\mcitedefaultendpunct}{\mcitedefaultseppunct}\relax
\EndOfBibitem
\bibitem[Hall(1952)]{hall1952electron}
Hall,~R.~N. Electron-hole recombination in germanium. \emph{Physical review}
  \textbf{1952}, \emph{87}, 387\relax
\mciteBstWouldAddEndPuncttrue
\mciteSetBstMidEndSepPunct{\mcitedefaultmidpunct}
{\mcitedefaultendpunct}{\mcitedefaultseppunct}\relax
\EndOfBibitem
\bibitem[Bousse \latin{et~al.}(1983)Bousse, De~Rooij, and
  Bergveld]{bousse1983influence}
Bousse,~L.; De~Rooij,~N.~F.; Bergveld,~P. The influence of counter-ion
  adsorption on the $\psi$0/pH characteristics of insulator surfaces.
  \emph{Surface science} \textbf{1983}, \emph{135}, 479--496\relax
\mciteBstWouldAddEndPuncttrue
\mciteSetBstMidEndSepPunct{\mcitedefaultmidpunct}
{\mcitedefaultendpunct}{\mcitedefaultseppunct}\relax
\EndOfBibitem
\bibitem[Protas \latin{et~al.}(2004)Protas, Bewley, and
  Hagen]{ProtasBewleyHagen2004}
Protas,~B.; Bewley,~T.~R.; Hagen,~G. A Computational Framework for the
  Regularization of Adjoint Analysis in Multiscale PDE Systems. \emph{J.
  Comput. Phys.} \textbf{2004}, \emph{195}, 49--89\relax
\mciteBstWouldAddEndPuncttrue
\mciteSetBstMidEndSepPunct{\mcitedefaultmidpunct}
{\mcitedefaultendpunct}{\mcitedefaultseppunct}\relax
\EndOfBibitem
\bibitem[Mistry \latin{et~al.}(2021)Mistry, Verma, Sripad, Ciez, Sulzer,
  Brosa~Planella, Timms, Zhang, Kurchin, Dechent, Li, Greenbank, Ahmad,
  Krishnamurthy, Fenton~Jr., Tenny, Patel, Juarez~Robles, Gasper, Colclasure,
  Baskin, Scown, Subramanian, Khoo, Allu, Howey, DeCaluwe, Roberts, and
  Viswanathan]{mistry2021minimal}
Mistry,~A.; Verma,~A.; Sripad,~S.; Ciez,~R.; Sulzer,~V.; Brosa~Planella,~F.;
  Timms,~R.; Zhang,~Y.; Kurchin,~R.; Dechent,~P. \latin{et~al.}  A Minimal
  Information Set To Enable Verifiable Theoretical Battery Research. \emph{ACS
  Energy Letters} \textbf{2021}, \emph{6}, 3831--3835\relax
\mciteBstWouldAddEndPuncttrue
\mciteSetBstMidEndSepPunct{\mcitedefaultmidpunct}
{\mcitedefaultendpunct}{\mcitedefaultseppunct}\relax
\EndOfBibitem
\end{mcitethebibliography}

\providecommand{\latin}[1]{#1}
\makeatletter
\providecommand{\doi}
  {\begingroup\let\do\@makeother\dospecials
  \catcode`\{=1 \catcode`\}=2 \doi@aux}
\providecommand{\doi@aux}[1]{\endgroup\texttt{#1}}
\makeatother
\providecommand*\mcitethebibliography{\thebibliography}
\csname @ifundefined\endcsname{endmcitethebibliography}
  {\let\endmcitethebibliography\endthebibliography}{}

% \newpage

% \section{TOC graphic}

% \begin{figure}
% \includegraphics[height=4.75cm,width=8cm]{iotas_all_cases_June12_2023.eps}
% \end{figure}

\end{document}